\documentclass[journal]{vgtc}              % review (journal style)
\onlineid{8006}

\vgtccategory{Research}

\usepackage[utf8]{inputenc}
\usepackage[svgnames]{xcolor}
\usepackage{tikz}
\usepackage{amssymb}
\usepackage{amsmath}
\usepackage{mathtools}
\usepackage[disable]{todonotes}
\usepackage[linesnumbered,noend,ruled]{algorithm2e}
\usepackage{enumitem}
\usepackage{appendix}
\usepackage{amssymb}% http://ctan.org/pkg/amssymb
\usepackage{pifont}% http://ctan.org/pkg/pifont
\usepackage{amsthm}
\usepackage{float}
\usetikzlibrary{arrows.meta,arrows,calc}

\DeclareMathOperator{\troot}{root}

\DeclareMathOperator{\depth}{depth}
\DeclareMathOperator{\cost}{c}

\newcommand{\odedist}{\delta_{\text{0}}}
\newcommand{\edist}{\delta_{\text{E}}}
\newcommand{\laedist}{\delta_h}
\DeclareMathOperator{\sces}{SCES}

\colorlet{revision}{black}

\vgtcpapertype{algorithm/technique}

\title{Accelerating Computation of Stable Merge Tree Edit Distances\\using Parameterized Heuristics}

\author{
  \authororcid{Florian Wetzels}{0000-0002-5526-7138},
  \authororcid{Heike Leitte}{0000-0002-7112-2190}, and
  \authororcid{Christoph Garth}{0000-0003-1669-8549}
}

\authorfooter{
  \item
  	Florian Wetzels, Heike Leitte and Christoph Garth are with University of Kaiserslautern-Landau.\\
  	E-mail: wetzels@cs.uni-kl.de, \{leitte\,$|$\,garth\}@rptu.de.
}

\abstract{In this paper, we present a novel heuristic algorithm for the stable but NP-complete deformation-based edit distance on merge trees.
Our key contribution is the introduction of a user-controlled look-ahead parameter that allows to trade off accuracy and computational cost.
We achieve a fixed parameter tractable running time that is polynomial in the size of the input but exponential in the look-ahead value.
This extension unlocks the potential of the deformation-based edit distance in handling saddle swaps, while maintaining feasible computation times.
Experimental results demonstrate the computational efficiency and effectiveness of this approach in handling specific perturbations.%
}


\teaser{
  \centering
  \includegraphics[width=0.95\linewidth]{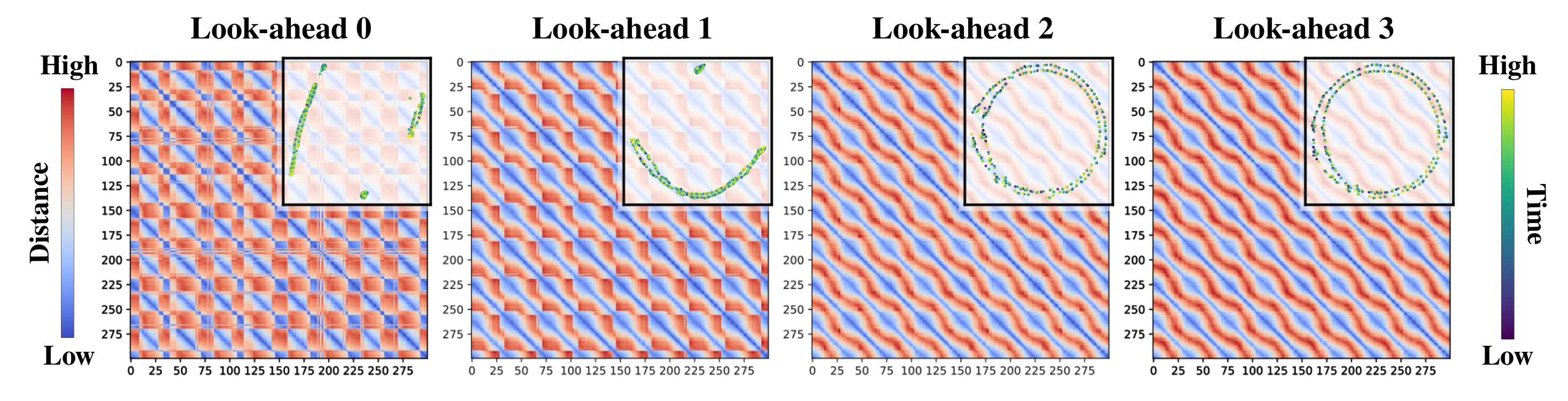}
  \caption{Four distance matrices \textcolor{revision}{(entry $(i,j)$ is the distance between $i$-th and $j$-th tree, shown as heatmaps)} of the vortex street dataset (axes ordered by time) for different look-ahead values 
  together with corresponding t-SNE embeddings. The color of the t-SNE points encodes time. The periodic behavior becomes more apparent in the smoother matrices with higher look-ahead values. The t-SNE embeddings also show cleaner periodicity: instead of separate clusters we get a smooth circle.}
  \label{fig:teaser}
}

\graphicspath{{figs/}{figures/}{pictures/}{images/}{./}} % where to search for the images

\usepackage{tabu}                      % only used for the table example
\usepackage{booktabs}                  % only used for the table example
\usepackage{lipsum}                    % used to generate placeholder text
\usepackage{mwe}                       % used to generate placeholder figures

\usepackage{mathptmx}                  % use matching math font

\begin{document}

%%%%%%%%%%%%%%%%%%%%%%%%%%%%%%%%%%%%%%%%%%%%%%%%%%%%%%%%%%%%%%%%
%%%%%%%%%%%%%%%%%%%%%% START OF THE PAPER %%%%%%%%%%%%%%%%%%%%%%
%%%%%%%%%%%%%%%%%%%%%%%%%%%%%%%%%%%%%%%%%%%%%%%%%%%%%%%%%%%%%%%%

%% The ``\maketitle'' command must be the first command after the
%% ``\begin{document}'' command. It prepares and prints the title block.
%% the only exception to this rule is the \firstsection command
\firstsection{Introduction}

\maketitle

\setlength{\abovedisplayskip}{3pt}
\setlength{\belowdisplayskip}{3pt}

\textcolor{revision}{Merge trees are a popular abstraction tool for scalar fields in scientific visualization and data analysis.
They represent equivalence classes of contours, their connectivity, and  critical points of the scalar field, see~\cite{DBLP:conf/soda/CarrSA00,DBLP:conf/focs/EdelsbrunnerLZ00,DBLP:conf/ppopp/MorozovW13} for a detailed introduction.
\emph{Edit distances} on merge trees have been successfully used} in clustering tasks~\cite{DBLP:journals/tvcg/PontVDT22,DBLP:journals/tvcg/WetzelsPTG24}, outlier or periodicity detection~\cite{DBLP:journals/tvcg/SridharamurthyM20,wetzels2022branch}, finding self-similarity~\cite{DBLP:journals/cgf/SaikiaSW14,ThomasN13,DBLP:journals/tvcg/SridharamurthyN23}, and summarization of ensembles~\cite{DBLP:journals/tvcg/PontVDT22,DBLP:journals/tvcg/PontVT23,DBLP:journals/cgf/LohfinkWLWG20,DBLP:journals/tvcg/WetzelsPTG24} or time series~\cite{DBLP:journals/tvcg/PontVDT22,DBLP:journals/tvcg/PontVT23,DBLP:journals/tvcg/WetzelsPTG24,DBLP:conf/visualization/LohfinkGWVG21}.
A variety of edit distances have been defined for merge trees in recent years~\cite{DBLP:journals/cgf/SaikiaSW14,DBLP:journals/tvcg/SridharamurthyM20,DBLP:journals/cgf/LohfinkWLWG20,wetzels2022branch,DBLP:journals/tvcg/PontVT23,wetzels2022path}, which differ significantly in complexity and expressiveness.

An important property for distances 
% metrics 
on topological abstractions is stability, which indicates that small changes in the scalar field only induce small distances between their abstractions.
Most edit distances for merge trees with efficient runtimes are inherently unstable.
They are usually unable to handle specific perturbations of the data, the so-called horizontal instabilities
\textcolor{revision}{which are caused by saddles changing their ancestor-relation}.
However, the deformation-based edit distance defined by Wetzels et al.~\cite{wetzels2022path,taming} has been experimentally shown to be stable even in the presence of such saddle swaps~\cite{taming}, though a theoretic result on stability is still work-in-progress.
Unfortunately, Wetzels et al.\ also show NP-hardness of the deformation-based edit distance.
The provided mixed integer linear programming 
\textcolor{revision}{(MIP)} 
implementation becomes infeasible for merge trees of more than 30 vertices.
Thus, the only available (to the best of our knowledge) stable edit-distance for merge trees remains infeasible to compute in many practical situations.
A constrained variant of the deformation-based edit distance (prohibiting insertions and deletions of inner nodes), is called the path mapping distance~\cite{wetzels2022path}.
It has a quartic time algorithm, but is not stable against saddle swaps.
However, it can handle basically any other form of instability, including so-called vertical instabilities \textcolor{revision}{(see~Section~\ref{sec:background})}.

In this paper, we introduce a heuristic algorithm for the deformation-based edit distance, which can handle horizontal instabilities up to a certain degree.
As 
stated above, the reason for other distances being unstable are saddle swaps.
Efficiently computable edit distances cannot handle those, 
due to their constraints on the allowed edit operations.
We avoid this issue by extending the path mapping distance  
by a user-controlled look-ahead parameter.
This parameter determines how many consecutive saddle swaps are allowed in the distance.
The running time is polynomial in the size of the input trees, but exponential in the look-ahead value.
This yields a running time strongly resembling fixed-parameter tractability (FPT), though the parameter is user-chosen, not depending on the input.
The user can thereby balance between running time and accuracy of the distance through a single integer parameter.
The resulting distance measure ranges from the original path mapping distance (look-ahead~$0$) to the unconstrained deformation-based edit distance (look-ahead set to the depth of the trees).

We evaluate the improved stability qualitatively and quantitatively, and study the runtime performance in comparison to other distance metrics.
Our experiments show vastly improved stability over the path mapping distance, even for moderate look-ahead values.
In terms of runtime, the unconstrained deformation-based edit distance is outperformed by several orders of magnitude.
Figure~\ref{fig:teaser} illustrates the improved stability with increasing look-ahead parameter on an established periodic dataset.

\smallskip\par\noindent
\textbf{Contribution.}
Our specific contributions are: (1) a novel heuristic algorithm for the deformation-based edit distance; (2) the corresponding edit distance model describing what kind of edit sequences are captured by the heuristic solution; (3) an experimental evaluation of the stability and \textcolor{revision}{time} complexity on established datasets. 
\textcolor{revision}{We also provide an open source implementation in the topology toolkit (TTK).}

\section{Related Work}

Topological abstractions are a key tool in scientific visualization in general (an introduction can be found in the survey by Heine et al.~\cite{heine16}), as well as the comparison of scalar fields (see the survey by Yan et al.~\cite{surveyComparison2021}).
In this paper, we consider \emph{edit distances} between \emph{merge trees}, see~\cite{treeEditSurvey} for a survey on edit distances between general rooted trees.

Several edit distances have been defined for merge trees: the merge tree edit distance by Sridharamurthy et al.~\cite{DBLP:journals/tvcg/SridharamurthyM20}, the merge tree Wasserstein distance by Pont et al.~\cite{DBLP:journals/tvcg/PontVDT22}, the extended branch decomposition graph method by Saikia et al.~\cite{DBLP:journals/cgf/SaikiaSW14}, as well as deformation-based edit distances and the branch mapping distance by Wetzels et al.~\cite{wetzels2022branch,wetzels2022path,taming}.
Other work on merge tree edit distances includes more advanced analysis and visualization methods based on the underlying edit mappings~\cite{DBLP:journals/cgf/LohfinkWLWG20,DBLP:journals/tvcg/PontVDT22,DBLP:journals/tvcg/SridharamurthyN23}.
Many other distances (either in the form of edit distances or not) exist for various topological descriptors: merge trees~\cite{BeketayevYMWH14,morozov14,DBLP:journals/tvcg/BollenTL23}, persistence diagrams~\cite{interleaving_distance,Cohen-Steiner2007,edelsbrunner09}, or Reeb graphs~\cite{DBLP:conf/3dor/BauerFL16,DBLP:journals/dcg/FabioL16}.
Alternative distance measures combine topological and geometrical similarity~\cite{YanWMGW20,Yan_geometry_aware,intrinsicMTdistance,ThomasN13,DBLP:conf/apvis/NarayananTN15}.

The focus of this paper is to improve stability, while retaining feasible runtimes.
Most merge tree edit distances lack formal stability results, whereas such results exist for distances on other topological descriptors~\cite{Cohen-Steiner2007,DBLP:journals/entcs/FabioL12}.
However, some experimental observations for merge trees exist.
Branch decomposition-based methods are known to be susceptible against vertical instabilities~\cite{DBLP:journals/cgf/SaikiaSW14,wetzels2022branch}.
In contrast, vertical stability was observed experimentally for the branch decomposition-independent methods by Wetzels et al.~\cite{wetzels2022branch,wetzels2022path}, including the path mapping distance.
The unconstrained deformation-based edit distance~\cite{taming,wetzels2022path} has been experimentally shown to be fully stable, even for horizontal instabilities.
A technique to improve stability for basically any merge tree distance is the so-called $\epsilon$-preprocessing~\cite{DBLP:journals/tvcg/SridharamurthyM20,DBLP:journals/tvcg/PontVDT22} which simplifies the trees prior to the distance computation.

\section{Background}
\label{sec:background}

The \emph{merge tree} of a scalar field 
\textcolor{revision}{$f: \mathbb{X} \rightarrow \mathbb{R}$}
is a tree structure representing the connectivity of
superlevel sets.
The nodes are the critical points of
$f$
whereas the edges represent classes of superlevel sets.
A detailed introduction can be found in \cite{DBLP:conf/soda/CarrSA00,DBLP:conf/focs/EdelsbrunnerLZ00,DBLP:conf/ppopp/MorozovW13}.
For ease of arguments, we only consider \emph{abstract split trees}, which capture the essential properties of split trees of compact $d$-manifolds with $d>1$, and revisit the definitions given in~\cite{wetzels2022branch,wetzels2022path,taming}.
All arguments can be adapted for join trees.

A rooted, unordered tree $T$ is a connected, directed graph, without undirected cycles and  with a unique sink, its root $\troot(T)$.
We denote the sets of its vertices/edges by $V(T)$/$E(T)$ \textcolor{revision}{and the empty tree by $\bot$}.
Given an edge $(c,p) \in E(T)$, we call $c$ the child of $p$ and $p$ the parent of $c$ (we use parent pointers).
Given a node $v \in V(T)$, we denote its children by $C_T(v)$, its parent by $P_T(v)$.
The degree $\deg_T(v) \coloneqq |C_T(v)|$ of a node $v$ in a tree $T$ is the number of its children, the degree $\deg(T)$ of $T$ is the maximum degree of a node in $T$.

A path of length $k$ in a rooted tree $T$ is a sequence of vertices $p=v_1 \dots v_k \in V(T)^k$ with $(v_{i},v_{i-1}) \in E(T)$ for all $2 \leq i \leq k$.
Note the strict root-to-leaf direction of the vertex sequence.
For many operators, we use paths as edge sets or vertex sets, i.e.\ we write $v \in p$, $e \in p$, $p \subseteq V$ or $p \subseteq E$ for $v \in V(T), e \in E(T), V \subseteq V(T), E \subseteq E(T)$.
We say that a path $p$ connects nodes $v$ and $u$ if $p = v \dots u$.
In this case, $u$ is a descendant of $v$ and $v$ an ancestor of $u$.
We use the same notation for edges, too.
The depth of a node, denoted $\depth_T(v)$, is the length of the unique path $\troot(T) \dots v$.
For a node $v \in V(T)$, the subtree rooted in $v$ is denoted by $T[v]$.

Merge trees are labeled trees - nodes inherit the scalar values of the domain's critical points - and can be interpreted as edge-labeled trees: labels represent the length
of the scalar range of the edge.
An unordered, rooted tree $T$ with edge labels $\ell:V(T) \rightarrow \mathbb{R}_{>0}$ is an \emph{Abstract Merge Tree} if the following properties hold:
\begin{itemize}
 \setlength\itemsep{0.2em}
 \setlength\topsep{0.2em}
 \item The root node has degree one, $\deg_T( \troot (T) ) = 1$
 \item All inner nodes have a degree of at least two,\\ $\deg_T(v) \neq 1$ for all $v \in V(T)$ with $v \neq \troot (T)$
\end{itemize}
\noindent
The edge label function $\ell$ of a merge tree $T$ can be lifted to paths as follows: $\ell(v_1 \dots v_k) = \sum_{2 \leq i \leq k} \ell((v_i,v_{i-1}))$.

Since the root of an abstract merge tree always has degree one and inner nodes do not, subtrees rooted in nodes are not abstract merge trees themselves.
Hence, we also identify subtrees by root \emph{edges}:
the subtree rooted in $(c,p) \in E(T)$, denoted by $T[(c,p)]$ or $T[c,p]$, is $T[c]$ extended by the node $p$ and the edge $(c,p)$.
A subtree rooted in a path $p \dots c$ can be defined analogously: it contains the tree $T[c]$, the node $p$ and the ``imaginary'' edge $(c,p)$.

Next, we also recap the definition of the \emph{deformation-based edit distance} as given in~\cite{wetzels2022path,taming}.
It is a merge tree-tailored adaptation of the well-established edit distance on unordered trees by Zhang~\cite{DBLP:journals/ipl/ZhangSS92}.
It uses the following three edit operations to transform one tree into another:
the \emph{relabel} operation changes the length 
of an edge;
the \emph{deletion} contracts an edge, i.e.\ for a node $v$ with children $c_0...c_k$ and parent $v$, we remove $(v,p)$ from the tree and change each edge $(c_i,v)$ to $(c_i,p)$;
the \emph{insertion} adds a new edge to the tree, i.e.\ it is the inverse operation to a deletion.
In contrast to classic tree edit distances, if a deletion leaves a node of degree one, we prune it, merging its two incident edges and adding their lengths.
I.e.\ if \emph{after} contracting an edge $(v,p)$ there is only one remaining child $c$ of $p$, we remove $p$ from the tree and replace the edges $(c,p),(p,p')$ by $(c,p')$ with $\ell((c,p')) \coloneqq \ell((c,p)) + \ell((p,p'))$.
Illustrations of these edit operations can be found in~\cite{wetzels2022path,taming}.
We use the terms edge contractions, edge collapse and and edge deletion interchangeably as well as inverse edge contractions and insertions.

If a sequence of edit operations $s$ transforms a tree $T_1$ into $T_2$, we denote this by $T_1 \xrightarrow{\scriptscriptstyle s} T_2$.
The edit operations are assigned a cost function which compares edges.
Since we use $\mathbb{R}_{>0}$ as the label set for abstract merge trees with edge labels, we use~$0$ as the blank symbol, i.e.\ the label of an empty or non-existing edge.
We use the euclidean distance on $\mathbb{R}_{\geq 0}$: $ \cost(l_1,l_2) = |l_1-l_2| $ for all $l_1,l_2 \in \mathbb{R}_{\geq 0}$.
The cost of an edit sequence is then the sum of all edit operation costs: $ \cost(s_1 \dots s_k) = \sum_{1 \leq i \leq k} |\cost(s_i)| $, where $\cost(s_i)$ is $\cost(l_1,l_2)$ if and edge is relabeled from $l_1$ to $l_2$, $\cost(0,l_2)$ in case of an insertion and $\cost(l_1,0)$ in case of a deletion.
The deformation-based edit distance between two trees $T_1,T_2$ is defined as the cost of a cost-optimal sequence transforming $T_1$ into $T_2$:
\[ \edist(T_1,T_2) = \min\{ \cost(s) \mid T_1 \xrightarrow{\scriptscriptstyle s} T_2 \}. \]

The \emph{path mapping distance}~\cite{wetzels2022path} is a constrained variant of the deformation-based edit distance where insertions and deletions are only allowed on edges to leaves.
It is based on the so-called one-degree edit distance by Selkow~\cite{DBLP:journals/ipl/Selkow77}.
We call $s$ a one-degree edit sequence, if all insertions and deletions happen on edges $(c,p)$ where $c$ is a leaf.
The path mapping distance is defined as
\[ \odedist(T_1,T_2) = \min \{ \cost(s) \mid T_1 \xrightarrow{\scriptscriptstyle s} T_2,\ s \text{ is one-degree} \}. \]

The unconstrained deformation-based edit distance $\edist$ is known to be NP-complete~\cite{taming}.
In contrast, the path mapping distance $\odedist$ is polynomial time computable~\cite{wetzels2022path}: there is a recursive formulation of $\odedist$ which can be computed in $\mathcal{O}(n^4)$ time for bounded degree trees through a dynamic programming approach.
The algorithm does not compute an optimal edit sequence directly, but rather the optimal \emph{path mapping}.
One-degree edit sequences and path mappings can be converted into each other in a straight-forward manner~\cite{wetzels2022path}.
See~\cite{wetzels2022path} for a detailed introduction and definition of path mappings.

The recursive algorithm for $\odedist$ on two abstract merge trees with root edges, $T_1[(n_1,p_1)],T_2[(n_2,p_2)]$, works as follows.
See App.~\ref{sec:app_rec}, suppl.\ material, or~\cite{wetzels2022path} for more details and pseudocode.
If $T_1[(n_1,p_1)]$ is empty, then we just return the the sum of all edge deletion costs in $T_2[(n_2,p_2)]$.
If $T_1[(n_1,p_1)]$ is just one edge, then we find the best matching path in $T_2[(n_2,p_2)]$ and delete all other edges.
If $T_2[(n_2,p_2)]$ is empty or just one edge, the cases are analogous.
Otherwise, if both $T_1[(n_1,p_1)],T_2[(n_2,p_2)]$ are non-trivial,
we return the minimum cost one of the following three cases:
we delete all but one children of $n_1$ and their subtrees and continue recursively;
we delete all but one children of $n_2$ and their subtrees and continue recursively;
or we match $(n_1,p_1)$ to $(n_2,p_2)$ and find the optimal assignment between the subtrees below $n_1,n_2$.
Formally, we return the minimum of the following three terms:
\begin{itemize}
 \setlength\itemsep{0.2em}
    \item \textcolor{revision}{$\min_{c_2 \in C_{T_2}(n_2)} \{ \sum_{c_2' \in C_{T_2}(n_2)} \odedist(\bot,T_2[c_2',n_2])$\\
    \centerline{$+\odedist(T_1[n_1,p_1],T_2[c_2,p_2]) - \odedist(\bot,T_2[c_2,n_2]) \} $;}}
  \item \textcolor{revision}{$\min_{c_1 \in C_{T_1}(n_1)} \{  \sum_{c_1' \in C_{T_1}(n_1)} \odedist(T_1[c_1',n_1],\bot)$\\
  \centerline{$ + \odedist(T_1[c_1,p_1],T_2[n_2,p_2]) - \odedist(T_1[c_1,n_1],\bot) \} $;}}
  \item $\min\{ \cost(m,\laedist) \mid m \in PM(\bigcup_{x \in C_{T_1}(n_1)} (x,n_1),\bigcup_{x \in C_{T_2}(n_2)} (x,n_2)) \}$.
\end{itemize}
Here, $PM(C_1,C_2)$ denotes the set of partial functions between two sets of root edges $C_1 \subseteq E(T_1),C_2 \subseteq E(T_2)$.
Given a partial mapping $m \in PM(C_1,C_2)$, and a distance function $\delta$ between abstract merge trees, the cost of $m$ under $\delta$ is defined as
$ \cost(m,\delta) \coloneqq \sum_{e_1 \notin m^{-1}(C_2)} \delta(T_1[e_1],\bot) + \sum_{e_2 \notin m(C_1)} \delta(\bot,T_2[e_2]) + \sum_{(e_1,e_2) \in m} \delta(T_1[e_1],T_2[e_2]) $.

To solve the partial mapping instance, the TTK implementation allows to choose between the Hungarian method~\cite{munkres} and the auction solver~\cite{DBLP:journals/mp/Bertsekas81} (it does so for any edit distance).
Since the auction solver only computes a heuristic solution, we only discuss theoretic runtime bounds for the Hungarian algorithm.
Note that the original algorithm for the constrained edit distance by Zhang~\cite{DBLP:journals/algorithmica/Zhang96} solved it through a minimum cost maximum flow instance, which has a better asymptotic runtime.
However, for ease of implementation, we chose to stick with the methods already present in TTK.

\textcolor{revision}{Other (conceptually different) merge tree edit distances that we consider are the \emph{merge tree Wasserstein distance} by Pont et al.\ (see~\cite{DBLP:journals/tvcg/PontVDT22} for details), denoted $\delta_W$, and the original \emph{merge tree edit distance} by Sridharamurthy et al.\ (see~\cite{DBLP:journals/tvcg/SridharamurthyM20} for details), denoted $\delta_S$.}

The focus of this paper is stability of merge tree distances.
Instabilities in merge trees are often categorized into vertical and horizontal ones.
Upon small perturbations of the data, the former describe changes in the persistence-based branching hierarchy, the latter saddle swaps (features change their nesting in the tree structure, like between $T_1,T_2$ in Figure~\ref{fig:swaps_lookahead}).
In~\cite{taming}, Wetzels et al.\ describe these phenomena in more detail with illustrations and provide experimental evidence that the unconstrained deformation-based edit distance is stable against both types of perturbations, while the path mapping distance only handles vertical instability.

\section{Method}
\label{sec:method}

In this section, we describe our novel heuristic algorithm for the deformation-based edit distance.
We extend the polynomial-time algorithm for the path mapping distance~\cite{wetzels2022path} summarized above by a user-chosen look-ahead parameter.
After describing the algorithm, our implementation and several optimizations we applied, we analyze important properties of the new method.
We discuss which edit sequences are represented by the new mapping model, the runtime complexity and monotonicity of the look-ahead parameter.
We denote the distance computed by the adapted recursion with look-ahead~$h$ by $\laedist$.

\subsection{Lookahead Algorithm}
\label{sec:alg}

Recall the recursive path mapping algorithm as given in Section~\ref{sec:background}.
It computes the optimal path mapping between all pairs of subtrees $T_1[(n_1,p_1)],T_2[(n_2,p_2)]$ recursively.
A case where an inner edge is deleted without its full subtree is never considered.
Therefore, we extend the recursion for each such pair by allowing to collapse a set of edges below 
\textcolor{revision}{(in terms of descendant)} 
the nodes $n_1$ or $n_2$ and then computing the optimal mapping between the resulting trees plus the costs of collapsing the edges.
In particular, we add a fourth case for two non-trivial trees.

Like for the matching case, we first relabel 
$p_1 \dots n_1$ to $p_2 \dots n_2$, meaning they are again matched.
However, instead of simply picking the optimal mapping between the subtrees directly below $n_1$ and $n_2$, we consider additional cases where specific edges below $n_1$ and $n_2$ are deleted.
Then, we compute the optimal mapping between the subtrees below $n_1,n_2$ \emph{after} those deletions.
We only consider deletions of consecutive \textcolor{revision}{(meaning incident)} edges, since other deletions are covered in the recursive cases.
Furthermore, we restrict to deletions of edges within a fixed but user-chosen look-ahead around the nodes $n_1,n_2$, to avoid exponential overhead.
See App.~\ref{sec:app_rec}, suppl.\ material, for exemplary illustrations of the extended recursive cases.

The added case returns the optimal costs of such deletions and corresponding mappings of subtrees.
In particular, we simply add the term $\texttt{OptCollapse}(T_1,T_2,n_1,n_2,h)$.
The core procedure here is to iterate all possible ways of collapsing consecutive edges below $n_1$ and $n_2$.
Thus, we now define the sets of edges to consider.
Given a tree $T$, an edge set $E \subseteq E(T)$ is strongly connected if for all $e_1,e_2 \in E$ there is a path $p$ connecting $e_1$ and $e_2$ such that $p \subseteq E$.
Given a tree $T$, a node $v \in V(T)$ and a look-ahead number $h$, an edge set $E \subseteq E(T)$ is within look-ahead $h$ of $v$, if there is an edge $(x,y) \in E$ with $y = v$ as well as for all $(x,y) \in E$ it holds that $\depth_T(y)>\depth_T(v)$ and $\depth_T(x)-\depth_T(v)<=h$.
For an edge set, we call the minimally required look-ahead from a node $v$ the distance between the node and the set, denoted
\[ d_T(v,E) \coloneqq \min\{ h \mid E \text{ is within look-ahead } h \text{ of } v \}. \]
For a node $v$ of a merge tree, we consider all strongly connected edge sets within look-ahead range of $v$.
We denote these sets by
\[ \sces_h(v,T) \coloneqq \{ E \mid d_T(v,E) \leq h, E \text{ is strongly connected} \}. \]

Given a tree $T$, a node $v \in V(T)$ and an edge set $E \in \sces_h(v,T)$, we define the boundary nodes of $E$ to be those nodes of which not all children are contained in $E$:
\[ B_T(E) \coloneqq \{ x \in V(T) \mid \exists c,p \in V(T): (x,p) \in E, (c,x) \in E(T) \setminus E \}. \]
We call the nodes outside of but connected to $E$ the leaves of $E$:
\[ L_T(E) \coloneqq \{ x \in V(T) \mid \exists y \in B_T(E): (x,y) \in E(T), (x,y) \not \in E \}. \]

The function \texttt{OptCollapse} (Algorithm~\ref{alg:collapse}) computes the optimal set of edges to collapse by trying each combination of $\sces_h(n_1,T_1)$ and $\sces_h(n_2,T_2)$.
For each combination $(E_1,E_2) \in \sces_h(n_1,T_1) \times \sces_h(n_2,T_2)$, the induced costs are the sum of the cost of contracting each edge in $E_1,E_2$ and the costs of all recursive mappings.
The subtrees to consider are those rooted in $L_{T_1}(E_1)$ and $L_{T_2}(E_2)$.
Like for the relabel case in the original recursion, we find the optimal mapping between the subtrees through the Hungarian method based on the results from the recursive cases.
Algorithm~\ref{alg:collapse} returns the optimal costs.

\begin{algorithm}[!b]
\caption{Optimal edge collapse}
\label{alg:collapse}
\SetKwFunction{optCollapse}{OptCollapse}
\SetKwProg{Fn}{Function}{:}{}
\DontPrintSemicolon
\Fn{\optCollapse{$T_1,T_2,n_1,n_2,h$}}{
 $d_{\text{opt}} = \infty$\;
 \ForEach{$E_1 \in \sces_h(n_1,T_1)$}{
 $C_1 = \bigcup_{x \in L_T(E_1)} (x,P_T(x))$\;
 \ForEach{$E_2 \in \sces_h(n_2,T_2)$}{
  $C_2 = \bigcup_{x \in L_T(E_2)} (x,P_T(x))$\;
  $d = \sum_{e \in E_1} \cost(e,\bot) + \sum_{e \in E_2} \cost(\bot,e)$\\
  $\;\;\;\;\; + \min\{ \cost(m,\laedist) \mid m\in PM(C_1,C_2) \}$\;
  $d_{\text{opt}} = \min(d_{\text{opt}},d)$\;
 }
 }
 \Return $d_{\text{opt}}$\;
}
\end{algorithm}

\subsection{Implementation and Optimizations}
\label{sec:impl}

We implemented the algorithm described above based on the existing 
implementation of 
the path mapping distance in TTK~\cite{DBLP:journals/tvcg/TiernyFLGM18}.
Our code is provided in suppl.\ material
(see App.~\ref{sec:app_code} for a description) 
and we plan to integrate it properly into TTK upon publication.

The existing code solves the path mapping recursion using a 
bottom-up 
dynamic programming approach.
It iterates all combinations of nodes $n_1 \in V(T_1),n_2 \in V(T_2)$ with ancestors 
$p_1 \in V(T_1),p_2 \in V(T_2)$ in a bottom-up fashion.
It then computes the optimal recursive case based on previously computed results for the subtrees.
For bounded-degree, 
the runtime is $\Theta(|T_1|\cdot |T_2| \cdot \depth(T_1) \cdot \depth(T_2))$.
For unbounded-degree trees, we get an additional factor of $\min(\deg(T_1),\deg(T_2)) \cdot \deg(T_1) \cdot \deg(T_2)$ for solving the optimal assignment instance.
For readability, we will assume bounded-degree trees in our runtime analysis, which is a reasonable assumption for merge trees.

We extend the existing algorithm by generating, for each node combination, the sets $\sces_h(n_1,T_1)$ and $\sces_h(n_2,T_2)$.
We do so through a
worklist-based procedure.
We maintain a set of candidate sets, which we extend by choosing for the next edge whether to collapse it or not.
Both cases yield new candidates, which are put back into the worklist.
Once every edge in the look-ahead area is handled, we are done.
Details are given in App.~\ref{sec:app_alg}, suppl.\ material.

This gives us a complete running time of 
$\mathcal{O}(|T_1|\cdot |T_2| \cdot \depth(T_1) \cdot \depth(T_2) \cdot f(h))$ where $f$ is some exponential function (details in Section~\ref{sec:complexity}).
For a fixed parameter~$h$, the runtime stays polynomial.
\textcolor{revision}{This is similar to a concept usually referred to as} fixed parameter tractability (FPT, see~\cite{DBLP:series/txtcs/FlumG06} for an introduction).
Nonetheless, practical runtimes become infeasible rather quickly in a naive implementation.
The bottleneck seems to be the high number of optimal assignment instances (line~8, Algorithm~\ref{alg:collapse}).
Hence, reducing the amount of assignment instances is essential for practical applicability. We prioritized this approach over optimizing the implementation of the generation procedure itself.
In particular, we focused on reducing the size of the generated sets $\sces_h(n_1,T_1)$ and $\sces_h(n_2,T_2)$, as well as the number of node combinations to perform Algorithm~\ref{alg:collapse} on.
We now go through the different optimizations we applied.

\begin{figure*}
    \centering
    \captionsetup[subfigure]{aboveskip=0pt,belowskip=0pt}
    
    \raisebox{9pt}{\includegraphics[angle=90,width=0.012\linewidth]{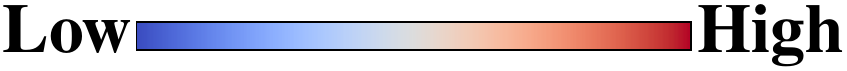}}
    \begin{subfigure}[t]{0.2\linewidth}
    \centering
    \includegraphics[width=0.9\linewidth]{figures/tosca/dm_mted}
    \caption{Merge Tree Edit Distance}
    \end{subfigure}
    \begin{subfigure}[t]{0.2\linewidth}
    \centering
    \includegraphics[width=0.9\linewidth]{figures/tosca/dm_pmd0}
    \caption{Path Mapping Distance}
    \end{subfigure}
    \begin{subfigure}[t]{0.2\linewidth}
    \centering
    \includegraphics[width=0.9\linewidth]{figures/tosca/dm_uted}
    \caption{Unconstr.\ Deformation Distance}
    \end{subfigure}
    \begin{subfigure}[t]{0.2\linewidth}
    \centering
    \includegraphics[width=0.9\linewidth]{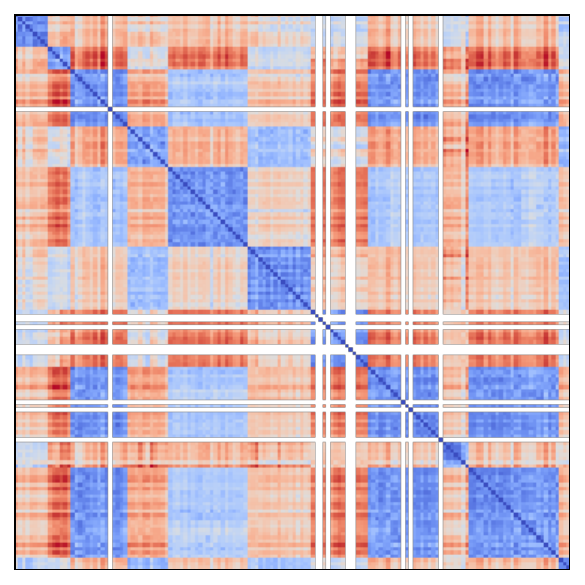}
    \caption{Unconstr.\ Deformation Distance}
    \end{subfigure}

    \raisebox{9pt}{\includegraphics[angle=90,width=0.012\linewidth]{figures/legend_dist.pdf}}
    \begin{subfigure}[t]{0.2\linewidth}
    \centering
    \includegraphics[width=0.9\linewidth]{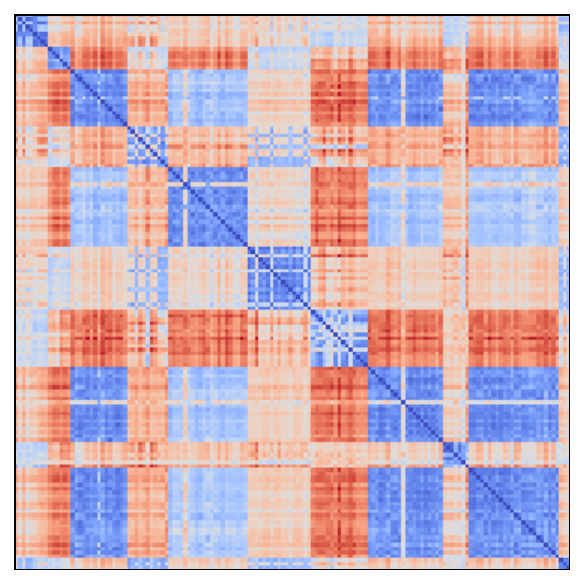}
    \caption{Look-ahead 1}
    \end{subfigure}
    \begin{subfigure}[t]{0.2\linewidth}
    \centering
    \includegraphics[width=0.9\linewidth]{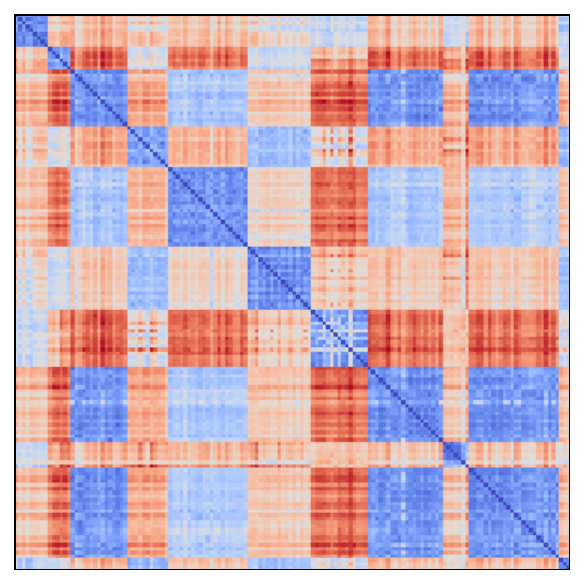}
    \caption{Look-ahead 2}
    \end{subfigure}
    \begin{subfigure}[t]{0.2\linewidth}
    \centering
    \includegraphics[width=0.9\linewidth]{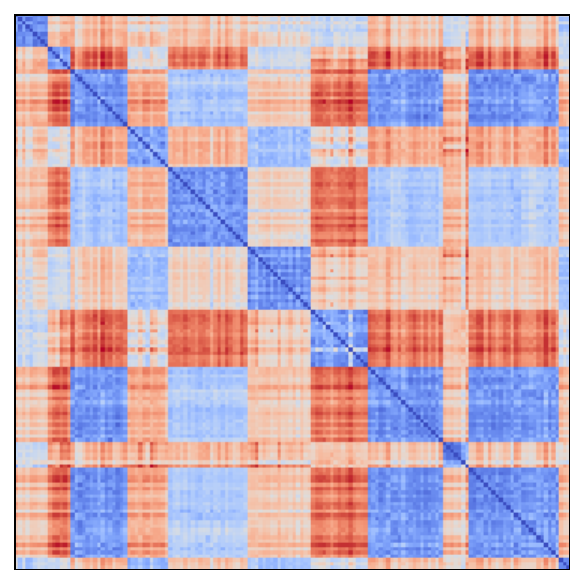}
    \caption{Look-ahead 4}
    \end{subfigure}
    \begin{subfigure}[t]{0.2\linewidth}
    \centering
    \includegraphics[width=0.9\linewidth]{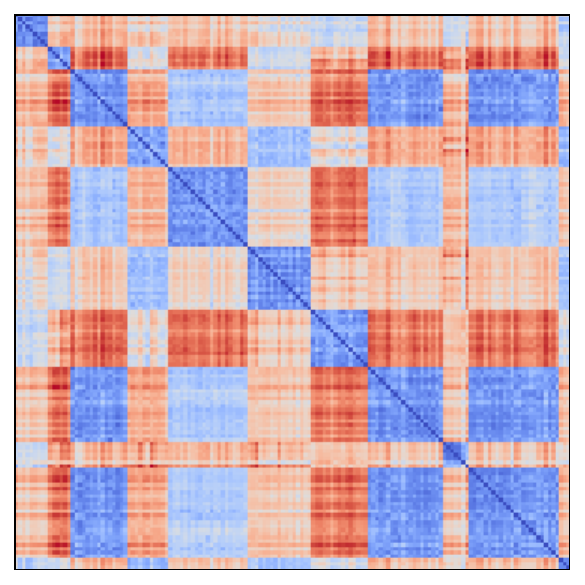}
    \caption{Look-ahead 8}
    \end{subfigure}
    \caption{Distance matrices (axes ordered by shape \textcolor{revision}{and pose lexicographically}) on the TOSCA ensemble using metrics from previous work and the look-ahead heuristic. The matrices in (a,b) exhibit many instabilities, whereas (c,d) show clean clusters. Note that for (c) the ensemble had to be filtered for small merge trees to achieve feasible runtimes. Thus, (d) shows the full size matrix with filtered rows/columns omitted. In (e-h), the higher the look-ahead value, the less instabilities can be seen. The matrix in (h) is barely distinguishable from (c/d).}
    \label{fig:tosca_mat1}
\end{figure*}

\smallskip\par\noindent
\textbf{Dropping Leaves.}
To reduce the size of $\sces_h(n_1,T_1)$ and $\sces_h(n_2,T_2)$, we  remove any set containing leaf edges.
Consider an edge $(x,y)$ within look-ahead $h$ of $n_1$.
If $x$ is a leaf of $T_1$, then collapsing $(x,y)$ does not contribute a subtree to $L_{T_1}(E_1)$.
Instead, we can skip the deletion of $(x,y)$, making $y$ part of the boundary and $x$ part of the leaves of $E_1$.
The deletion of $(x,y)$ would be equivalently covered through the optimal matching in line~8 of Algorithm~\ref{alg:collapse}, since there mapping a subtree to the empty tree is always a valid option.
The same can be done for
$n_2$.
This reduces the size of the generated sets and thus the number of executions of \mbox{lines 7-9} in Algorithm~\ref{alg:collapse}.

\smallskip\par\noindent
\textbf{Dynamic Programming.}
Next, we consider optimizations to reduce the number of node tuples for which we have to compute the sets $\sces_h(n_1,T_1)$ and $\sces_h(n_2,T_2)$, i.e.\ the number of executions of the fourth case and Algorithm~\ref{alg:collapse}.
For a node tuple $(n_1,p_1,n_2,p_2)$, the computation of the optimal edge sets to collapse and their associated cost is independent of $p_1$ and $p_2$.
Thus, it suffices to compute them only once for each pair $(n_1,n_2)$.
It is possible to do so by again utilizing memoization.
After computing $\texttt{OptCollapse}(n_1,n_2,h)$, we can access the stored result for each $4$-tuple $(n_1,p_1,n_2,p_2)$.
This reduces the running time from $n^4 \cdot f(h)$ to $n^4 + n^2 \cdot f(h)$.
Due to the exponential nature of $f(h)$, this optimization had the largest impact out of all those we applied.
Indeed, it is the only one changing the asymptotic worst-case running time.

\smallskip\par\noindent
\textbf{Upper Bound Optimization.}
The last algorithmic optimization 
reduces the number of node tuples as well as the size of the edge sets.

In many cases, the subtrees represented by the tuple differ significantly in their size, such that they will never be mapped in an optimal solution.
It does not make sense to compare them in such a case.
We compute an upper bound for the total distance with look-ahead~$h$ in a preprocessing step: the normal path mapping distance $\odedist(T_1,T_2)$.
For a given tuple $(n_1,n_2)$, we then compute a trivial lower bound for any deformation-based edit distance, the difference $d_P(n_1,n_2)$
\textcolor{revision}{between the total sum of edge weights}
of $T_1[n_1]$ and $T_2[n_2]$.
If $d_P(n_1,n_2)>\odedist(T_1,T_2)$ holds, the optimal mapping cannot match the subtrees rooted in $n_1$ and $n_2$ onto each other.
We can therefore skip such tuples.
This strategy is closely related to the optimization in the original MIP solution for the unconstrained deformation-based edit distance, see~\cite{taming}.

Furthermore, the same reasoning can be applied for each pair of edge sets $E_1,E_2$.
We can skip the optimal assignment computation between $C_1,C_2$ \mbox{(lines 7-9)}, if a local lower bound is already higher than the upper bound given by $\odedist$.
We used the following lower bound: the cost of collapsing $E_1$ and $E_2$, together with the difference between the total 
\textcolor{revision}{sums of edge weights}
of $C_1$ and $C_2$.

\smallskip\par\noindent
\textbf{Parallelization \& Optimal Assignment Algorithm.}
To achieve low runtimes, we made heavy use of parallelization.
TTK already enables parallel computation of distance matrices through OpenMP~\cite{openmp08,dagum1998openmp}.
We 
slightly 
adapted the existing code as follows.
Previously, each OMP task was assigned a full line of the distance matrix, which is sufficient in most cases.
However, our distance algorithm itself is not parallelized (in the setting of distance matrices, this is the inferior method anyway).
Thus, in some cases, we need more aggressive parallelization \textcolor{revision}{on the matrix level} to achieve high occupancy.
Therefore, we assigned a single pair of trees to each task.

Furthermore, one of the main hotspots of the computation is the optimal assignment problem for each case of collapsed edges.
In contrast to the original path mapping distance, we have a large number of elements (subtrees) to assign.
Therefore, we use the approximate auction solver\cite{DBLP:journals/mp/Bertsekas81}, which is already implemented in TTK, instead of an optimal solution, e.g.\ through the Hungarian method~\cite{munkres} or hard coded cases.
However, initial experiments showed that using different solvers for different sizes of the assignment problem could be advantageous.
This should be studied in future work.

\subsection{Analysis}
\label{sec:complexity}

Next, we briefly study theoretic properties of the defined distance.

\smallskip\par\noindent
\textbf{Edit Operation Model.}
We first discuss what kind of edit distance is represented by the mappings considered by~$\laedist$, in terms of which edit operations or sequences are allowed.
We restrict to an intuitive description.
Consider the fourth recursive case added in the heuristic method, on subtrees $T_1[n_1 \dots p_1],T_2[n_2 \dots p_2]$.
In essence, we compute the optimal path mapping between trees $T_1',T_2'$ derived from $T_1[n_1 \dots p_1],T_2[n_2 \dots p_2]$ by contracting strongly connected edge sets $E_1,E_2$ below $n_1,n_2$.
Since we can do such contractions in recursive cases as well, the computed optimal edit sequence can contain those \textcolor{revision}{directly} below any mapped path.
Since we can always pull deletions to the front of an edit sequnce and insertions to the back~\cite{wetzels2022path}, we can reformulate the distance as follows.
An edit sequence is of look-ahead~$h$, if it is of the form $S_1S_2S_3$ where $S_1$ contains only deletions and never contracts more than~$h$ consecutive edges, $S_2$ is a one-degree sequence, and $S_3$ contains only insertions and never inserts consectuive paths of more than~$h$ edges.
The distance $\laedist(T_1,T_2)$ then is the cost of an optimal sequence $s$ of look-ahead~$h$ with $T_1 \xrightarrow{\scriptscriptstyle s} T_2$.
Note that if we set the look-ahead parameter to the depth of the tree, arbitrary deletions are allowed.
Thus, we obtain the unconstrained distance.
Figure~\ref{fig:swaps_lookahead} shows an example of how the look-ahead heuristic can or cannot handle certain kinds of saddle swaps.

\begin{figure}[!b]
  \centering
  \resizebox{\linewidth}{!}{
  \begin{tikzpicture}[scale=0.4]%xscale=0.55,yscale=0.4]
  %\draw[help lines, color=gray!30, dashed] (-3.5,-2.5) grid (3.5+7.2,15.5);
  \pgfmathsetmacro{\trianglew}{0.55*2}
  \pgfmathsetmacro{\triangleh}{1*2}

  \begin{scope}[shift={(-20,0)}]

  \node[draw=none,fill=none,circle] at (-4, -1) (dummy) {};
  \node[draw=none,fill=none,circle] at (11, -1) (dummy) {};
  \node[draw=none,fill=none,circle] at (-4, 15) (dummy) {};
  \node[draw=none,fill=none,circle] at (11, 15) (dummy) {};
  
  \node[draw,circle,fill=gray!100] at (0, 0) (root_1) {};
  \node[draw,circle,fill=gray!100] at (0, 2) (s1_1) {};
  \node[draw,circle,fill=gray!100] at (-0.75, 3) (s2_1) {};
  \node[draw,circle,fill=gray!100] at (-1.5, 4) (s3_1) {};
  
  \node[] at (-3.5, 14.33) (m1_1l) {$A$};
  \node[inner sep = 0, outer sep = 0, minimum size = 0] at (-3.5, 13) (m1_1) {};
  \draw[draw=black] (-3.5,13) -- (-3.5 -\trianglew,13 +\triangleh) -- (-3.5 + \trianglew,13 +\triangleh) -- cycle;
  
  \node[] at (3.5, 14.33) (m2_1l) {$B$};
  \node[inner sep = 0, outer sep = 0, minimum size = 0] at (3.5, 13) (m2_1) {};
  \draw[draw=black] (3.5,13) -- (3.5 -\trianglew,13 +\triangleh) -- (3.5 + \trianglew,13 +\triangleh) -- cycle;
  
  \node[] at (-1, 12.33) (m3_1l) {$C$};
  \node[inner sep = 0, outer sep = 0, minimum size = 0] at (-1, 11) (m3_1) {};
  \draw[draw=black] (-1,11) -- (-1 -\trianglew,11 +\triangleh) -- (0-1+ \trianglew,11 +\triangleh) -- cycle;
  
  \node[] at (0.5, 10.33) (m4_1l) {$D$};
  \node[inner sep = 0, outer sep = 0, minimum size = 0] at (0.5, 9) (m4_1) {};
  \draw[draw=black] (0.5,9) -- (0.5 -\trianglew,9 +\triangleh) -- (0.5 + \trianglew,9 +\triangleh) -- cycle;

  \draw[gray,very thick] (root_1) -- (s1_1);
  \draw[gray,very thick] (s1_1) -- (m2_1);
  \draw[gray,very thick] (s1_1) -- (s2_1);
  \draw[gray,very thick] (s2_1) -- (s3_1);
  \draw[gray,very thick] (s2_1) -- (m4_1);
  \draw[gray,very thick] (s3_1) -- (m1_1);
  \draw[gray,very thick] (s3_1) -- (m3_1);

  \node[] at (0, -2) (t1_l) {\huge$T_1$};

  \end{scope}

  \begin{scope}[shift={(-10,0)}]

  \node[draw=none,fill=none,circle] at (-4, -1) (dummy) {};
  \node[draw=none,fill=none,circle] at (11, -1) (dummy) {};
  \node[draw=none,fill=none,circle] at (-4, 15) (dummy) {};
  \node[draw=none,fill=none,circle] at (11, 15) (dummy) {};
  
  \node[draw,circle,fill=gray!100] at (0, 0) (root_2) {};
  \node[draw,circle,fill=gray!100] at (0, 2) (s1_2) {};
  \node[draw,circle,fill=gray!100] at (-1.5, 3) (s3_2) {};
  
  \node[] at (-3.5, 14.33) (m1_2l) {$A$};
  \node[inner sep = 0, outer sep = 0, minimum size = 0] at (-3.5, 13) (m1_2) {};
  \draw[draw=black] (-3.5,13) -- (-3.5 -\trianglew,13 +\triangleh) -- (-3.5 + \trianglew,13 +\triangleh) -- cycle;
  
  \node[] at (3.5, 14.33) (m2_2l) {$B$};
  \node[inner sep = 0, outer sep = 0, minimum size = 0] at (3.5, 13) (m2_2) {};
  \draw[draw=black] (3.5,13) -- (3.5 -\trianglew,13 +\triangleh) -- (3.5 + \trianglew,13 +\triangleh) -- cycle;
  
  \node[] at (-1, 11.33) (m3_2l) {$C$};
  \node[inner sep = 0, outer sep = 0, minimum size = 0] at (-1, 10) (m3_2) {};
  \draw[draw=black] (-1,10) -- (-1 -\trianglew,10 +\triangleh) -- (0-1+ \trianglew,10 +\triangleh) -- cycle;
  
  \node[] at (0.75, 9.33) (m4_2l) {$D$};
  \node[inner sep = 0, outer sep = 0, minimum size = 0] at (0.75, 8) (m4_2) {};
  \draw[draw=black] (0.75,8) -- (0.75 -\trianglew,8 +\triangleh) -- (0.75 + \trianglew,8 +\triangleh) -- cycle;

  \draw[gray,very thick] (root_2) -- (s1_2);
  \draw[gray,very thick] (s1_2) -- (m2_2);
  \draw[gray,very thick] (s1_2) -- (s3_2);
  \draw[gray,very thick] (s1_2) -- (m4_2);
  \draw[gray,very thick] (s3_2) -- (m1_2);
  \draw[gray,very thick] (s3_2) -- (m3_2);

  \node[] at (0, -2) (t2_l) {\huge$T_1'$};

  \end{scope}

  \begin{scope}[shift={(0,0)}]
    
  \node[draw,circle,fill=gray!100] at (0, 0) (root_3) {};
  \node[draw,circle,fill=gray!100] at (0, 2) (s1_3) {};
  
  \node[] at (-3.5, 14.33) (m1_3l) {$A$};
  \node[inner sep = 0, outer sep = 0, minimum size = 0] at (-3.5, 13) (m1_3) {};
  \draw[draw=black] (-3.5,13) -- (-3.5 -\trianglew,13 +\triangleh) -- (-3.5 + \trianglew,13 +\triangleh) -- cycle;
  
  \node[] at (3.5, 14.33) (m2_3l) {$B$};
  \node[inner sep = 0, outer sep = 0, minimum size = 0] at (3.5, 13) (m2_3) {};
  \draw[draw=black] (3.5,13) -- (3.5 -\trianglew,13 +\triangleh) -- (3.5 + \trianglew,13 +\triangleh) -- cycle;
  
  \node[] at (1, 12.33-1.5) (m3_3l) {$C$};
  \node[inner sep = 0, outer sep = 0, minimum size = 0] at (1, 11-1.5) (m3_3) {};
  \draw[draw=black] (1,11-1.5) -- (1 -\trianglew,11-1.5 +\triangleh) -- (1+ \trianglew,11-1.5 +\triangleh) -- cycle;
  
  \node[] at (-0.5, 10.33-1.5) (m4_3l) {$D$};
  \node[inner sep = 0, outer sep = 0, minimum size = 0] at (-0.5, 9-1.5) (m4_3) {};
  \draw[draw=black] (-0.5,9-1.5) -- (-0.5 -\trianglew,9-1.5 +\triangleh) -- (-0.5 + \trianglew,9-1.5 +\triangleh) -- cycle;

  \draw[gray,very thick] (root_3) -- (s1_3);
  \draw[gray,very thick] (s1_3) -- (m1_3);
  \draw[gray,very thick] (s1_3) -- (m4_3);
  \draw[gray,very thick] (s1_3) -- (m2_3);
  \draw[gray,very thick] (s1_3) -- (m3_3);

  \node[] at (0, -2) (t3_l) {\huge$T_{1/2}''$};

  \end{scope}

  \begin{scope}[shift={(10,0)}]
    
  \node[draw,circle,fill=gray!100] at (0, 0) (root_4) {};
  \node[draw,circle,fill=gray!100] at (0, 2) (s1_4) {};
  \node[draw,circle,fill=gray!100] at (1.5, 3) (s3_4) {};
  
  \node[] at (-3.5, 14.33) (m1_4l) {$A$};
  \node[inner sep = 0, outer sep = 0, minimum size = 0] at (-3.5, 13) (m1_4) {};
  \draw[draw=black] (-3.5,13) -- (-3.5 -\trianglew,13 +\triangleh) -- (-3.5 + \trianglew,13 +\triangleh) -- cycle;
  
  \node[] at (3.5, 14.33) (m2_4l) {$B$};
  \node[inner sep = 0, outer sep = 0, minimum size = 0] at (3.5, 13) (m2_4) {};
  \draw[draw=black] (3.5,13) -- (3.5 -\trianglew,13 +\triangleh) -- (3.5 + \trianglew,13 +\triangleh) -- cycle;
  
  \node[] at (1, 11.33) (m3_4l) {$C$};
  \node[inner sep = 0, outer sep = 0, minimum size = 0] at (1, 10) (m3_4) {};
  \draw[draw=black] (1,10) -- (1 -\trianglew,10 +\triangleh) -- (1+ \trianglew,10 +\triangleh) -- cycle;
  
  \node[] at (-0.75, 9.33) (m4_4l) {$D$};
  \node[inner sep = 0, outer sep = 0, minimum size = 0] at (-0.75, 8) (m4_4) {};
  \draw[draw=black] (-0.75,8) -- (-0.75 -\trianglew,8 +\triangleh) -- (-0.75 + \trianglew,8 +\triangleh) -- cycle;

  \draw[gray,very thick] (root_4) -- (s1_4);
  \draw[gray,very thick] (s1_4) -- (m1_4);
  \draw[gray,very thick] (s1_4) -- (s3_4);
  \draw[gray,very thick] (s1_4) -- (m4_4);
  \draw[gray,very thick] (s3_4) -- (m2_4);
  \draw[gray,very thick] (s3_4) -- (m3_4);

  \node[] at (0, -2) (t4_l) {\huge$T_2'$};

  \end{scope}

  \begin{scope}[shift={(20,0)}]
    
  \node[draw,circle,fill=gray!100] at (0, 0) (root_5) {};
  \node[draw,circle,fill=gray!100] at (0, 2) (s1_5) {};
  \node[draw,circle,fill=gray!100] at (0.75, 3) (s2_5) {};
  \node[draw,circle,fill=gray!100] at (1.5, 4) (s3_5) {};
  
  \node[] at (-3.5, 14.33) (m1_5l) {$A$};
  \node[inner sep = 0, outer sep = 0, minimum size = 0] at (-3.5, 13) (m1_5) {};
  \draw[draw=black] (-3.5,13) -- (-3.5 -\trianglew,13 +\triangleh) -- (-3.5 + \trianglew,13 +\triangleh) -- cycle;
  
  \node[] at (3.5, 14.33) (m2_5l) {$B$};
  \node[inner sep = 0, outer sep = 0, minimum size = 0] at (3.5, 13) (m2_5) {};
  \draw[draw=black] (3.5,13) -- (3.5 -\trianglew,13 +\triangleh) -- (3.5 + \trianglew,13 +\triangleh) -- cycle;
  
  \node[] at (1, 12.33) (m3_5l) {$C$};
  \node[inner sep = 0, outer sep = 0, minimum size = 0] at (1, 11) (m3_5) {};
  \draw[draw=black] (1,11) -- (1 -\trianglew,11 +\triangleh) -- (1+ \trianglew,11 +\triangleh) -- cycle;
  
  \node[] at (-0.5, 10.33) (m4_5l) {$D$};
  \node[inner sep = 0, outer sep = 0, minimum size = 0] at (-0.5, 9) (m4_5) {};
  \draw[draw=black] (-0.5,9) -- (-0.5 -\trianglew,9 +\triangleh) -- (-0.5 + \trianglew,9 +\triangleh) -- cycle;

  \draw[gray,very thick] (root_5) -- (s1_5);
  \draw[gray,very thick] (s1_5) -- (m1_5);
  \draw[gray,very thick] (s1_5) -- (s2_5);
  \draw[gray,very thick] (s2_5) -- (s3_5);
  \draw[gray,very thick] (s2_5) -- (m4_5);
  \draw[gray,very thick] (s3_5) -- (m2_5);
  \draw[gray,very thick] (s3_5) -- (m3_5);

  \node[] at (0, -2) (t5_l) {\huge$T_2$};

  \end{scope}
  
  \draw[ultra thick,-{Stealth[length=6mm, width=4mm]},dotted,violet] (s1_1) to[bend right] node[midway,below] {LA$1$-contraction} (s1_2);
  \draw[ultra thick,-{Stealth[length=6mm, width=4mm]},dotted,violet] (s1_2) to[bend right] node[midway,below] {LA$2$-contraction} (s1_3);
  \draw[ultra thick,-{Stealth[length=6mm, width=4mm]},dotted,violet] (s1_5) to[bend left] node[midway,below] {LA$1$-contraction} (s1_4);
  \draw[ultra thick,-{Stealth[length=6mm, width=4mm]},dotted,violet] (s1_4) to[bend left] node[midway,below] {LA$2$-contraction} (s1_3);
  
  \end{tikzpicture}
  }
  \caption{Example for the look-ahead method: $T_1$ and $T_2$ differ by a horizontal instability. Thus, using a constrained edit distance, correctly matching any two of the features $A,B,C,D$, requires deletion of the other two. In contrast, if we collapse the saddle of feature $D$ ($T_1'$ and $T_2'$), we can always match $D$ and two other features, but have to delete one. If we also collapse the saddle of $C$ ($T_1''$ and $T_2''$), we can match all four features. A look-ahead of~$1$ allows mappings between $T_1',T_2'$, a look-ahead of~$2$ also mappings between $T_1'',T_2''$.}
  \label{fig:swaps_lookahead}
\end{figure}

\smallskip\par\noindent
\textbf{Complexity.} As described above, the running time of the adapted recursion with look-ahead parameter $h$ using the dynamic programming optimization can be described as
\begin{equation}
\label{eq:runtime}
    \mathcal{O}(|T_1|\cdot |T_2| \cdot \depth(T_1) \cdot \depth(T_2) + |T_1|\cdot |T_2| \cdot f(h))
\end{equation}
wher $f$ is an exponential function.
A detailed discussion on the composition of $f$ is given at the end of this section.

For a fixed look-ahead value, this running time is polynomial, thus FPT.
Note however, that it does not precisely fit the typical definitions of FPT algorithms or parameterized complexity.
The look-ahead parameter is user-controlled and changes the distance measure, so it is not a property of the input or output.
We believe though that this term best captures the essence of our approach.

Note that, in practice, runtimes depend heavily on the given problem instance.
The given worst-case upper bound is only a very rough estimate.
There are two main factors for this.
First, the shapes of the two trees have a large influence on the effectiveness of the leave dropping optimization.
Second, the upper bound optimization strongly depends on the path mapping distance and how it relates to the actual look-ahead distance, since a low path mapping distance will lead to more node tuples being dropped.
We deem a proper analysis (either formal or empirical) beyond the scope of this paper.

We conclude this section with a more detailed description of the factor $f(h)$ in the runtime upper bound.
We can express $f(h)$ as $f'(h)^2 \cdot \textcolor{revision}{d^{3h}}$ with $f'$ describing the maximum size of the look-ahead region, i.e.\ $\max \{ |\sces_h(n,T_i)| \mid 1 \leq i \leq 2, n \in V(T_i) \}$, and $d = \max(\deg(T_1),\deg(T_2))$ (assumed to be bounded).
To understand the two factors, note the following.
We have to perform a cubic Hungarian method for each pair of collapsed sets.
For each of the two trees, there are $f'(h)$ many sets.
Each set has at most $d^h$ (a very rough upper bound) many leaves.

\begin{figure}[!b]
  \centering
  \includegraphics[width=0.82\linewidth]{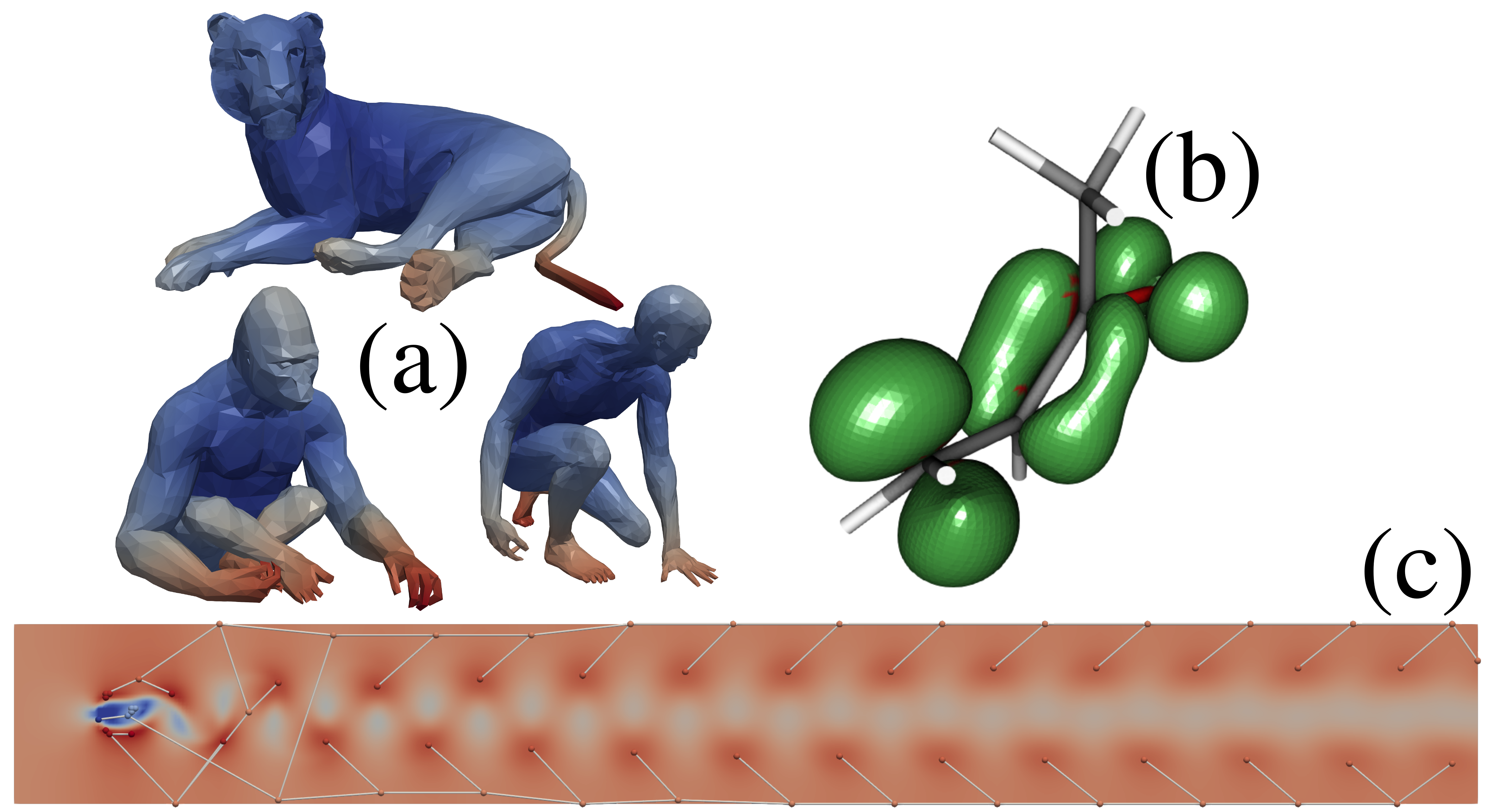}
  \caption{Example members of (a) the TOSCA ensemble, (b) the MVK time series and (c) the vortex street (merge tree embedded).}
  \label{fig:tosca_shapes}
\end{figure}

Next, consider the size of $f'(h)$.
Each strongly connected edge set corresponds to a specific tree shape below a given node $n \in V(T_1) \cup V(T_2)$.
In the worst case, the considered region below the given node is a complete $d$-ary tree, meaning that any tree $d$-ary tree shape is possible.
Thus, in the worst case, $f'$ describes the number of $d$-ary trees of depth $h$, where $d = \max(\deg(T_1),\deg(T_2))$.
To the best of our knowledge, there is no closed form for this function.

\begin{figure*}
    \centering
    \captionsetup[subfigure]{aboveskip=0pt,belowskip=0pt}

    \raisebox{9pt}{\includegraphics[angle=90,width=0.012\linewidth]{figures/legend_dist.pdf}}
    % \hspace{0.013\linewidth}
    \begin{subfigure}[t]{0.2\linewidth}
    \centering
    \includegraphics[width=0.9\linewidth]{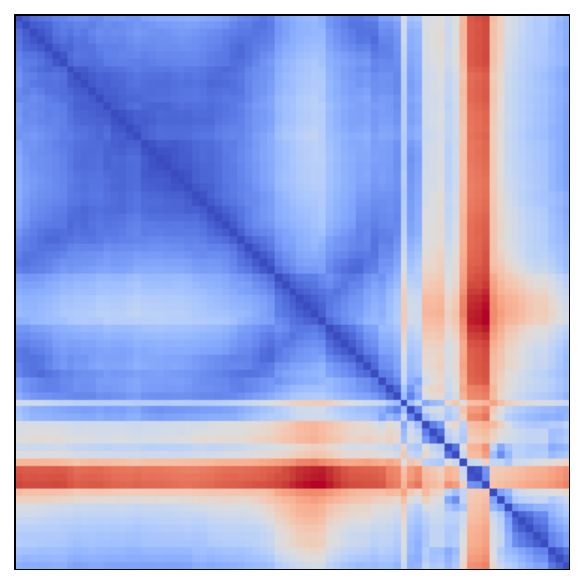}
    \caption{Path Mapping Dist., hole field}
    \end{subfigure}
    \begin{subfigure}[t]{0.2\linewidth}
    \centering
    \includegraphics[width=0.9\linewidth]{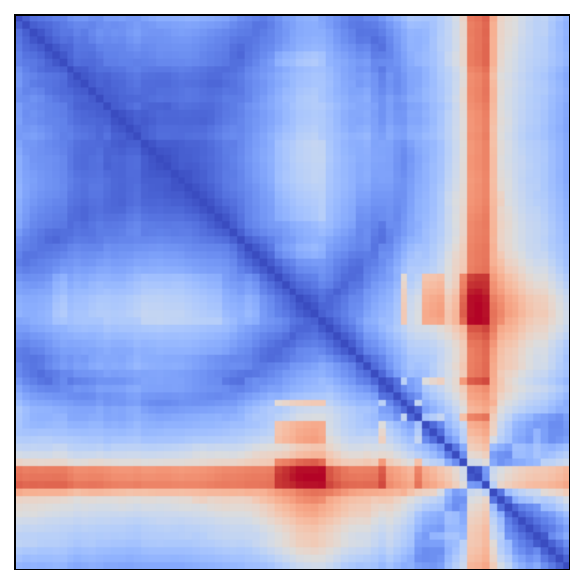}
    \caption{Look-ahead 1, hole field}
    \end{subfigure}
    \begin{subfigure}[t]{0.2\linewidth}
    \centering
    \includegraphics[width=0.9\linewidth]{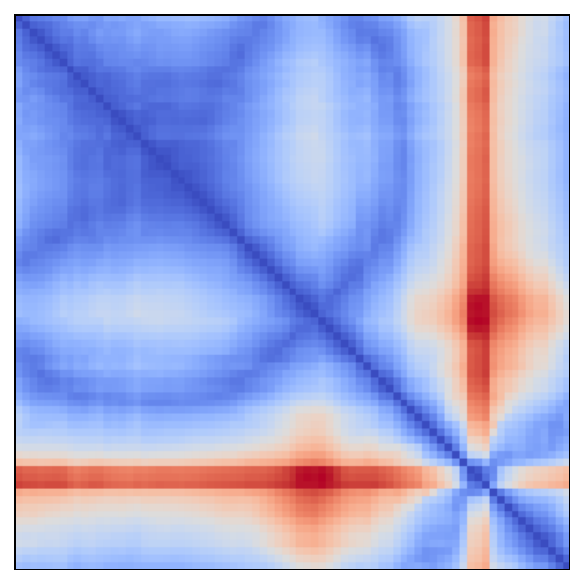}
    \caption{Look-ahead 4, hole field}
    \end{subfigure}
    \begin{subfigure}[t]{0.2\linewidth}
    \centering
    \includegraphics[width=0.9\linewidth]{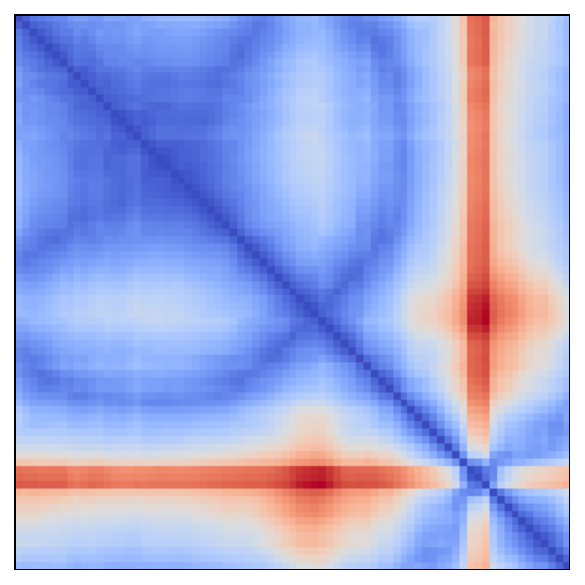}
    \caption{Unconstr.\ Deform. Dist., hole field}
    \end{subfigure}
    
    \raisebox{9pt}{\includegraphics[angle=90,width=0.012\linewidth]{figures/legend_dist.pdf}}
    \begin{subfigure}[t]{0.2\linewidth}
    \centering
    \includegraphics[width=0.9\linewidth]{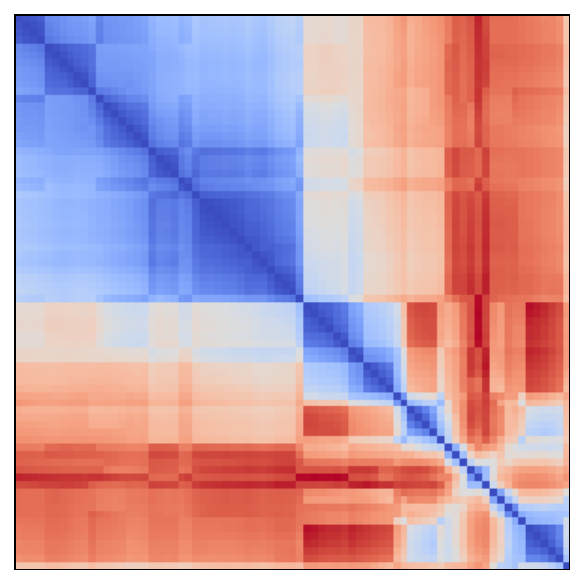}
    \caption{Path Mapping Dist., particle field}
    \end{subfigure}
    \begin{subfigure}[t]{0.2\linewidth}
    \centering
    \includegraphics[width=0.9\linewidth]{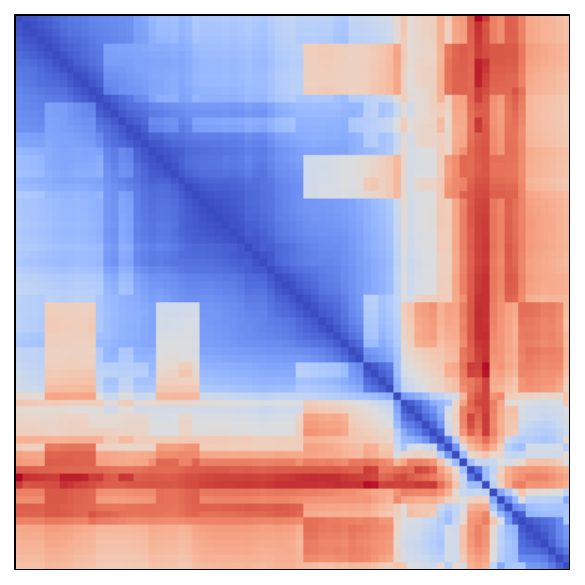}
    \caption{Look-ahead 1, particle field}
    \end{subfigure}
    \begin{subfigure}[t]{0.2\linewidth}
    \centering
    \includegraphics[width=0.9\linewidth]{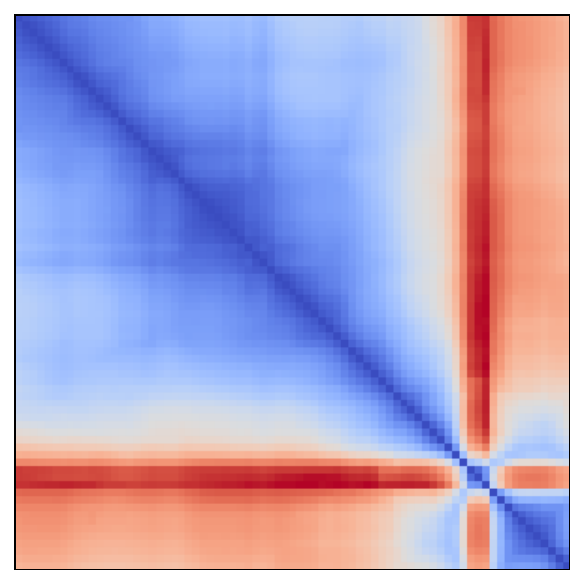}
    \caption{Look-ahead 6, particle field}
    \end{subfigure}
    \begin{subfigure}[t]{0.2\linewidth}
    \centering
    \includegraphics[width=0.9\linewidth]{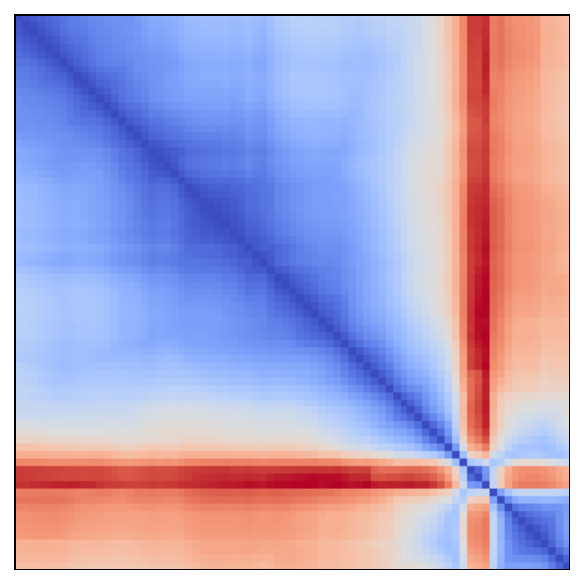}
    \caption{Unconstr.\ Deform. Dist., particle field}
    \end{subfigure}
    \caption{
    A selection of distance matrices (axes ordered by time) for the $S_1$ hole field (top) and the $S_1$ particle field (bottom) of the MVK molecule. Stable behavior is reached at look-ahead~$4$ and~$6$, respectively.}
    \label{fig:mvk_mat2}
\end{figure*}

\smallskip\par\noindent
\textbf{Monotonicity.}
A key property of the
look-ahead parameter is its monotonicity.
For any two merge trees $T_1,T_2$ and look-ahead values $h > k$, $\delta_h(T_1,T_2) \leq \delta_k(T_1,T_2)$ holds, which is straight-forward to argue: any mapping considered for $\delta_k$ is also a valid mapping for $\delta_h$.
Since the recursive algorithm finds the minimal cost mapping, we can conclude that $\delta_h(T_1,T_2) \leq \delta_k(T_1,T_2)$.
The same argument can be done using the edit operation model discussed above: any edit sequence of look-ahead~$k$ is also an edit sequence of look-ahead~$h$.

Monotonicity improves usability significantly, as a user can simply increase the look-ahead parameter as far as the computational capacities allow.
This also distinguishes the look-ahead parameter from another form of stability parameter, the so-called $\epsilon$-preprocessing~\cite{DBLP:journals/tvcg/SridharamurthyM20,DBLP:journals/tvcg/PontVDT22}.
Here, close saddles are merged until no inner edges below a certain threshold length exist.
During our experiments, we noticed that this parameter is indeed not monotonic: we encountered instances where more instabilities were introduced with higher $\epsilon$.
However, this happens only for distances working on merge trees directly, like
the merge tree edit distance 
$\delta_S$
and all deformation-based edit distances.
In contrast, we did not observe this behavior for
the merge tree Wasserstein distance
$\delta_W$.
As a combination of the look-ahead and the $\epsilon$-parameter could have high potential, this issue should be considered in future work.

\section{Experiments}

In this section, we verify enhanced stability and feasible runtimes experimentally on established datasets, \textcolor{revision}{which are either ensembles for clustering tasks or time series}.
We use data that has been analyzed using the path mapping or the deformation-based edit distance.
We perform two kinds of experiments.
First, we reproduce (qualitatively) results
of
the MIP implementation of the deformation-based edit distance.
Second, on datasets where the deformation-based edit distance cannot be computed in feasible time, we show improved results in comparison to the path mapping distance.
\textcolor{revision}{We mostly demonstrate this on distance matrices like in Figure~\ref{fig:teaser}.}

To also quantitatively verify the improved stability of the novel distances, we applied two measures on selected datasets.
\textcolor{revision}{On one dataset (which is used in a clustering setting)}, we applied the silhouette index to evaluate the quality of the distance metric.
On two datasets for which results with the unconstrained distance exist, we also analyze convergence for increasing look-ahead values.

Distance metrics are also critical in data analysis tasks such as classification, clustering, and outlier detection, which often rely on embedding the data into a lower-dimensional space.
The key to effective embedding lies in selecting a distance metric that emphasizes meaningful similarities and differences between data points.
To assess the impact of distance stability on embedding quality, we compared the performance of Multidimensional Scaling (MDS) and t-SNE using stable and unstable distances
measures.
Our results indicate that unstable distances lead to more error-prone embeddings in the case studies examined.
\textcolor{revision}{To account for parameter sensitivity of t-SNE, we provide results for varying parameters in supplementary material.}

Our experiments were executed on a workstation with two AMD EPYC 7453 28-Core processors (56 physical cores and 112 logical cores through hyperthreading) and 500GB of RAM.
We used up to 100 threads to match the setup in~\cite{taming} for direct comparison.
All preprocessing steps were performed with TTK.

\subsection{TOSCA Ensemble}

The first experiment uses the TOSCA dataset~\cite{DBLP:series/mcs/BronsteinBK09}, a shape matching ensemble consisting of human and animal shapes in varying poses.
A scalar field on the vertices represents the average geodesic distance~\cite{HilagaSKK01} to other vertices (example members are shown in Figure~\ref{fig:tosca_shapes}).
The data has been used in previous work~\cite{DBLP:journals/tvcg/SridharamurthyM20,taming} and we use the same topological simplification, a threshold of 6\% of the scalar range, as in~\cite{taming} and computed the split trees.

Previous results are shown in Figure~\ref{fig:tosca_mat1}(a-d).
Distance matrices \emph{should} show a clear cluster for each of the shapes.
\textcolor{revision}{The ordering by shape means the matrix should show clearly separated, consecutive regions of low distance}.
Although the clusters are vaguely detectable in the matrix computed with $\delta_S$ and $\delta_0$, many instabilities disturb the resulting image.
This is due to horizontal instabilities and also vertical instabilities in the case of~$\delta_S$ ($\delta_0$ is vertically stable and has a cleaner overall appearance, but outliers remain).
In contrast, the unconstrained deformation-based edit distance shows the clusters without any noise.

Figure~\ref{fig:tosca_mat1} then compares the previous results with a selection of heuristic solutions for look-aheads between~$1$ and~$8$.
We observe that the look-ahead heuristic improves the results quickly.
With a look-ahead value of~$1$, the image is significantly cleaner and only few instabilities remain.
With a look-ahead value larger than~$2$, almost no significant instabilities are visually detectable.
The matrix for look-ahead~$8$ looks, in essence, the same as the one computed with the unconstrained deformation-based edit distance.
We should note that the ensemble used with the unconstrained distance had to be reduced to make the distance matrix computation feasible.
All members with more than 26 vertices were removed.

For distance matrices for \emph{all} distances (including 
$\delta_W$
and all look-ahead values), see App.~\ref{sec:app_tosca}, suppl.\ material.
There, we also provide MDS embeddings of each distance matrix to further highlight the improved quality of the new method, 
\textcolor{revision}{as well as results for lower simplification thresholds (not possible with the MIP solution)}.

\begin{table}[]
\caption{Relative difference between heuristic solutions and the unconstrained distance on the TOSCA and MVK datasets. Since the instabilities appear infrequently, we show the mean, maximum, minimum and standard deviation for each comparison.}
\label{tab:convergence}
\centering
\resizebox{0.7\linewidth}{!}{
\begin{tabular}{l||c|c|c|c}
 & Mean & Max & Min & $\sigma$ \\
\hline
\hline
TOSCA,$\delta_0$ & $49\%$ & $789\%$ & $<0.01\%$ & $80\%$ \\
TOSCA,$\delta_1$ & $8\%$ & $501\%$ & $<0.01\%$ & $23\%$ \\
TOSCA,$\delta_2$ & $0.8\%$ & $157\%$ & $<0.01\%$ & $5.6\%$ \\
TOSCA,$\delta_3$ & $0.4\%$ & $130\%$ & $<0.01\%$ & $4.2\%$ \\
TOSCA,$\delta_4$ & $0.1\%$ & $76\%$ & $<0.01\%$ & $1.8\%$ \\
TOSCA,$\delta_5$ & $0.03\%$ & $47\%$ & $<0.01\%$ & $0.6\%$ \\
TOSCA,$\delta_6$ & $0.01\%$ & $21\%$ & $<0.01\%$ & $0.3\%$ \\
TOSCA,$\delta_7$ & $<0.01\%$ & $1\%$ & $<0.01\%$ & $0.04\%$ \\
TOSCA,$\delta_8$ & $<0.01\%$ & $1\%$ & $<0.01\%$ & $0.04\%$ \\
\hline
MVK,$\delta_0$ & $22\%$ & $410\%$ & $<0.01\%$ & $37\%$ \\
MVK,$\delta_1$ & $6.9\%$ & $388\%$ & $<0.01\%$ & $17\%$ \\
MVK,$\delta_2$ & $2.7\%$ & $80\%$ & $<0.01\%$ & $5.9\%$ \\
MVK,$\delta_3$ & $1.6\%$ & $35\%$ & $<0.01\%$ & $2.7\%$ \\
MVK,$\delta_4$ & $1.4\%$ & $29\%$ & $<0.01\%$ & $2\%$ \\
MVK,$\delta_5$ & $1.3\%$ & $13\%$ & $<0.01\%$ & $1.7\%$ \\
MVK,$\delta_6$ & $1.29\%$ & $12\%$ & $<0.01\%$ & $1.6\%$ \\
\end{tabular}
}
\end{table}

To also quantitatively verify improved stability, we applied two measures.
First, we computed the difference between the heuristic solution and the unconstrained distance for each look-ahead value and each pair of ensemble members.
The results can be found in Table~\ref{tab:convergence}.
It shows clear convergence towards the unconstrained distance for increasing look-ahead.
Second, we computed the silhouette score~\cite{ROUSSEEUW198753} on the annotated clusters with different distance matrices.
Each input mesh has the represented shape/animal annotated, such that these clusters can be considered the ground truth.
Thus, good or meaningful distance metrics should achieve a high clustering score.
We merged all human shapes into one cluster.
The path mapping distance yields a score of~$0.187$, the look-ahead~$1$ distance a score of~$0.376$.
All other distances give a score of $>0.43$, with the unconstrained distance and the look-ahead~$8$ distance scoring almost identical at~$0.448$.
The TOSCA ensemble is the only dataset on which we could compute such a measure, since it is the only one with a ground truth clustering at hand.

\smallskip\par\noindent
\textbf{Runtime.}
Computing the full
distance matrix took only up to a few seconds, depending on the look-ahead parameter (details \textcolor{revision}{for $\delta_1$-$\delta_6$} are shown in Table~\ref{tab:runtime}).
Until a look-ahead value of~$3$, the matrix could be computed within a second.
The completely stable approximation with look-ahead value~$8$ only needed 3~seconds.
In contrast, the computation on the \emph{reduced} distance matrix with the unconstrained distance took five days, performed on equivalent hardware using the same amount of threads.
In fact, even when only using commercial of-the-shelf hardware, the look-ahead distance matrix was computable in the range of seconds, whereas the computation of unconstrained distance depends on advanced hardware with heavy parallelization.
Furthermore, the heuristic method enables us to compute almost stable distance matrices for lower simplification thresholds \textcolor{revision}{(runtimes also in Table~\ref{tab:runtime})}.
See App.~\ref{sec:app_tosca} in suppl.\ material for more details and the distance matrices.

\begin{figure*}
    \centering
    \includegraphics[width=0.88\linewidth]{figures/ionization/figure_instabilities3.pdf}
    \caption{Ionization front time series: example members of the different phases (mid) and distance matrices for look-ahead~$0$ and~$3$ together with corresponding t-SNE embeddings. The phases are also annotated in the matrices and the embeddings are colored accordingly. Instabilities in the distance (highlighted in the left) lead to embeddings that are clustered rather than a smooth path.}
    \label{fig:ionization_mat}
\end{figure*}

\subsection{MVK Time Series}

The MVK dataset~\cite{doi:10.1021/acs.jpca.3c02582} is a time-dependent 3D scalar field describing electron density around a molecule.
It consists of 75 time steps, each a $115\times121\times83$ regular grid with four scalar fields attached, which describe electron density in various states or configurations.
In~\cite{10.2312:evs.20241069}, Wetzels et al.\
computed the distance matrix using the unconstrained deformation-based edit distance for all four fields.
Since the study was focused on maxima in absolute density, they used split trees, which were simplified with a relative threshold of 1\%.
For all four fields, the merge trees have 20-30 vertices with an average of~22.

To achieve feasible runtimes and semantically meaningful mappings, Wetzels et al.\ added geometric constraints on the edit mappings.
Two leaf nodes can only be mapped onto each other if they belong to the same atom segment in the molecule,
based on a Voronoi tessellation of the domain.
Since this reduces the search space for the MIP instance significantly, computation times of 20 minutes up to an hour were achieved for the full matrix.

To ensure comparability, we added the same geometric constraints into our
implementation.
We then computed distance matrices based on the new heuristic for all four fields and varying look-ahead values.
Figure~\ref{fig:mvk_mat2} shows a comparison of some example matrices.
The full set of matrices is provided in suppl.\ material, App.~\ref{sec:app_mvk}.

The most prominent feature in the unconstrained distance matrices are the outlier time steps~61-63.
Abrupt changes for these time steps are expected as they correspond to a phenomenon called electronic state crossings~\cite{10.2312:evs.20241069}.
Thus, recognizing the outliers is of practical importance.
For small look-ahead values or the original path mapping distance, the outlier behavior is overshadowed to different degrees.
For larger look-ahead values, the distance matrix is indistinguishable from the unconstrained one.

We performed the same convergence analysis as on the TOSCA ensemble (see Table~\ref{tab:convergence}), a clear convergence can be observed.

\smallskip\par\noindent
\textbf{Runtime.}
For three of the four fields, computing the full matrix with look-ahead~$6$ took less than two seconds.
For the $S_2$ hole field, it took 6~seconds, which interestingly does not correspond to any increase in size of the trees.
Computing the full distance matrix with the adapted unconstrained distance took 20~minutes up to an hour.
Table~\ref{tab:runtime} contains summarized/averaged runtimes.

\subsection{2D Ionization Front}

The ionization front dataset, originally used in the SciVis contest 2008~\cite{scivis2008}, is available both as a 3D or 2D time-dependent scalar field (2D derived through slicing).
Each member is a $600 \times 248$ regular grid representing one of 200 time steps of a simulation of ionization front propagation at universe formation.
We preprocessed each scalar field with normalization and topological simplification using a relative threshold of 5\%.
As ionization flares are captured by maxima of the ion concentration, we compute split trees.
The merge trees consist of~4 to~70 nodes, \textcolor{revision}{41}~on average.
The same preprocessed data has been used in~\cite{DBLP:journals/tvcg/PontVDT22,wetzels2022branch,DBLP:journals/tvcg/WetzelsPTG24} to compare merge tree edit mappings in the context of feature tracking.

Distance matrices for the full time series and look-aheads~$0-3$ can be found in App.~\ref{sec:app_ionization}, suppl. material.
Figure~\ref{fig:ionization_mat} shows two of them.
They all exhibit a consistent structure, characterized by a small outlier cluster in the first five time steps, followed by three clusters corresponding to different phases of the simulation.
The first one spans approximately~100 time steps, with a small subcluster (steps~25 to~39) showing larger distances to the rest.
Afterwards,
two additional clusters are evident, though less pronounced: inner similarity drops of towards the edges of the clusters.
The clusters are not strictly separated; instead, transitions between clusters are smooth, with a small radius of low distances surrounding the diagonal.

The different phases are illustrated in Figure~\ref{fig:ionization_mat}.
The initial outlier steps feature a large, dominant minimum on the left.
Time steps 5-105 have an overall similar structure, which can be subdivided into three subphases: 
first, the central maximum is dominant (steps 5–25); next, the side peaks grow in prominence, reducing the distinction of the central maximum (steps 25–40); then, the central maximum becomes dominant again, splitting into two peaks (steps 40–105).
Beyond step 105, many peaks begin to split further, with new peaks emerging on the far left and the side peaks becoming the primary features~(105-160).
In the final phase (steps 160–199), several main features, most notably the central peaks, gradually move out of the considered range and disappear.
Transitions such as the continuous splitting of the central maximum occur smoothly.

The
matrices vary in how clearly they represent these clusters.
The path mapping matrix exhibits several outlier stripes of differing intensity and thickness, along with occasional discontinuities within the clusters.
With a look-ahead value of~$1$, outliers and jumps are reduced significantly, with a value of~$2$ only two subtle outliers remain (see suppl.\ material).
Finally, the matrix with a look-ahead value of~$3$ (Figure~\ref{fig:ionization_mat}) presents a smooth structure with no visually significant outliers.

In addition, we computed t-SNE embeddings from the distance matrices.
A selection is shown in Figure~\ref{fig:ionization_mat}.
The points are colored according to a rough estimate of the different phases (clusters transition smoothly, so clear distinctions are challenging to define).
For higher look-ahead values, the embedded points form an almost continuous curve, except the initial outlier steps.
Clusters emerge in specific areas but gradually fade into one another through path-like structures.
In contrast, for lower look-ahead values, the embedding becomes more fragmented and discontinuous.

Given that t-SNE is highly sensitive to parameter settings, we tested various values of \emph{perplexity} and \emph{early exaggeration} (App.~\ref{sec:app_ionization}, suppl.\ material).
For higher look-ahead values, the overall curve-like structure remained consistent, while for lower look-ahead values, the embedding consistently showed large jumps or discontinuities, even when the overall structure appeared path-like.
Since we expect the embedding to reflect the smooth transitions observed in the original data, such discontinuities indicate a misrepresentation.
Therefore, unstable distances should be used with caution in these types of methods, especially when the parameter space is not thoroughly explored.
In contrast, the look-ahead distance improves stability and produces more consistent results.
All embeddings and a comparison with other edit distances is provided in suppl.\ material, App.~\ref{sec:app_ionization}.

\smallskip\par\noindent
\textbf{Runtime.}
Computation times for the full $200\times200$ distance matrix (Table~\ref{tab:runtime}) were in the range of seconds for all considered look-ahead values until stable behavior is reached.
Using the original path mapping distance, it took less than a second.
For a look-ahead value of~$3$, it took 25 seconds in total.
For a look-ahead value of~$4$, it took three minutes, however, the matrix is basically indistinguishable from the matrix for look-ahead~$3$ (see suppl.\ material, App.~\ref{sec:app_ionization}).
Even for look-aheads of~$5$ or~$6$, the full matrix took less than one hour.

\subsection{2D Vortex Street}

The last dataset we consider is a time-varying scalar field representing the velocity magnitude of flow around a cylinder on a 400 × 50 regular 2D grid.
It was simulated by Weinkauf~\cite{weinkauf10c} using \emph{Gerris Flow Solver}~\cite{gerrisflowsolver} and consists of 1001 time steps chosen from an interval with a fully developed von Kármán vortex street.
Accordingly, the time series exhibits strongly periodic behavior, which has been observed using distances on topological descriptors in several other works~\cite{DBLP:conf/apvis/NarayananTN15,DBLP:journals/cgf/SaikiaSW14,DBLP:journals/tvcg/SridharamurthyM20,wetzels2022branch,wetzels2022path}.
More precisely, other distance metrics identified a half period of~38 and a full period of~75.

We consider split trees where 
numbers of vertices 
range from~62 to~82 (average~68).
We computed distance matrices with the look-ahead heuristic and the metrics from previous works.
Computing the unconstrained deformation-based edit distance is unrealistic, as the limit for feasible runtimes is around~30 vertices.
The first matrix in Figure~\ref{fig:teaser} shows a previous result achieved with the path mapping distance for the first~300 time steps.
Very similar images are produced by $\delta_S$ and $\delta_W$, see suppl.\ material, App.~\ref{sec:app_vortex}.
We then computed the matrix for the same interval using the look-ahead heuristic with look-ahead values of~$1-3$, which can also be found in Figure~\ref{fig:teaser}.

Looking at the different matrices, the periodicity is detectable in all four of them.
However, the first matrix shows big jumps in the distance.
It looks like there are mainly two different phases between which the time series jumps back and forth.
Without previous knowledge about the dataset, one might assume that it consists of two clusters and the matrix is just poorly ordered.
In contrast, when we increase the look-ahead parameter, the matrix gets smoother.
With look-ahead~$1$, some jumps are still present, but the overall look is more periodic and less clustered.
For look-ahead values larger than~$1$, all transitions are smooth.
We can see a continuous periodic pattern in the matrix, no clusters are visible.
This fits much better to the original data, which is a continuous time series.

To confirm this observation, we also performed a t-SNE embedding of the~300 time steps based on the different distance matrices, which can also be found in Figure~\ref{fig:teaser}.
The original path mapping distance gives three clusters, one small and two larger ones.
A look-ahead value of~$1$ yields two clusters.
Larger look-ahead values do not show any clusters, but yield a circle-shaped embedding, which perfectly fits the intuition of a periodic pattern.
To verify that this observation is not due to parameter choice for t-SNE, we computed the embedding for different values of perplexity (early exaggeration does not influence the result significantly on this dataset).
Furthermore, we computed embeddings based on MDS.
Unstable distance matrices consistently lead to clusters, whereas they appear only for very small perplexity values when using stable matrices.
The results are presented in suppl.\ material, App.~\ref{sec:app_vortex}.

Furthermore, we were able to compute the full distance matrix for all 1000 time steps up to a look-ahead of~$6$.
However, stable behavior is already reached with look-ahead~$2$, results remain qualitatively equal beyond that value.
A stable full matrix is also shown in App.~\ref{sec:app_vortex}.

\smallskip\par\noindent
\textbf{Runtime.}
Computing the full matrix (Table~\ref{tab:runtime}) with the original path mapping distance took 89 seconds.
Using a look-ahead of up to~$3$
(where stable matrices are achieved),
computation times went up to several minutes (6 minutes for look-ahead~3).
The same holds for other polynomial time edit distances where we observed computation times of 4~to~5 minutes.
When 
increasing the look-ahead 
value 
to~$4$, we observed a big jump, showing the exponential nature of the look-ahead: the full matrix took half an hour.
Times went up to three hours for a look-ahead of~$6$.

\begin{table}[]
\caption{Runtimes for the full distance matrix computation for different distance metrics and datasets. $|T|$ describes the average tree size, $n_T$ the number of merge trees in the ensemble.}
\label{tab:runtime}
\centering
\setlength{\tabcolsep}{3pt} % Default value: 6pt
\resizebox{0.82\linewidth}{!}{
\begin{tabular}{l|c|c|c|c|c|c}
 & TOSCA & TOSCA & TOSCA & MVK & Ion. Front & Vortex Str. \\
 & (6\%) & (4\%) & (2\%) &  &  &  \\
\hline
\hline
$|T|$ & 20 & 26 & 40 & 22 & 41 & 68 \\
$n_T$ & 148 & 148 & 148 & 75 & 200 & 1000 \\
\hline
$t(\delta_W)$ & 3.7s & 4.6s & 5.7s & 0.4s & 7.6s & 91.8s \\
$t(\delta_S)$ & 4.6s & 3.2s & 8.4s & 0.6s & 12.1s & 279s \\
$t(\odedist)$ & 0.3s & 0.2s & 1.9s & 0.1s & 0.9s & 88.9s \\
$t(\delta_1)$ & 0.2s & 0.3s & 3.7s & 0.1s & 2.0s & 181s \\
$t(\delta_2)$ & 0.3s & 0.4s & 4.1s & 0.1s & 4.5s & 205s \\
$t(\delta_3)$ & 0.5s & 1.1s & 6.5s & 0.3s & 25.5s & 362s \\
$t(\delta_4)$ & 1.1s & 2.8s & 17.7s & 0.7s & 171s & 30m \\
$t(\delta_5)$ & 1.8s & 5.7s & 47.3s & 1.5s & 860s & 103m \\
$t(\delta_6)$ & 2.3s & 9.0s & 90.4s & 2.5s & 49m & 240m \\
$t(\edist)$ & 5d & - & - & 40m & - & - \\
\end{tabular}
}
\end{table}

\subsection{Runtime and Scaling}

We now provide details on the runtimes in our experiments.
The full matrix computation times for each dataset and considered distance are summarized in Table~\ref{tab:runtime}.
As expected, the look-ahead distance is slower than constrained distances $\delta_W,\delta_S,\delta_0$, especially for higher look-ahead values.
However, it is faster than the unconstrained deformation-based edit distance by several orders of magnitude.
It thereby enables computation of stable distance matrices on medium size datasets (up to 100 nodes), where it was previously impossible using exact implementations of NP-complete distances.

\begin{figure}[!b]
    \centering
    \includegraphics[width=\linewidth]{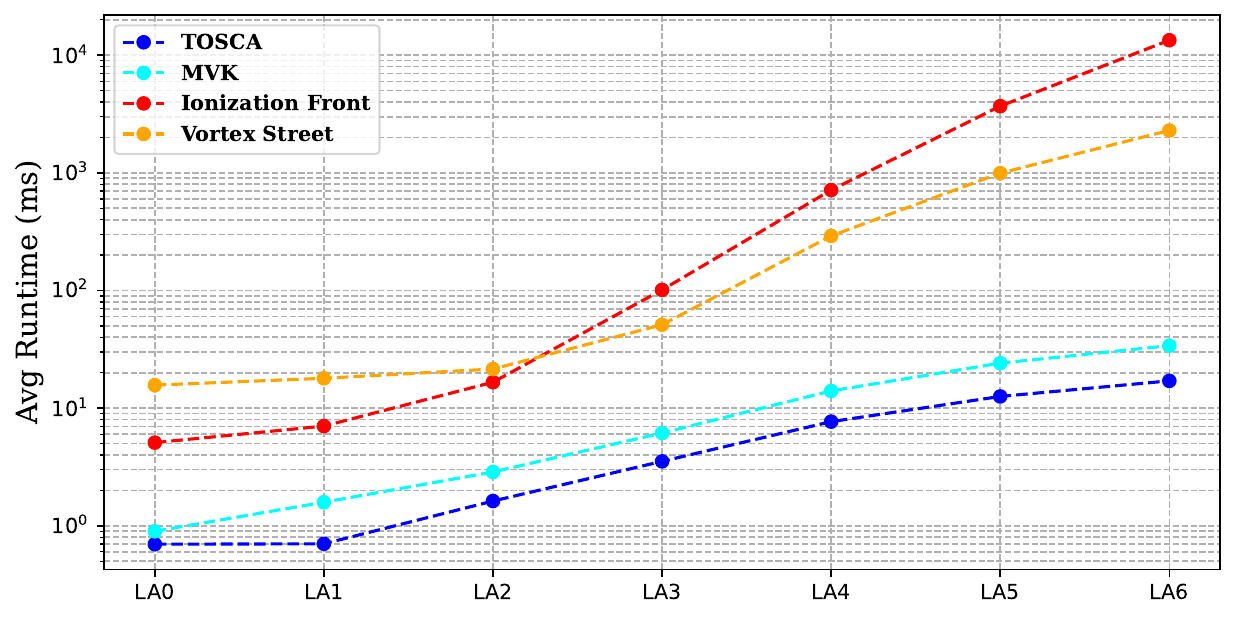}
    \caption{Average runtimes for different datasets and look-ahead values in log scale. The runtime grows exponentially with increasing look-ahead. Note the drop-off in gradient towards the right: the look-ahead ``saturates'' the tree depth from a certain point on, such that the runtimes do no longer increase exponentially.}
    \label{fig:runtimes_per_la}
\end{figure}

\smallskip\par\noindent
\textbf{Look-ahead Scaling.} Figure~\ref{fig:runtimes_per_la} shows average runtimes per single distance computation for increasing look-ahead.
Runtimes grow exponentially with look-ahead at first (we hypothesize that early increase in slope is due to the asymptotics taking effect late), but saturation effects can be observed: On small instances, the exponential nature only shows up to a certain look-ahead value.
Overall, they remain feasible for all look-ahead values that are necessary to achieve stable matrices on reasonably-sized merge trees and ensembles.

The observation shows that actual features rarely perform 
a lot of
saddle swaps between similar members in smooth, practical datasets.
In contrast, when working on noisy data, higher look-ahead values were necessary to achieve stable matrices (see App.~\ref{sec:app_tosca}, suppl. material).
This
indeed 
fits the intuition of the
heuristic edit distance: 
given a saddle swap in a merge tree, scalar noise potentially introduces many vertices between the swapped saddles.
It can therefore only be captured by high look-ahead values.

\begin{figure}
    \centering
    \includegraphics[width=\linewidth]{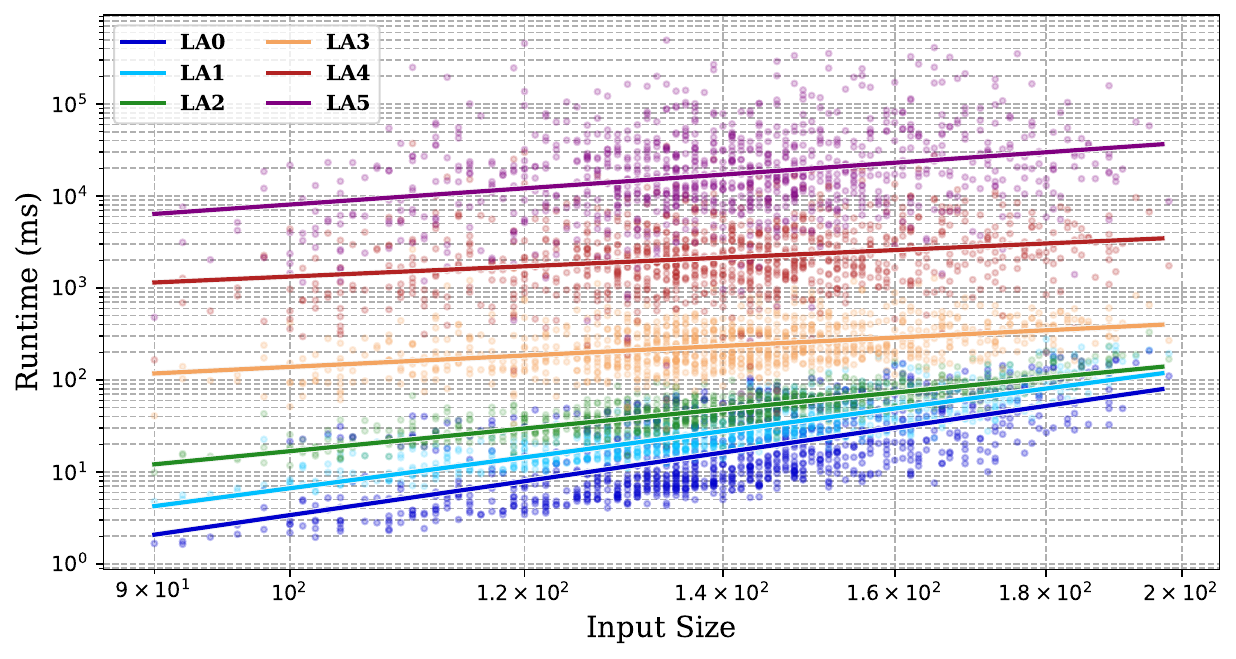}
    \caption{Runtimes for different look-ahead values as a function of tree size ($|T_1|+|T_2|$). Both axes are in log scale. The scatter plot shows individual distance computations, the overlaid line a linear fit. The runtime grows polynomial 
    with
    input size. We removed trees of low depth, to account for saturation effect similar to Figure~\ref{fig:runtimes_per_la}. If we look at the left end of the lines, we can still observe initial saturation: the increase in runtime becomes slower for high look-ahead values.}
    \label{fig:runtimes_per_size}
\end{figure}

\smallskip\par\noindent
\textbf{Input Size Scaling.} Figure~\ref{fig:runtimes_per_size} shows, for increasing look-ahead values, runtimes of single distance computations as a function on inputs size.
For a fixed look-ahead value, runtimes grow polynomial (linear in log log scale), as expected, and thereby follow the theoretic bounds.
We observe a decrease in gradient for higher look-ahead values, due to the quadratic factor becoming dominant over the quartic one on the input sizes considered (cf.\ Equation~\ref{eq:runtime}).
This experiment was performed on a single dataset (ionization front) using varying simplification thresholds (2-5\%).
Using multiple datasets for the scaling with size yields unrealistic results, due to many optimizations depending on the tree shape and the saturation effect shown in Figure~\ref{fig:runtimes_per_la}.

\section{Conclusion}

In this paper, we presented a novel heuristic algorithm for the deformation-based edit distance.
We described how the heuristic defines a less constrained variant of the deformation-based edit distance that places somewhere in between the path mapping distance and the unconstrained version.
It allows the user to choose a look-ahead parameter, for which higher values increase the stability monotonically but also the computational complexity.

Experiments based on our open-source implementation demonstrate that the described trade-off between stability and time shows up in practice.
The heuristic approach allows accessing the superior properties of the unconstrained deformation-based edit distance for a large variety of practically relevant cases.
Look-ahead values of up to~$4$ already yield very stable behavior on practical data, keeping the increase in runtime moderate.

\smallskip\par\noindent
\textbf{Limitations.}
Even though our method is polynomial (FPT) and outperforms the unconstrained deformation-based edit distance by a large margin, runtime 
can be a limiting factor on datasets with extreme structural complexity.
For feature-rich data (where simplification to small 
numbers of vertices 
is not possible), the quartic complexity (same as the path mapping distance) increases runtimes infeasibly when going above a few hundred vertices.
On noisy data, large look-ahead values can be necessary, exponentially increasing the runtime.

A further issue to be considered is result verification.
It would be useful to guide users in choosing a stability parameter, or to provide an automated measure of result quality, which appears unrealistic.
However, this holds for essentially all  edit distances for merge trees proposed thus far.
In our case, monotonicity (in terms of stability) of the parameter is ensured.

\smallskip\par\noindent
\textbf{Future Work.}
Based on the limitations listed above, we want to consider several optimization possibilities in the future.
It should be studied whether the quartic time algorithm is actually optimal (similar to the quadratic lower bound for classic constrained edit distances) or whether there are better methods.
Furthermore, as discussed in Section~\ref{sec:impl}, alternative methods for the optimal assignment instances should be considered.
Possible options are application of a minimum cost maximum flow solver or the choosing the correct algorithm depending on the size of the instance.
\textcolor{revision}{A formal study of stability guarantees or the metric property is another area of interest.}

Furthermore, integration into advanced analysis frameworks like merge tree barycenters~\cite{DBLP:journals/tvcg/PontVDT22}, dataset summarization methods~\cite{DBLP:journals/cgf/LohfinkWLWG20,DBLP:journals/tvcg/PontVDT22,DBLP:journals/tvcg/PontVT23}, auto-encoders~\cite{DBLP:journals/tvcg/PontT24} or other machine learning techniques should be studied, as well as the possibility to integrate the $\epsilon$-preprocessing in a monotonic form.

\acknowledgments{
The authors wish to thank Marvin Petersen and Jonas Lukasczyk for valuable discussions and in particular Markus Anders for initiating the idea behind this work. Furthermore, many thanks to the anonymous reviewers for their helpful input, to Mathieu Pont for providing implementation support, to Raghavendra Sridharamurthy for providing the pre-processed TOSCA dataset, as well as Nanna Holmgaard List and Talha Bin Masood for providing the MVK dataset. This work is funded by the Deutsche Forschungsgemeinschaft (DFG, German Research Foundation) – 442077441.}

\section*{Supplementary Material}

This manuscript is accompanied by supplementary material:

\begin{itemize}
    \item We provide a supplementary PDF that contains additional experiments and discussions.
    \item The publicly available source code~\cite{repository_lookahead} is provided together with detailed instructions to compile it and reproduce the images shown in~\autoref{fig:teaser}. This implementation will be contributed as open source to TTK in the future.
    
    The companion ZIP file contains an archive of the repository.
\end{itemize}

\bibliographystyle{abbrv-doi-hyperref}

\bibliography{main}

\begin{thebibliography}{10}

\bibitem{DBLP:conf/3dor/BauerFL16}
U.~Bauer, B.~D. Fabio, and C.~Landi.
\newblock An edit distance for reeb graphs.
\newblock In A.~Ferreira, A.~Giachetti, and D.~Giorgi, eds., {\em 9th
  Eurographics Workshop on 3D Object Retrieval, 3DOR@Eurographics 2016, Lisbon,
  Portugal, May 8, 2016}. Eurographics Association, 2016.
  \href{https://doi.org/10.2312/3dor.20161084}
{doi: {{%
10\hspace{.1pt}\discretionary{.}{%
}{.}\hspace{.4pt}2312\discretionary{/}{%
}{/}3dor\hspace{.1pt}\discretionary{.}{%
}{.}\hspace{.4pt}20161084}}}


\bibitem{BeketayevYMWH14}
K.~Beketayev, D.~Yeliussizov, D.~Morozov, G.~H. Weber, and B.~Hamann.
\newblock Measuring the distance between merge trees.
\newblock In P.~Bremer, I.~Hotz, V.~Pascucci, and R.~Peikert, eds., {\em
  Topological Methods in Data Analysis and Visualization III, Theory,
  Algorithms, and Applications}, pp. 151--165. Springer, 2014.
  \href{https://doi.org/10.1007/978-3-319-04099-8_10}
{doi: {{%
10\hspace{.1pt}\discretionary{.}{%
}{.}\hspace{.4pt}1007\discretionary{/}{%
}{/}978\discretionary{%
}{-}{-}3\discretionary{%
}{-}{-}319\discretionary{%
}{-}{-}04099\discretionary{%
}{-}{-}8\_10}}}


\bibitem{DBLP:journals/mp/Bertsekas81}
D.~P. Bertsekas.
\newblock A new algorithm for the assignment problem.
\newblock {\em Math. Program.}, 21(1):152--171, 1981.
  \href{https://doi.org/10.1007/BF01584237}
{doi: {{%
10\hspace{.1pt}\discretionary{.}{%
}{.}\hspace{.4pt}1007\discretionary{/}{%
}{/}BF01584237}}}


\bibitem{treeEditSurvey}
P.~Bille.
\newblock A survey on tree edit distance and related problems.
\newblock {\em Theoretical Computer Science}, 337(1-3):217--239, 2005.
  \href{https://doi.org/10.1016/j.tcs.2004.12.030}
{doi: {{%
10\hspace{.1pt}\discretionary{.}{%
}{.}\hspace{.4pt}1016\discretionary{/}{%
}{/}j\hspace{.1pt}\discretionary{.}{%
}{.}\hspace{.4pt}tcs\hspace{.1pt}\discretionary{.}{%
}{.}\hspace{.4pt}2004\hspace{.1pt}\discretionary{.}{%
}{.}\hspace{.4pt}12\hspace{.1pt}\discretionary{.}{%
}{.}\hspace{.4pt}030}}}


\bibitem{DBLP:journals/tvcg/BollenTL23}
B.~Bollen, P.~Tennakoon, and J.~A. Levine.
\newblock Computing a stable distance on merge trees.
\newblock {\em {IEEE} Trans. Vis. Comput. Graph.}, 29(1):1168--1177, 2023.
  \href{https://doi.org/10.1109/TVCG.2022.3209395}
{doi: {{%
10\hspace{.1pt}\discretionary{.}{%
}{.}\hspace{.4pt}1109\discretionary{/}{%
}{/}TVCG\hspace{.1pt}\discretionary{.}{%
}{.}\hspace{.4pt}2022\hspace{.1pt}\discretionary{.}{%
}{.}\hspace{.4pt}3209395}}}


\bibitem{DBLP:series/mcs/BronsteinBK09}
A.~M. Bronstein, M.~M. Bronstein, and R.~Kimmel.
\newblock {\em Numerical Geometry of Non-Rigid Shapes}.
\newblock Monographs in Computer Science. Springer, 2009.
  \href{https://doi.org/10.1007/978-0-387-73301-2}
{doi: {{%
10\hspace{.1pt}\discretionary{.}{%
}{.}\hspace{.4pt}1007\discretionary{/}{%
}{/}978\discretionary{%
}{-}{-}0\discretionary{%
}{-}{-}387\discretionary{%
}{-}{-}73301\discretionary{%
}{-}{-}2}}}


\bibitem{DBLP:conf/soda/CarrSA00}
H.~A. Carr, J.~Snoeyink, and U.~Axen.
\newblock Computing contour trees in all dimensions.
\newblock In D.~B. Shmoys, ed., {\em Proceedings of the Eleventh Annual
  {ACM-SIAM} Symposium on Discrete Algorithms, January 9-11, 2000, San
  Francisco, CA, {USA}}, pp. 918--926. {ACM/SIAM}, 2000.

\bibitem{doi:10.1021/acs.jpca.3c02582}
P.~Chakraborty, R.~C. Couto, and N.~H. List.
\newblock Deciphering methylation effects on s2 ($\pi$$\pi$*) internal
  conversion in the simplest linear $\alpha$, $\beta$-unsaturated carbonyl.
\newblock {\em The Journal of Physical Chemistry A}, 127(25):5360--5373, 2023.
\newblock PMID: 37331016. \href{https://doi.org/10.1021/acs.jpca.3c02582}
{doi: {{%
10\hspace{.1pt}\discretionary{.}{%
}{.}\hspace{.4pt}1021\discretionary{/}{%
}{/}acs\hspace{.1pt}\discretionary{.}{%
}{.}\hspace{.4pt}jpca\hspace{.1pt}\discretionary{.}{%
}{.}\hspace{.4pt}3c02582}}}


\bibitem{interleaving_distance}
F.~Chazal, D.~Cohen{-}Steiner, M.~Glisse, L.~J. Guibas, and S.~Oudot.
\newblock Proximity of persistence modules and their diagrams.
\newblock In J.~Hershberger and E.~Fogel, eds., {\em Proceedings of the 25th
  {ACM} Symposium on Computational Geometry, Aarhus, Denmark, June 8-10, 2009},
  pp. 237--246. {ACM}, 2009. \href{https://doi.org/10.1145/1542362.1542407}
{doi: {{%
10\hspace{.1pt}\discretionary{.}{%
}{.}\hspace{.4pt}1145\discretionary{/}{%
}{/}1542362\hspace{.1pt}\discretionary{.}{%
}{.}\hspace{.4pt}1542407}}}


\bibitem{Cohen-Steiner2007}
D.~Cohen{-}Steiner, H.~Edelsbrunner, and J.~Harer.
\newblock Stability of persistence diagrams.
\newblock {\em Discret. Comput. Geom.}, 37(1):103--120, 2007.
  \href{https://doi.org/10.1007/s00454-006-1276-5}
{doi: {{%
10\hspace{.1pt}\discretionary{.}{%
}{.}\hspace{.4pt}1007\discretionary{/}{%
}{/}s00454\discretionary{%
}{-}{-}006\discretionary{%
}{-}{-}1276\discretionary{%
}{-}{-}5}}}


\bibitem{dagum1998openmp}
L.~Dagum and R.~Menon.
\newblock {OpenMP: An Industry Standard API for Shared-Memory Programming}.
\newblock {\em Computational Science \& Engineering, IEEE}, 5(1):46--55, 1998.

\bibitem{edelsbrunner09}
H.~Edelsbrunner and J.~Harer.
\newblock {\em Computational Topology - an Introduction}.
\newblock American Mathematical Society, 2010.

\bibitem{DBLP:conf/focs/EdelsbrunnerLZ00}
H.~Edelsbrunner, D.~Letscher, and A.~Zomorodian.
\newblock Topological persistence and simplification.
\newblock In {\em 41st Annual Symposium on Foundations of Computer Science,
  {FOCS} 2000, 12-14 November 2000, Redondo Beach, California, {USA}}, pp.
  454--463. {IEEE} Computer Society, 2000.
  \href{https://doi.org/10.1109/SFCS.2000.892133}
{doi: {{%
10\hspace{.1pt}\discretionary{.}{%
}{.}\hspace{.4pt}1109\discretionary{/}{%
}{/}SFCS\hspace{.1pt}\discretionary{.}{%
}{.}\hspace{.4pt}2000\hspace{.1pt}\discretionary{.}{%
}{.}\hspace{.4pt}892133}}}


\bibitem{DBLP:journals/entcs/FabioL12}
B.~D. Fabio and C.~Landi.
\newblock Stability of reeb graphs of closed curves.
\newblock In L.~Fajstrup, E.~Goubault, and M.~Raussen, eds., {\em Proceedings
  of the workshop on Geometric and Topological Methods in Computer Science,
  {GETCO} 2010, Aalborg, Denmark, January 11-15, 2010}, vol. 283 of {\em
  Electronic Notes in Theoretical Computer Science}, pp. 71--76. Elsevier,
  2010. \href{https://doi.org/10.1016/J.ENTCS.2012.05.006}
{doi: {{%
10\hspace{.1pt}\discretionary{.}{%
}{.}\hspace{.4pt}1016\discretionary{/}{%
}{/}J\hspace{.1pt}\discretionary{.}{%
}{.}\hspace{.4pt}ENTCS\hspace{.1pt}\discretionary{.}{%
}{.}\hspace{.4pt}2012\hspace{.1pt}\discretionary{.}{%
}{.}\hspace{.4pt}05\hspace{.1pt}\discretionary{.}{%
}{.}\hspace{.4pt}006}}}


\bibitem{DBLP:journals/dcg/FabioL16}
B.~D. Fabio and C.~Landi.
\newblock The edit distance for reeb graphs of surfaces.
\newblock {\em Discret. Comput. Geom.}, 55(2):423--461, 2016.
  \href{https://doi.org/10.1007/s00454-016-9758-6}
{doi: {{%
10\hspace{.1pt}\discretionary{.}{%
}{.}\hspace{.4pt}1007\discretionary{/}{%
}{/}s00454\discretionary{%
}{-}{-}016\discretionary{%
}{-}{-}9758\discretionary{%
}{-}{-}6}}}


\bibitem{DBLP:series/txtcs/FlumG06}
J.~Flum and M.~Grohe.
\newblock {\em Parameterized Complexity Theory}.
\newblock Texts in Theoretical Computer Science. An {EATCS} Series. Springer,
  2006. \href{https://doi.org/10.1007/3-540-29953-X}
{doi: {{%
10\hspace{.1pt}\discretionary{.}{%
}{.}\hspace{.4pt}1007\discretionary{/}{%
}{/}3\discretionary{%
}{-}{-}540\discretionary{%
}{-}{-}29953\discretionary{%
}{-}{-}X}}}


\bibitem{intrinsicMTdistance}
E.~Gasparovic, E.~Munch, S.~Oudot, K.~Turner, B.~Wang, and Y.~Wang.
\newblock Intrinsic interleaving distance for merge trees.
\newblock {\em CoRR}, 1908.00063, 2019.

\bibitem{heine16}
C.~Heine, H.~Leitte, M.~Hlawitschka, F.~Iuricich, L.~D. Floriani,
  G.~Scheuermann, H.~Hagen, and C.~Garth.
\newblock A survey of topology-based methods in visualization.
\newblock {\em Comput. Graph. Forum}, 35(3):643--667, 2016.
  \href{https://doi.org/10.1111/cgf.12933}
{doi: {{%
10\hspace{.1pt}\discretionary{.}{%
}{.}\hspace{.4pt}1111\discretionary{/}{%
}{/}cgf\hspace{.1pt}\discretionary{.}{%
}{.}\hspace{.4pt}12933}}}


\bibitem{HilagaSKK01}
M.~Hilaga, Y.~Shinagawa, T.~Komura, and T.~L. Kunii.
\newblock {Topology matching for fully automatic similarity estimation of 3D
  shapes}.
\newblock In {\em ACM SIGGRAPH}, 2001.

\bibitem{DBLP:conf/visualization/LohfinkGWVG21}
A.~P. Lohfink, F.~Gartzky, F.~Wetzels, L.~Vollmer, and C.~Garth.
\newblock Time-varying fuzzy contour trees.
\newblock In {\em 2021 {IEEE} Visualization Conference, {IEEE} {VIS} 2021 -
  Short Papers, New Orleans, LA, USA, October 24-29, 2021}, pp. 86--90. {IEEE},
  2021. \href{https://doi.org/10.1109/VIS49827.2021.9623286}
{doi: {{%
10\hspace{.1pt}\discretionary{.}{%
}{.}\hspace{.4pt}1109\discretionary{/}{%
}{/}VIS49827\hspace{.1pt}\discretionary{.}{%
}{.}\hspace{.4pt}2021\hspace{.1pt}\discretionary{.}{%
}{.}\hspace{.4pt}9623286}}}


\bibitem{DBLP:journals/cgf/LohfinkWLWG20}
A.~P. Lohfink, F.~Wetzels, J.~Lukasczyk, G.~H. Weber, and C.~Garth.
\newblock Fuzzy contour trees: Alignment and joint layout of multiple contour
  trees.
\newblock {\em Comput. Graph. Forum}, 39(3):343--355, 2020.
  \href{https://doi.org/10.1111/cgf.13985}
{doi: {{%
10\hspace{.1pt}\discretionary{.}{%
}{.}\hspace{.4pt}1111\discretionary{/}{%
}{/}cgf\hspace{.1pt}\discretionary{.}{%
}{.}\hspace{.4pt}13985}}}


\bibitem{morozov14}
D.~Morozov, K.~Beketayev, and G.~H. Weber.
\newblock Interleaving distance between merge trees.
\newblock In {\em TopoInVis}. 2014.

\bibitem{DBLP:conf/ppopp/MorozovW13}
D.~Morozov and G.~H. Weber.
\newblock Distributed merge trees.
\newblock In A.~Nicolau, X.~Shen, S.~P. Amarasinghe, and R.~W. Vuduc, eds.,
  {\em {ACM} {SIGPLAN} Symposium on Principles and Practice of Parallel
  Programming, PPoPP '13, Shenzhen, China, February 23-27, 2013}, pp. 93--102.
  {ACM}, 2013. \href{https://doi.org/10.1145/2442516.2442526}
{doi: {{%
10\hspace{.1pt}\discretionary{.}{%
}{.}\hspace{.4pt}1145\discretionary{/}{%
}{/}2442516\hspace{.1pt}\discretionary{.}{%
}{.}\hspace{.4pt}2442526}}}


\bibitem{munkres}
J.~Munkres.
\newblock Algorithms for the assignment and transportation problems.
\newblock {\em Journal of the Society for Industrial and Applied Mathematics},
  5(1):32--38, 1957. \href{https://doi.org/10.1137/0105003}
{doi: {{%
10\hspace{.1pt}\discretionary{.}{%
}{.}\hspace{.4pt}1137\discretionary{/}{%
}{/}0105003}}}


\bibitem{DBLP:conf/apvis/NarayananTN15}
V.~Narayanan, D.~M. Thomas, and V.~Natarajan.
\newblock Distance between extremum graphs.
\newblock In S.~Liu, G.~Scheuermann, and S.~Takahashi, eds., {\em 2015 {IEEE}
  Pacific Visualization Symposium, PacificVis 2015, Hangzhou, China, April
  14-17, 2015}, pp. 263--270. {IEEE} Computer Society, 2015.
  \href{https://doi.org/10.1109/PACIFICVIS.2015.7156386}
{doi: {{%
10\hspace{.1pt}\discretionary{.}{%
}{.}\hspace{.4pt}1109\discretionary{/}{%
}{/}PACIFICVIS\hspace{.1pt}\discretionary{.}{%
}{.}\hspace{.4pt}2015\hspace{.1pt}\discretionary{.}{%
}{.}\hspace{.4pt}7156386}}}


\bibitem{openmp08}
{OpenMP Architecture Review Board}.
\newblock {OpenMP Application Program Interface Version 3.0}, May 2008.

\bibitem{DBLP:journals/tvcg/PontT24}
M.~Pont and J.~Tierny.
\newblock Wasserstein auto-encoders of merge trees (and persistence diagrams).
\newblock {\em {IEEE} Trans. Vis. Comput. Graph.}, 30(9):6390--6406, 2024.
  \href{https://doi.org/10.1109/TVCG.2023.3334755}
{doi: {{%
10\hspace{.1pt}\discretionary{.}{%
}{.}\hspace{.4pt}1109\discretionary{/}{%
}{/}TVCG\hspace{.1pt}\discretionary{.}{%
}{.}\hspace{.4pt}2023\hspace{.1pt}\discretionary{.}{%
}{.}\hspace{.4pt}3334755}}}


\bibitem{DBLP:journals/tvcg/PontVDT22}
M.~Pont, J.~Vidal, J.~Delon, and J.~Tierny.
\newblock Wasserstein distances, geodesics and barycenters of merge trees.
\newblock {\em {IEEE} Trans. Vis. Comput. Graph.}, 28(1):291--301, 2022.
  \href{https://doi.org/10.1109/TVCG.2021.3114839}
{doi: {{%
10\hspace{.1pt}\discretionary{.}{%
}{.}\hspace{.4pt}1109\discretionary{/}{%
}{/}TVCG\hspace{.1pt}\discretionary{.}{%
}{.}\hspace{.4pt}2021\hspace{.1pt}\discretionary{.}{%
}{.}\hspace{.4pt}3114839}}}


\bibitem{DBLP:journals/tvcg/PontVT23}
M.~Pont, J.~Vidal, and J.~Tierny.
\newblock Principal geodesic analysis of merge trees (and persistence
  diagrams).
\newblock {\em {IEEE} Trans. Vis. Comput. Graph.}, 29(2):1573--1589, 2023.
  \href{https://doi.org/10.1109/TVCG.2022.3215001}
{doi: {{%
10\hspace{.1pt}\discretionary{.}{%
}{.}\hspace{.4pt}1109\discretionary{/}{%
}{/}TVCG\hspace{.1pt}\discretionary{.}{%
}{.}\hspace{.4pt}2022\hspace{.1pt}\discretionary{.}{%
}{.}\hspace{.4pt}3215001}}}


\bibitem{gerrisflowsolver}
S.~Popinet.
\newblock Free computational fluid dynamics.
\newblock {\em ClusterWorld}, 2(6), 2004.

\bibitem{ROUSSEEUW198753}
P.~J. Rousseeuw.
\newblock Silhouettes: A graphical aid to the interpretation and validation of
  cluster analysis.
\newblock {\em Journal of Computational and Applied Mathematics}, 20:53--65,
  1987. \href{https://doi.org/10.1016/0377-0427(87)90125-7}
{doi: {{%
10\hspace{.1pt}\discretionary{.}{%
}{.}\hspace{.4pt}1016\discretionary{/}{%
}{/}0377\discretionary{%
}{-}{-}0427\discretionary{%
}{(}{(}87\discretionary{)}{%
}{)}90125\discretionary{%
}{-}{-}7}}}


\bibitem{DBLP:journals/cgf/SaikiaSW14}
H.~Saikia, H.~Seidel, and T.~Weinkauf.
\newblock Extended branch decomposition graphs: Structural comparison of scalar
  data.
\newblock {\em Comput. Graph. Forum}, 33(3):41--50, 2014.
  \href{https://doi.org/10.1111/cgf.12360}
{doi: {{%
10\hspace{.1pt}\discretionary{.}{%
}{.}\hspace{.4pt}1111\discretionary{/}{%
}{/}cgf\hspace{.1pt}\discretionary{.}{%
}{.}\hspace{.4pt}12360}}}


\bibitem{DBLP:journals/ipl/Selkow77}
S.~M. Selkow.
\newblock The tree-to-tree editing problem.
\newblock {\em Inf. Process. Lett.}, 6(6):184--186, 1977.
  \href{https://doi.org/10.1016/0020-0190(77)90064-3}
{doi: {{%
10\hspace{.1pt}\discretionary{.}{%
}{.}\hspace{.4pt}1016\discretionary{/}{%
}{/}0020\discretionary{%
}{-}{-}0190\discretionary{%
}{(}{(}77\discretionary{)}{%
}{)}90064\discretionary{%
}{-}{-}3}}}


\bibitem{DBLP:journals/tvcg/SridharamurthyM20}
R.~Sridharamurthy, T.~B. Masood, A.~Kamakshidasan, and V.~Natarajan.
\newblock Edit distance between merge trees.
\newblock {\em {IEEE} Trans. Vis. Comput. Graph.}, 26(3):1518--1531, 2020.
  \href{https://doi.org/10.1109/TVCG.2018.2873612}
{doi: {{%
10\hspace{.1pt}\discretionary{.}{%
}{.}\hspace{.4pt}1109\discretionary{/}{%
}{/}TVCG\hspace{.1pt}\discretionary{.}{%
}{.}\hspace{.4pt}2018\hspace{.1pt}\discretionary{.}{%
}{.}\hspace{.4pt}2873612}}}


\bibitem{DBLP:journals/tvcg/SridharamurthyN23}
R.~Sridharamurthy and V.~Natarajan.
\newblock Comparative analysis of merge trees using local tree edit distance.
\newblock {\em {IEEE} Trans. Vis. Comput. Graph.}, 29(2):1518--1530, 2023.
  \href{https://doi.org/10.1109/TVCG.2021.3122176}
{doi: {{%
10\hspace{.1pt}\discretionary{.}{%
}{.}\hspace{.4pt}1109\discretionary{/}{%
}{/}TVCG\hspace{.1pt}\discretionary{.}{%
}{.}\hspace{.4pt}2021\hspace{.1pt}\discretionary{.}{%
}{.}\hspace{.4pt}3122176}}}


\bibitem{scivis2008}
R.~Taylor, A.~Chourasia, D.~Whalen, and M.~L. Norman.
\newblock {The IEEE SciVis Contest}.
\newblock \url{http://sciviscontest.ieeevis.org/2008/}, 2008.

\bibitem{ThomasN13}
D.~M. Thomas and V.~Natarajan.
\newblock Detecting symmetry in scalar fields using augmented extremum graphs.
\newblock {\em {IEEE} Trans. Vis. Comput. Graph.}, 19(12):2663--2672, 2013.
  \href{https://doi.org/10.1109/TVCG.2013.148}
{doi: {{%
10\hspace{.1pt}\discretionary{.}{%
}{.}\hspace{.4pt}1109\discretionary{/}{%
}{/}TVCG\hspace{.1pt}\discretionary{.}{%
}{.}\hspace{.4pt}2013\hspace{.1pt}\discretionary{.}{%
}{.}\hspace{.4pt}148}}}


\bibitem{DBLP:journals/tvcg/TiernyFLGM18}
J.~Tierny, G.~Favelier, J.~A. Levine, C.~Gueunet, and M.~Michaux.
\newblock The topology toolkit.
\newblock {\em {IEEE} Trans. Vis. Comput. Graph.}, 24(1):832--842, 2018.
  \href{https://doi.org/10.1109/TVCG.2017.2743938}
{doi: {{%
10\hspace{.1pt}\discretionary{.}{%
}{.}\hspace{.4pt}1109\discretionary{/}{%
}{/}TVCG\hspace{.1pt}\discretionary{.}{%
}{.}\hspace{.4pt}2017\hspace{.1pt}\discretionary{.}{%
}{.}\hspace{.4pt}2743938}}}


\bibitem{weinkauf10c}
T.~Weinkauf and H.~Theisel.
\newblock Streak lines as tangent curves of a derived vector field.
\newblock {\em IEEE Transactions on Visualization and Computer Graphics
  (Proceedings Visualization 2010)}, 16(6):1225--1234, November - December
  2010.

\bibitem{taming}
F.~Wetzels, M.~Anders, and C.~Garth.
\newblock Taming horizontal instability in merge trees: On the computation of a
  comprehensive deformation-based edit distance.
\newblock In {\em 2023 Topological Data Analysis and Visualization
  (TopoInVis)}, pp. 82--92, 2023.
  \href{https://doi.org/10.1109/TopoInVis60193.2023.00015}
{doi: {{%
10\hspace{.1pt}\discretionary{.}{%
}{.}\hspace{.4pt}1109\discretionary{/}{%
}{/}TopoInVis60193\hspace{.1pt}\discretionary{.}{%
}{.}\hspace{.4pt}2023\hspace{.1pt}\discretionary{.}{%
}{.}\hspace{.4pt}00015}}}


\bibitem{wetzels2022path}
F.~Wetzels and C.~Garth.
\newblock A deformation-based edit distance for merge trees.
\newblock In {\em 2022 Topological Data Analysis and Visualization
  (TopoInVis)}, pp. 29--38, 2022.
  \href{https://doi.org/10.1109/TopoInVis57755.2022.00010}
{doi: {{%
10\hspace{.1pt}\discretionary{.}{%
}{.}\hspace{.4pt}1109\discretionary{/}{%
}{/}TopoInVis57755\hspace{.1pt}\discretionary{.}{%
}{.}\hspace{.4pt}2022\hspace{.1pt}\discretionary{.}{%
}{.}\hspace{.4pt}00010}}}


\bibitem{wetzels2022branch}
F.~Wetzels, H.~Leitte, and C.~Garth.
\newblock Branch decomposition-independent edit distances for merge trees.
\newblock {\em Computer Graphics Forum}, 41(3):367--378, 2022.
  \href{https://doi.org/10.1111/cgf.14547}
{doi: {{%
10\hspace{.1pt}\discretionary{.}{%
}{.}\hspace{.4pt}1111\discretionary{/}{%
}{/}cgf\hspace{.1pt}\discretionary{.}{%
}{.}\hspace{.4pt}14547}}}


\bibitem{repository_lookahead}
F.~Wetzels, H.~Leitte, and C.~Garth.
\newblock Accelerating computation of stable merge tree edit distances using
  parameterized heuristics (supplementary source code).
\newblock \url{https://github.com/scivislab/Path-Mappings-with-Lookahead},
  2025.

\bibitem{10.2312:evs.20241069}
F.~Wetzels, T.~B. Masood, N.~H. List, I.~Hotz, and C.~Garth.
\newblock {Exploring Electron Density Evolution using Merge Tree Mappings}.
\newblock In C.~Tominski, M.~Waldner, and B.~Wang, eds., {\em EuroVis 2024 -
  Short Papers}. The Eurographics Association, 2024.
  \href{https://doi.org/10.2312/evs.20241069}
{doi: {{%
10\hspace{.1pt}\discretionary{.}{%
}{.}\hspace{.4pt}2312\discretionary{/}{%
}{/}evs\hspace{.1pt}\discretionary{.}{%
}{.}\hspace{.4pt}20241069}}}


\bibitem{DBLP:journals/tvcg/WetzelsPTG24}
F.~Wetzels, M.~Pont, J.~Tierny, and C.~Garth.
\newblock Merge tree geodesics and barycenters with path mappings.
\newblock {\em {IEEE} Trans. Vis. Comput. Graph.}, 30(1):1095--1105, 2024.
  \href{https://doi.org/10.1109/TVCG.2023.3326601}
{doi: {{%
10\hspace{.1pt}\discretionary{.}{%
}{.}\hspace{.4pt}1109\discretionary{/}{%
}{/}TVCG\hspace{.1pt}\discretionary{.}{%
}{.}\hspace{.4pt}2023\hspace{.1pt}\discretionary{.}{%
}{.}\hspace{.4pt}3326601}}}


\bibitem{Yan_geometry_aware}
L.~Yan, T.~Bin~Masood, F.~Rasheed, I.~Hotz, and B.~Wang.
\newblock Geometry aware merge tree comparisons for time-varying data with
  interleaving distances.
\newblock {\em IEEE Transactions on Visualization and Computer Graphics}, pp.
  1--1, 2022. \href{https://doi.org/10.1109/TVCG.2022.3163349}
{doi: {{%
10\hspace{.1pt}\discretionary{.}{%
}{.}\hspace{.4pt}1109\discretionary{/}{%
}{/}TVCG\hspace{.1pt}\discretionary{.}{%
}{.}\hspace{.4pt}2022\hspace{.1pt}\discretionary{.}{%
}{.}\hspace{.4pt}3163349}}}


\bibitem{surveyComparison2021}
L.~Yan, T.~B. Masood, R.~Sridharamurthy, F.~Rasheed, V.~Natarajan, I.~Hotz, and
  B.~Wang.
\newblock Scalar field comparison with topological descriptors: Properties and
  applications for scientific visualization.
\newblock {\em Comput. Graph. Forum}, 40(3):599--633, 2021.
  \href{https://doi.org/10.1111/cgf.14331}
{doi: {{%
10\hspace{.1pt}\discretionary{.}{%
}{.}\hspace{.4pt}1111\discretionary{/}{%
}{/}cgf\hspace{.1pt}\discretionary{.}{%
}{.}\hspace{.4pt}14331}}}


\bibitem{YanWMGW20}
L.~Yan, Y.~Wang, E.~Munch, E.~Gasparovic, and B.~Wang.
\newblock A structural average of labeled merge trees for uncertainty
  visualization.
\newblock {\em {IEEE} Trans. Vis. Comput. Graph.}, 26(1):832--842, 2020.
  \href{https://doi.org/10.1109/TVCG.2019.2934242}
{doi: {{%
10\hspace{.1pt}\discretionary{.}{%
}{.}\hspace{.4pt}1109\discretionary{/}{%
}{/}TVCG\hspace{.1pt}\discretionary{.}{%
}{.}\hspace{.4pt}2019\hspace{.1pt}\discretionary{.}{%
}{.}\hspace{.4pt}2934242}}}


\bibitem{DBLP:journals/algorithmica/Zhang96}
K.~Zhang.
\newblock A constrained edit distance between unordered labeled trees.
\newblock {\em Algorithmica}, 15(3):205--222, 1996.
  \href{https://doi.org/10.1007/BF01975866}
{doi: {{%
10\hspace{.1pt}\discretionary{.}{%
}{.}\hspace{.4pt}1007\discretionary{/}{%
}{/}BF01975866}}}


\bibitem{DBLP:journals/ipl/ZhangSS92}
K.~Zhang, R.~Statman, and D.~E. Shasha.
\newblock On the editing distance between unordered labeled trees.
\newblock {\em Inf. Process. Lett.}, 42(3):133--139, 1992.
  \href{https://doi.org/10.1016/0020-0190(92)90136-J}
{doi: {{%
10\hspace{.1pt}\discretionary{.}{%
}{.}\hspace{.4pt}1016\discretionary{/}{%
}{/}0020\discretionary{%
}{-}{-}0190\discretionary{%
}{(}{(}92\discretionary{)}{%
}{)}90136\discretionary{%
}{-}{-}J}}}


\end{thebibliography}

\clearpage

\appendix

\section{Algorithm Details}
\label{sec:app_alg}

In this section, we provide additional descriptions and illustrations for the algorithms described in Sections~\ref{sec:background} and~\ref{sec:method} of the main paper.

\subsection{Path Mapping and Look-ahead Recursion}
\label{sec:app_rec}

First, we add a more detailed description of the path mapping recursion and its adaptions for the look-ahead heuristic, as well as pseudocode and additional illustrations.

Figure~\ref{fig:recursion} shows four cases of the recursive structure.
The first two come from the original path mapping recursion.
There, we have overall two types of cases (intuitively speaking) for subtrees $T_1[(n_1,p_1)],T_2[(n_2,p_2)]$:
\begin{itemize}
    \item[(a)] We match the root paths $(n_1,p_1),(n_2,p_2)$ onto each other and continue with recursive cases for the subtrees rooted in the children of $n_1,n_2$. In this case, the optimal assignment happens on the two sets of subtrees directly below $n_1,n_2$.
    \item[(b)] We delete one of the subtrees rooted in the children of $n_1$ or $n_2$, e.g.\ the subtree rooted in $(v_{1,2},n_1)$ in Figure~\ref{fig:recursion}. In this case, we continue in one recursive case, $T_1[(v_{1,1},p_1)]$ (the remaining child of $n_1$) with $T_2[(n_2,p_2)]$ (the second tree unchanged).
\end{itemize}

In the extended variant with look-ahead, we also add cases for deletions of edges without deleting the full subtree below them.
These deletions come in the form of collapsing sets of consecutive edges below $n_1,n_2$.
In Figure~\ref{fig:recursion}~(c,d), two example deletions are shown.

The complete algorithm in psuedocode is given in Alogirthm~\ref{alg:edist}.
\textcolor{revision}{The code uses $\delta_h(T_1,T_2,n_1,p_1,n_2,p_2)$ synonym to $\delta_h(T_1[n1,p_1],T_2[n_2,p_2])$.}
The lines that have been added for the look-ahead heuristic (and are not part of the original path mapping algorithm) are highlighted in blue.
Lines~2 to~14 describe the trivial base-cases: one of the trees is either empty or has only one edge.
Lines~17 to~21 describe the three cases for non-trivial trees as discussed in the main paper and visualized in Figure~\ref{fig:recursion}(a,b).
Line~22 adds the fourth case for the look-ahead collapse procedure, illustrated in Figure~\ref{fig:recursion}(c,d).

\SetKwComment{Comment}{/* }{ */}
\setlength{\algomargin}{5pt}
\SetAlgoVlined
\begin{algorithm}[!ht]
\caption{Path mapping distance with look-ahead}
\label{alg:edist}
\SetKwFunction{pathDist}{$\laedist$}
\SetKwProg{Fn}{Function}{:}{}
\DontPrintSemicolon
\Fn{\pathDist{$T_1,T_2,n_1,p_1,n_2,p_2$}}{
\If{$n_1=\bot$ and $n_2$ is a leaf}{
  \Return $\cost(\bot,p_2...n_2)$\;
}
\If{$n_2=\bot$ and $n_1$ is a leaf}{
  \Return $\cost(p_1...n_1,\bot)$\;
}
\If{$n_1=\bot$ and $n_2$ is an inner node}{
  \Return~$\cost(\bot,p_2...n_2) + \sum_{c_2 \in C_{T_2}(n_2)} \pathDist(\bot,\bot,c_2,n_2)$\;
}
\If{$n_2=\bot$ and $n_1$ is an inner node}{
  \Return~$\cost(p_1...n_1,\bot) + \sum_{c_1 \in C_{T_1}(n_1)} \pathDist(c_1,n_1,\bot,\bot)$\;
}
\If{$n_1$ is a leaf and $n_2$ is a leaf}{
  \Return $\cost(p_1...n_1,p_2...n_2)$\;
}
\If{$n_1$ is a leaf and $n_2$ is an inner node}{
  \Return $ \min_{c_2 \in C_{T_2}(n_2)} \{ \pathDist(n_1,p_1,c_2,p_2) 
  - \pathDist(\bot,\bot,c_2,n_2)
  + \sum_{c_2' \in C_{T_2}(n_2)} \pathDist(\bot,\bot,c_2',n_2) \} $\;
}
\If{$n_2$ is a leaf and $n_1$ is an inner node}{
  \Return $ \min_{c_1 \in C_{T_1}(n_1)} \{ \pathDist(c_1,p_1,n_2,p_2) 
  - \pathDist(c_1,n_1,\bot,\bot)
  + \sum_{c_1' \in C_{T_1}(n_1)} \pathDist(c_1',n_1,\bot,\bot) \} $\;
}
\If{$n_1$ is an inner node and $n_2$ is an inner node}{
  $d_1 \coloneqq \min_{c_2 \in C_{T_2}(n_2)} \{ \pathDist(n_1,p_1,c_2,p_2) 
  - \pathDist(\bot,\bot,c_2,n_2)
  + \sum_{c_2' \in C_{T_2}(n_2)} \pathDist(\bot,\bot,c_2',n_2) \} $\;
  $d_2 \coloneqq \min_{c_1 \in C_{T_1}(n_1)} \{ \pathDist(c_1,p_1,n_2,p_2) 
  - \pathDist(c_1,n_1,\bot,\bot)
  + \sum_{c_1' \in C_{T_1}(n_1)} \pathDist(c_1',n_1,\bot,\bot) \} $\;
  $C_1 = \bigcup_{x \in C_{T_1}(n_1)} (x,n_1)$\;
  $C_2 = \bigcup_{x \in C_{T_2}(n_2)} (x,n_2)$\;
  $d_3 \coloneqq \min\{ \cost(m,\laedist) \mid m \in PM(C_1,C_2) \}$\;
  \textcolor{blue}{$d_4 \coloneqq \texttt{OptCollapse}(T_1,T_2,n_1,n_2,h)$}\;
  \Return~$\min(d_1,d_2,\min(d_3,\textcolor{blue}{d_4})+\cost(p_1...n_1,p_2...n_2))$\;
}
}
\end{algorithm}

\begin{figure*}
\centering

\begin{subfigure}[t]{0.8\linewidth}
    \centering
    \resizebox{0.99\linewidth}{!}{
    \begin{tikzpicture}[xscale=0.6,yscale=0.8]
    \node[draw=none,fill=none,circle] at (-16, -2.5) (dummy) {};
    \node[draw=none,fill=none,circle] at (16, -2.5) (dummy) {};
    \node[draw=none,fill=none,circle] at (-16, 0.5) (dummy) {};
    \node[draw=none,fill=none,circle] at (16, 0.5) (dummy) {};
    
    \begin{scope}[shift={(-6.5,0)}]
    \node[] at (0, 0) (p1) {\tiny$p_1$};
    \node[] at (0, -1) (n1) {\tiny$n_1$};
    \node[] at (-1,-2) (c11) {\tiny$v_{1,1}$};
    \node[] at (1, -2) (c12) {\tiny$v_{1,2}$};
    \draw [ultra thick,CornflowerBlue] (p1) -- (n1);
    \draw [ultra thick,OrangeRed] (n1) -- (c11);
    \draw [ultra thick,OrangeRed] (n1) -- (c12);
    \draw [very thick,black] (c11) -- (-1.5,-2.4) -- (-1.75,-2.75) -- (-1.25,-2.75) -- (-1.5,-2.4);
    \draw [very thick,black] (c11) -- (-0.5,-2.4) -- (-0.75,-2.75) -- (-0.25,-2.75) -- (-0.5,-2.4);
    \draw [very thick,black] (c12) -- (1.5,-2.4) -- (1.75,-2.75) -- (1.25,-2.75) -- (1.5,-2.4);
    \draw [very thick,black] (c12) -- (0.5,-2.4) -- (0.75,-2.75) -- (0.25,-2.75) -- (0.5,-2.4);
    \end{scope}
    
    \begin{scope}[shift={(-2.5,0)}]
    \node[] at (0, 0) (p2) {\tiny$p_2$};
    \node[] at (0, -1) (n2) {\tiny$n_2$};
    \node[] at (-1, -2) (c21) {\tiny$v_{2,1}$};
    \node[] at (1, -2) (c22) {\tiny$v_{2,2}$};
    \draw [ultra thick,CornflowerBlue] (p2) -- (n2);
    \draw [ultra thick,OrangeRed] (n2) -- (c21);
    \draw [ultra thick,OrangeRed] (n2) -- (c22);
    \draw [very thick,black] (c21) -- (-1.5,-2.4) -- (-1.75,-2.75) -- (-1.25,-2.75) -- (-1.5,-2.4);
    \draw [very thick,black] (c21) -- (-0.5,-2.4) -- (-0.75,-2.75) -- (-0.25,-2.75) -- (-0.5,-2.4);
    \draw [very thick,black] (c22) -- (1.5,-2.4) -- (1.75,-2.75) -- (1.25,-2.75) -- (1.5,-2.4);
    \draw [very thick,black] (c22) -- (0.5,-2.4) -- (0.75,-2.75) -- (0.25,-2.75) -- (0.5,-2.4);
    \end{scope}
    
    \node[] at (-4.5,-0.6) (label) {\large\textbf{(a)}};
    
    \begin{scope}[shift={(2.5,0)}]
    \node[] at (0, 0) (p2) {\tiny$p_1$};
    \node[] at (0, -1) (n2) {\tiny$n_1$};
    \node[] at (-1, -2) (c21) {\tiny$v_{1,1}$};
    \node[] at (1, -2) (c22) {\tiny$v_{1,2}$};
    \draw [ultra thick,CornflowerBlue] (p2) -- (n2);
    \draw [ultra thick,CornflowerBlue] (n2) -- (c21);
    \draw [ultra thick,gray!40] (n2) -- (c22);
    \draw [very thick,black] (c21) -- (-1.5,-2.4) -- (-1.75,-2.75) -- (-1.25,-2.75) -- (-1.5,-2.4);
    \draw [very thick,black] (c21) -- (-0.5,-2.4) -- (-0.75,-2.75) -- (-0.25,-2.75) -- (-0.5,-2.4);
    \draw [very thick,black,gray!40] (c22) -- (1.5,-2.4) -- (1.75,-2.75) -- (1.25,-2.75) -- (1.5,-2.4);
    \draw [very thick,black,gray!40] (c22) -- (0.5,-2.4) -- (0.75,-2.75) -- (0.25,-2.75) -- (0.5,-2.4);
    \end{scope}
    
    \begin{scope}[shift={(6.5,0)}]
    \node[] at (0, 0) (p2) {\tiny$p_2$};
    \node[] at (0, -1) (n2) {\tiny$n_2$};
    \node[] at (-1, -2) (c21) {\tiny$v_{2,1}$};
    \node[] at (1, -2) (c22) {\tiny$v_{2,2}$};
    \draw [ultra thick,CornflowerBlue] (p2) -- (n2);
    \draw [ultra thick] (n2) -- (c21);
    \draw [ultra thick] (n2) -- (c22);
    \draw [very thick,black] (c21) -- (-1.5,-2.4) -- (-1.75,-2.75) -- (-1.25,-2.75) -- (-1.5,-2.4);
    \draw [very thick,black] (c21) -- (-0.5,-2.4) -- (-0.75,-2.75) -- (-0.25,-2.75) -- (-0.5,-2.4);
    \draw [very thick,black] (c22) -- (1.5,-2.4) -- (1.75,-2.75) -- (1.25,-2.75) -- (1.5,-2.4);
    \draw [very thick,black] (c22) -- (0.5,-2.4) -- (0.75,-2.75) -- (0.25,-2.75) -- (0.5,-2.4);
    \end{scope}
    
    \node[] at (4.5,-0.6) (label) {\large\textbf{(b)}};
    
    \end{tikzpicture}
    }
    \caption*{
    	\centering
    	Two cases from the original recursion without look-ahead: In (a) we match $(p_1 \dots n_1)$ to $(p_2 \dots n_2)$; collapse no edges; and compute the optimal assignment between $\{T_1[(v_{1,1},n_1)],T_1[(v_{1,2},n_1)]\}$ and $\{T_2[(v_{2,1},n_2)],T_2[(v_{2,2},n_2)]\}$.
    	\mbox{In (b) we collapse no edges; delete $T_1[(v_{1,2},n_1)]$; and continue recursively in $T_1[(v_{1,1},p_1)]$ and $T_1[(n_2,p_2)]$.}}
\end{subfigure}

\begin{subfigure}[t]{0.8\linewidth}
    \centering
    \resizebox{0.99\linewidth}{!}{
    \begin{tikzpicture}[xscale=0.6,yscale=0.8]
    \node[draw=none,fill=none,circle] at (-16, -4.5) (dummy) {};
    \node[draw=none,fill=none,circle] at (16, -4.5) (dummy) {};
    \node[draw=none,fill=none,circle] at (-16, 0.5) (dummy) {};
    \node[draw=none,fill=none,circle] at (16, 0.5) (dummy) {};
    
    \begin{scope}[shift={(-8,0)}]
    \node[] at (0, 0) (p1) {\small$p_1$};
    \node[] at (0, -1) (n1) {\small$n_1$};
    \node[] at (-4,-2) (v11) {\small$v_{1,1}$};
    \node[] at (4, -2) (v12) {\small$v_{1,2}$};
    \node[] at (-6, -3) (u11) {\small$u_{1,1}$};
    \node[] at (-2, -3) (u12) {\small$u_{1,2}$};
    \node[] at (2, -3) (u13) {\small$u_{1,3}$};
    \node[] at (6, -3) (u14) {\small$u_{1,4}$};
    \node[] at (-7, -4) (c11) {\small$c_{1,1}$};
    \node[] at (-5, -4) (c12) {\small$c_{1,2}$};
    \node[] at (-3, -4) (c13) {\small$c_{1,3}$};
    \node[] at (-1, -4) (c14) {\small$c_{1,4}$};
    \node[] at (1, -4) (c15) {\small$c_{1,5}$};
    \node[] at (3, -4) (c16) {\small$c_{1,6}$};
    \node[] at (5, -4) (c17) {\small$c_{1,7}$};
    \node[] at (7, -4) (c18) {\small$c_{1,8}$};
    \draw [ultra thick,CornflowerBlue] (p1) -- (n1);
    \draw [ultra thick,gray!40] (n1) -- (v11);
    \draw [ultra thick,gray!40] (n1) -- (v12);
    \draw [ultra thick,OrangeRed] (v11) -- (u11);
    \draw [ultra thick,OrangeRed] (v11) -- (u12);
    \draw [ultra thick,OrangeRed] (v12) -- (u13);
    \draw [ultra thick,OrangeRed] (v12) -- (u14);
    \draw [ultra thick] (u11) -- (c11);
    \draw [ultra thick] (u11) -- (c12);
    \draw [ultra thick] (u12) -- (c13);
    \draw [ultra thick] (u12) -- (c14);
    \draw [ultra thick] (u13) -- (c15);
    \draw [ultra thick] (u13) -- (c16);
    \draw [ultra thick] (u14) -- (c17);
    \draw [ultra thick] (u14) -- (c18);
    \draw [very thick,black] (c11) -- ($(c11) + (-0.5,-0.4)$) -- ($(c11) + (-0.75,-0.75)$) -- ($(c11) + (-0.25,-0.75)$) -- ($(c11) + (-0.5,-0.4)$);
    \draw [very thick,black] (c11) -- ($(c11) + (0.5,-0.4)$) -- ($(c11) + (0.25,-0.75)$) -- ($(c11) + (0.75,-0.75)$) -- ($(c11) + (0.5,-0.4)$);
    \draw [very thick,black] (c12) -- ($(c12) + (-0.5,-0.4)$) -- ($(c12) + (-0.75,-0.75)$) -- ($(c12) + (-0.25,-0.75)$) -- ($(c12) + (-0.5,-0.4)$);
    \draw [very thick,black] (c12) -- ($(c12) + (0.5,-0.4)$) -- ($(c12) + (0.25,-0.75)$) -- ($(c12) + (0.75,-0.75)$) -- ($(c12) + (0.5,-0.4)$);
    \draw [very thick,black] (c13) -- ($(c13) + (-0.5,-0.4)$) -- ($(c13) + (-0.75,-0.75)$) -- ($(c13) + (-0.25,-0.75)$) -- ($(c13) + (-0.5,-0.4)$);
    \draw [very thick,black] (c13) -- ($(c13) + (0.5,-0.4)$) -- ($(c13) + (0.25,-0.75)$) -- ($(c13) + (0.75,-0.75)$) -- ($(c13) + (0.5,-0.4)$);
    \draw [very thick,black] (c14) -- ($(c14) + (-0.5,-0.4)$) -- ($(c14) + (-0.75,-0.75)$) -- ($(c14) + (-0.25,-0.75)$) -- ($(c14) + (-0.5,-0.4)$);
    \draw [very thick,black] (c14) -- ($(c14) + (0.5,-0.4)$) -- ($(c14) + (0.25,-0.75)$) -- ($(c14) + (0.75,-0.75)$) -- ($(c14) + (0.5,-0.4)$);
    \draw [very thick,black] (c15) -- ($(c15) + (-0.5,-0.4)$) -- ($(c15) + (-0.75,-0.75)$) -- ($(c15) + (-0.25,-0.75)$) -- ($(c15) + (-0.5,-0.4)$);
    \draw [very thick,black] (c15) -- ($(c15) + (0.5,-0.4)$) -- ($(c15) + (0.25,-0.75)$) -- ($(c15) + (0.75,-0.75)$) -- ($(c15) + (0.5,-0.4)$);
    \draw [very thick,black] (c16) -- ($(c16) + (-0.5,-0.4)$) -- ($(c16) + (-0.75,-0.75)$) -- ($(c16) + (-0.25,-0.75)$) -- ($(c16) + (-0.5,-0.4)$);
    \draw [very thick,black] (c16) -- ($(c16) + (0.5,-0.4)$) -- ($(c16) + (0.25,-0.75)$) -- ($(c16) + (0.75,-0.75)$) -- ($(c16) + (0.5,-0.4)$);
    \draw [very thick,black] (c17) -- ($(c17) + (-0.5,-0.4)$) -- ($(c17) + (-0.75,-0.75)$) -- ($(c17) + (-0.25,-0.75)$) -- ($(c17) + (-0.5,-0.4)$);
    \draw [very thick,black] (c17) -- ($(c17) + (0.5,-0.4)$) -- ($(c17) + (0.25,-0.75)$) -- ($(c17) + (0.75,-0.75)$) -- ($(c17) + (0.5,-0.4)$);
    \draw [very thick,black] (c18) -- ($(c18) + (-0.5,-0.4)$) -- ($(c18) + (-0.75,-0.75)$) -- ($(c18) + (-0.25,-0.75)$) -- ($(c18) + (-0.5,-0.4)$);
    \draw [very thick,black] (c18) -- ($(c18) + (0.5,-0.4)$) -- ($(c18) + (0.25,-0.75)$) -- ($(c18) + (0.75,-0.75)$) -- ($(c18) + (0.5,-0.4)$);
    \end{scope}

    \begin{scope}[shift={(8,0)}]

    \node[] at (0, 0) (p2) {\small$p_2$};
    \node[] at (0, -1) (n2) {\small$n_2$};
    \node[] at (-4,-2) (v21) {\small$v_{2,1}$};
    \node[] at (4, -2) (v22) {\small$v_{2,2}$};
    \node[] at (-6, -3) (u21) {\small$u_{2,1}$};
    \node[] at (-2, -3) (u22) {\small$u_{2,2}$};
    \node[] at (2, -3) (u23) {\small$u_{2,3}$};
    \node[] at (6, -3) (u24) {\small$u_{2,4}$};
    \node[] at (-7, -4) (c21) {\small$c_{2,1}$};
    \node[] at (-5, -4) (c22) {\small$c_{2,2}$};
    \node[] at (-3, -4) (c23) {\small$c_{2,3}$};
    \node[] at (-1, -4) (c24) {\small$c_{2,4}$};
    \node[] at (1, -4) (c25) {\small$c_{2,5}$};
    \node[] at (3, -4) (c26) {\small$c_{2,6}$};
    \node[] at (5, -4) (c27) {\small$c_{2,7}$};
    \node[] at (7, -4) (c28) {\small$c_{2,8}$};
    \draw [ultra thick,CornflowerBlue] (p2) -- (n2);
    \draw [ultra thick,gray!40] (n2) -- (v21);
    \draw [ultra thick,gray!40] (n2) -- (v22);
    \draw [ultra thick,OrangeRed] (v21) -- (u21);
    \draw [ultra thick,OrangeRed] (v21) -- (u22);
    \draw [ultra thick,OrangeRed] (v22) -- (u23);
    \draw [ultra thick,OrangeRed] (v22) -- (u24);
    \draw [ultra thick] (u21) -- (c21);
    \draw [ultra thick] (u21) -- (c22);
    \draw [ultra thick] (u22) -- (c23);
    \draw [ultra thick] (u22) -- (c24);
    \draw [ultra thick] (u23) -- (c25);
    \draw [ultra thick] (u23) -- (c26);
    \draw [ultra thick] (u24) -- (c27);
    \draw [ultra thick] (u24) -- (c28);
    \draw [very thick,black] (c21) -- ($(c21) + (-0.5,-0.4)$) -- ($(c21) + (-0.75,-0.75)$) -- ($(c21) + (-0.25,-0.75)$) -- ($(c21) + (-0.5,-0.4)$);
    \draw [very thick,black] (c21) -- ($(c21) + (0.5,-0.4)$) -- ($(c21) + (0.25,-0.75)$) -- ($(c21) + (0.75,-0.75)$) -- ($(c21) + (0.5,-0.4)$);
    \draw [very thick,black] (c22) -- ($(c22) + (-0.5,-0.4)$) -- ($(c22) + (-0.75,-0.75)$) -- ($(c22) + (-0.25,-0.75)$) -- ($(c22) + (-0.5,-0.4)$);
    \draw [very thick,black] (c22) -- ($(c22) + (0.5,-0.4)$) -- ($(c22) + (0.25,-0.75)$) -- ($(c22) + (0.75,-0.75)$) -- ($(c22) + (0.5,-0.4)$);
    \draw [very thick,black] (c23) -- ($(c23) + (-0.5,-0.4)$) -- ($(c23) + (-0.75,-0.75)$) -- ($(c23) + (-0.25,-0.75)$) -- ($(c23) + (-0.5,-0.4)$);
    \draw [very thick,black] (c23) -- ($(c23) + (0.5,-0.4)$) -- ($(c23) + (0.25,-0.75)$) -- ($(c23) + (0.75,-0.75)$) -- ($(c23) + (0.5,-0.4)$);
    \draw [very thick,black] (c24) -- ($(c24) + (-0.5,-0.4)$) -- ($(c24) + (-0.75,-0.75)$) -- ($(c24) + (-0.25,-0.75)$) -- ($(c24) + (-0.5,-0.4)$);
    \draw [very thick,black] (c24) -- ($(c24) + (0.5,-0.4)$) -- ($(c24) + (0.25,-0.75)$) -- ($(c24) + (0.75,-0.75)$) -- ($(c24) + (0.5,-0.4)$);
    \draw [very thick,black] (c25) -- ($(c25) + (-0.5,-0.4)$) -- ($(c25) + (-0.75,-0.75)$) -- ($(c25) + (-0.25,-0.75)$) -- ($(c25) + (-0.5,-0.4)$);
    \draw [very thick,black] (c25) -- ($(c25) + (0.5,-0.4)$) -- ($(c25) + (0.25,-0.75)$) -- ($(c25) + (0.75,-0.75)$) -- ($(c25) + (0.5,-0.4)$);
    \draw [very thick,black] (c26) -- ($(c26) + (-0.5,-0.4)$) -- ($(c26) + (-0.75,-0.75)$) -- ($(c26) + (-0.25,-0.75)$) -- ($(c26) + (-0.5,-0.4)$);
    \draw [very thick,black] (c26) -- ($(c26) + (0.5,-0.4)$) -- ($(c26) + (0.25,-0.75)$) -- ($(c26) + (0.75,-0.75)$) -- ($(c26) + (0.5,-0.4)$);
    \draw [very thick,black] (c27) -- ($(c27) + (-0.5,-0.4)$) -- ($(c27) + (-0.75,-0.75)$) -- ($(c27) + (-0.25,-0.75)$) -- ($(c27) + (-0.5,-0.4)$);
    \draw [very thick,black] (c27) -- ($(c27) + (0.5,-0.4)$) -- ($(c27) + (0.25,-0.75)$) -- ($(c27) + (0.75,-0.75)$) -- ($(c27) + (0.5,-0.4)$);
    \draw [very thick,black] (c28) -- ($(c28) + (-0.5,-0.4)$) -- ($(c28) + (-0.75,-0.75)$) -- ($(c28) + (-0.25,-0.75)$) -- ($(c28) + (-0.5,-0.4)$);
    \draw [very thick,black] (c28) -- ($(c28) + (0.5,-0.4)$) -- ($(c28) + (0.25,-0.75)$) -- ($(c28) + (0.75,-0.75)$) -- ($(c28) + (0.5,-0.4)$);

    \end{scope}
    
    \node[] at (0,-1.6) (label) {\large\textbf{(c)}};
    
    \end{tikzpicture}
    }
    \caption*{\centering Example case: we match $(p_1 \dots n_1)$ to $(p_2 \dots n_2)$; collapse $\{(v_{1,1},n_1),(v_{1,2},n_1)\}$ and $\{(v_{2,1},n_2),(v_{2,1},n_2)\}$; and compute the optimal assignment between $\{T_1[(u_{1,1},v_{1,1})],T_1[(u_{1,2},v_{1,1})],T_1[(u_{1,3},v_{1,2})],T_1[(u_{1,4},v_{1,2})]\}$ and $\{T_2[(u_{2,1},v_{2,1})],T_2[(u_{2,2},v_{2,1})],T_2[(u_{2,3},v_{2,2})],T_2[(u_{2,4},v_{2,2})]\}$.}
\end{subfigure}

\begin{subfigure}[t]{0.8\linewidth}
    \centering
    \resizebox{0.99\linewidth}{!}{
    \begin{tikzpicture}[xscale=0.6,yscale=0.8]
    \node[draw=none,fill=none,circle] at (-16, -4.5) (dummy) {};
    \node[draw=none,fill=none,circle] at (16, -4.5) (dummy) {};
    \node[draw=none,fill=none,circle] at (-16, 0.5) (dummy) {};
    \node[draw=none,fill=none,circle] at (16, 0.5) (dummy) {};
    
    \begin{scope}[shift={(-8,0)}]
    \node[] at (0, 0) (p1) {\small$p_1$};
    \node[] at (0, -1) (n1) {\small$n_1$};
    \node[] at (-4,-2) (v11) {\small$v_{1,1}$};
    \node[] at (4, -2) (v12) {\small$v_{1,2}$};
    \node[] at (-6, -3) (u11) {\small$u_{1,1}$};
    \node[] at (-2, -3) (u12) {\small$u_{1,2}$};
    \node[] at (2, -3) (u13) {\small$u_{1,3}$};
    \node[] at (6, -3) (u14) {\small$u_{1,4}$};
    \node[] at (-7, -4) (c11) {\small$c_{1,1}$};
    \node[] at (-5, -4) (c12) {\small$c_{1,2}$};
    \node[] at (-3, -4) (c13) {\small$c_{1,3}$};
    \node[] at (-1, -4) (c14) {\small$c_{1,4}$};
    \node[] at (1, -4) (c15) {\small$c_{1,5}$};
    \node[] at (3, -4) (c16) {\small$c_{1,6}$};
    \node[] at (5, -4) (c17) {\small$c_{1,7}$};
    \node[] at (7, -4) (c18) {\small$c_{1,8}$};
    \draw [ultra thick,CornflowerBlue] (p1) -- (n1);
    \draw [ultra thick,gray!40] (n1) -- (v11);
    \draw [ultra thick,gray!40] (n1) -- (v12);
    \draw [ultra thick,gray!40] (v11) -- (u11);
    \draw [ultra thick,gray!40] (v11) -- (u12);
    \draw [ultra thick,OrangeRed] (v12) -- (u13);
    \draw [ultra thick,gray!40] (v12) -- (u14);
    \draw [ultra thick,OrangeRed] (u11) -- (c11);
    \draw [ultra thick,OrangeRed] (u11) -- (c12);
    \draw [ultra thick,OrangeRed] (u12) -- (c13);
    \draw [ultra thick,OrangeRed] (u12) -- (c14);
    \draw [ultra thick] (u13) -- (c15);
    \draw [ultra thick] (u13) -- (c16);
    \draw [ultra thick,OrangeRed] (u14) -- (c17);
    \draw [ultra thick,OrangeRed] (u14) -- (c18);
    \draw [very thick,black] (c11) -- ($(c11) + (-0.5,-0.4)$) -- ($(c11) + (-0.75,-0.75)$) -- ($(c11) + (-0.25,-0.75)$) -- ($(c11) + (-0.5,-0.4)$);
    \draw [very thick,black] (c11) -- ($(c11) + (0.5,-0.4)$) -- ($(c11) + (0.25,-0.75)$) -- ($(c11) + (0.75,-0.75)$) -- ($(c11) + (0.5,-0.4)$);
    \draw [very thick,black] (c12) -- ($(c12) + (-0.5,-0.4)$) -- ($(c12) + (-0.75,-0.75)$) -- ($(c12) + (-0.25,-0.75)$) -- ($(c12) + (-0.5,-0.4)$);
    \draw [very thick,black] (c12) -- ($(c12) + (0.5,-0.4)$) -- ($(c12) + (0.25,-0.75)$) -- ($(c12) + (0.75,-0.75)$) -- ($(c12) + (0.5,-0.4)$);
    \draw [very thick,black] (c13) -- ($(c13) + (-0.5,-0.4)$) -- ($(c13) + (-0.75,-0.75)$) -- ($(c13) + (-0.25,-0.75)$) -- ($(c13) + (-0.5,-0.4)$);
    \draw [very thick,black] (c13) -- ($(c13) + (0.5,-0.4)$) -- ($(c13) + (0.25,-0.75)$) -- ($(c13) + (0.75,-0.75)$) -- ($(c13) + (0.5,-0.4)$);
    \draw [very thick,black] (c14) -- ($(c14) + (-0.5,-0.4)$) -- ($(c14) + (-0.75,-0.75)$) -- ($(c14) + (-0.25,-0.75)$) -- ($(c14) + (-0.5,-0.4)$);
    \draw [very thick,black] (c14) -- ($(c14) + (0.5,-0.4)$) -- ($(c14) + (0.25,-0.75)$) -- ($(c14) + (0.75,-0.75)$) -- ($(c14) + (0.5,-0.4)$);
    \draw [very thick,black] (c15) -- ($(c15) + (-0.5,-0.4)$) -- ($(c15) + (-0.75,-0.75)$) -- ($(c15) + (-0.25,-0.75)$) -- ($(c15) + (-0.5,-0.4)$);
    \draw [very thick,black] (c15) -- ($(c15) + (0.5,-0.4)$) -- ($(c15) + (0.25,-0.75)$) -- ($(c15) + (0.75,-0.75)$) -- ($(c15) + (0.5,-0.4)$);
    \draw [very thick,black] (c16) -- ($(c16) + (-0.5,-0.4)$) -- ($(c16) + (-0.75,-0.75)$) -- ($(c16) + (-0.25,-0.75)$) -- ($(c16) + (-0.5,-0.4)$);
    \draw [very thick,black] (c16) -- ($(c16) + (0.5,-0.4)$) -- ($(c16) + (0.25,-0.75)$) -- ($(c16) + (0.75,-0.75)$) -- ($(c16) + (0.5,-0.4)$);
    \draw [very thick,black] (c17) -- ($(c17) + (-0.5,-0.4)$) -- ($(c17) + (-0.75,-0.75)$) -- ($(c17) + (-0.25,-0.75)$) -- ($(c17) + (-0.5,-0.4)$);
    \draw [very thick,black] (c17) -- ($(c17) + (0.5,-0.4)$) -- ($(c17) + (0.25,-0.75)$) -- ($(c17) + (0.75,-0.75)$) -- ($(c17) + (0.5,-0.4)$);
    \draw [very thick,black] (c18) -- ($(c18) + (-0.5,-0.4)$) -- ($(c18) + (-0.75,-0.75)$) -- ($(c18) + (-0.25,-0.75)$) -- ($(c18) + (-0.5,-0.4)$);
    \draw [very thick,black] (c18) -- ($(c18) + (0.5,-0.4)$) -- ($(c18) + (0.25,-0.75)$) -- ($(c18) + (0.75,-0.75)$) -- ($(c18) + (0.5,-0.4)$);
    \end{scope}

    \begin{scope}[shift={(8,0)}]

    \node[] at (0, 0) (p2) {\small$p_2$};
    \node[] at (0, -1) (n2) {\small$n_2$};
    \node[] at (-4,-2) (v21) {\small$v_{2,1}$};
    \node[] at (4, -2) (v22) {\small$v_{2,2}$};
    \node[] at (-6, -3) (u21) {\small$u_{2,1}$};
    \node[] at (-2, -3) (u22) {\small$u_{2,2}$};
    \node[] at (2, -3) (u23) {\small$u_{2,3}$};
    \node[] at (6, -3) (u24) {\small$u_{2,4}$};
    \node[] at (-7, -4) (c21) {\small$c_{2,1}$};
    \node[] at (-5, -4) (c22) {\small$c_{2,2}$};
    \node[] at (-3, -4) (c23) {\small$c_{2,3}$};
    \node[] at (-1, -4) (c24) {\small$c_{2,4}$};
    \node[] at (1, -4) (c25) {\small$c_{2,5}$};
    \node[] at (3, -4) (c26) {\small$c_{2,6}$};
    \node[] at (5, -4) (c27) {\small$c_{2,7}$};
    \node[] at (7, -4) (c28) {\small$c_{2,8}$};
    \draw [ultra thick,CornflowerBlue] (p2) -- (n2);
    \draw [ultra thick,gray!40] (n2) -- (v21);
    \draw [ultra thick,gray!40] (n2) -- (v22);
    \draw [ultra thick,OrangeRed] (v21) -- (u21);
    \draw [ultra thick,gray!40] (v21) -- (u22);
    \draw [ultra thick,OrangeRed] (v22) -- (u23);
    \draw [ultra thick,OrangeRed] (v22) -- (u24);
    \draw [ultra thick] (u21) -- (c21);
    \draw [ultra thick] (u21) -- (c22);
    \draw [ultra thick,OrangeRed] (u22) -- (c23);
    \draw [ultra thick,OrangeRed] (u22) -- (c24);
    \draw [ultra thick] (u23) -- (c25);
    \draw [ultra thick] (u23) -- (c26);
    \draw [ultra thick] (u24) -- (c27);
    \draw [ultra thick] (u24) -- (c28);
    \draw [very thick,black] (c21) -- ($(c21) + (-0.5,-0.4)$) -- ($(c21) + (-0.75,-0.75)$) -- ($(c21) + (-0.25,-0.75)$) -- ($(c21) + (-0.5,-0.4)$);
    \draw [very thick,black] (c21) -- ($(c21) + (0.5,-0.4)$) -- ($(c21) + (0.25,-0.75)$) -- ($(c21) + (0.75,-0.75)$) -- ($(c21) + (0.5,-0.4)$);
    \draw [very thick,black] (c22) -- ($(c22) + (-0.5,-0.4)$) -- ($(c22) + (-0.75,-0.75)$) -- ($(c22) + (-0.25,-0.75)$) -- ($(c22) + (-0.5,-0.4)$);
    \draw [very thick,black] (c22) -- ($(c22) + (0.5,-0.4)$) -- ($(c22) + (0.25,-0.75)$) -- ($(c22) + (0.75,-0.75)$) -- ($(c22) + (0.5,-0.4)$);
    \draw [very thick,black] (c23) -- ($(c23) + (-0.5,-0.4)$) -- ($(c23) + (-0.75,-0.75)$) -- ($(c23) + (-0.25,-0.75)$) -- ($(c23) + (-0.5,-0.4)$);
    \draw [very thick,black] (c23) -- ($(c23) + (0.5,-0.4)$) -- ($(c23) + (0.25,-0.75)$) -- ($(c23) + (0.75,-0.75)$) -- ($(c23) + (0.5,-0.4)$);
    \draw [very thick,black] (c24) -- ($(c24) + (-0.5,-0.4)$) -- ($(c24) + (-0.75,-0.75)$) -- ($(c24) + (-0.25,-0.75)$) -- ($(c24) + (-0.5,-0.4)$);
    \draw [very thick,black] (c24) -- ($(c24) + (0.5,-0.4)$) -- ($(c24) + (0.25,-0.75)$) -- ($(c24) + (0.75,-0.75)$) -- ($(c24) + (0.5,-0.4)$);
    \draw [very thick,black] (c25) -- ($(c25) + (-0.5,-0.4)$) -- ($(c25) + (-0.75,-0.75)$) -- ($(c25) + (-0.25,-0.75)$) -- ($(c25) + (-0.5,-0.4)$);
    \draw [very thick,black] (c25) -- ($(c25) + (0.5,-0.4)$) -- ($(c25) + (0.25,-0.75)$) -- ($(c25) + (0.75,-0.75)$) -- ($(c25) + (0.5,-0.4)$);
    \draw [very thick,black] (c26) -- ($(c26) + (-0.5,-0.4)$) -- ($(c26) + (-0.75,-0.75)$) -- ($(c26) + (-0.25,-0.75)$) -- ($(c26) + (-0.5,-0.4)$);
    \draw [very thick,black] (c26) -- ($(c26) + (0.5,-0.4)$) -- ($(c26) + (0.25,-0.75)$) -- ($(c26) + (0.75,-0.75)$) -- ($(c26) + (0.5,-0.4)$);
    \draw [very thick,black] (c27) -- ($(c27) + (-0.5,-0.4)$) -- ($(c27) + (-0.75,-0.75)$) -- ($(c27) + (-0.25,-0.75)$) -- ($(c27) + (-0.5,-0.4)$);
    \draw [very thick,black] (c27) -- ($(c27) + (0.5,-0.4)$) -- ($(c27) + (0.25,-0.75)$) -- ($(c27) + (0.75,-0.75)$) -- ($(c27) + (0.5,-0.4)$);
    \draw [very thick,black] (c28) -- ($(c28) + (-0.5,-0.4)$) -- ($(c28) + (-0.75,-0.75)$) -- ($(c28) + (-0.25,-0.75)$) -- ($(c28) + (-0.5,-0.4)$);
    \draw [very thick,black] (c28) -- ($(c28) + (0.5,-0.4)$) -- ($(c28) + (0.25,-0.75)$) -- ($(c28) + (0.75,-0.75)$) -- ($(c28) + (0.5,-0.4)$);

    \end{scope}
    
    \node[] at (0,-1.6) (label) {\large\textbf{(d)}};
    
    \end{tikzpicture}
    }
    \caption*{\centering Example case: we match $(p_1 \dots n_1)$ to $(p_2 \dots n_2)$; collapse $\{(v_{1,1},n_1),(v_{1,2},n_1),(u_{1,1},v_{1,1}),(u_{1,2},v_{1,1}),(u_{1,4},v_{1,2})\}$ and $\{(v_{2,1},n_2),(v_{2,1},n_2),(u_{2,2},v_{2,1})\}$; and compute the optimal assignment between $\{T_1[(c_{1,1},u_{1,1})],T_1[(c_{1,2},u_{1,1})],T_1[(c_{1,3},u_{1,2})],T_1[(c_{1,4},u_{1,2})],T_1[(u_{1,3},v_{1,2})],T_1[(c_{1,7},u_{1,4})],T_1[(c_{1,8},u_{1,4})]\}$ and $\{T_2[(u_{2,1},v_{2,1})],T_2[(c_{2,3},u_{2,2})],T_2[(c_{2,4},u_{2,2})],T_2[(u_{2,3},v_{2,2})],T_2[(u_{2,4},v_{2,2})]\}$.}
\end{subfigure}
\caption{Illustrations of example cases in the original (a,b) and  look-ahead (c,d) recursion. Fixed matchings are shown in blue, collapsed edges in gray, root edges of the subtrees in the optimal assignment instance in red.}
\label{fig:recursion}
\end{figure*}

\begin{figure*}
    \centering
    \captionsetup[subfigure]{aboveskip=-1pt,belowskip=-1pt}
    
    \begin{subfigure}[t]{0.2\linewidth}
    \includegraphics[width=\linewidth]{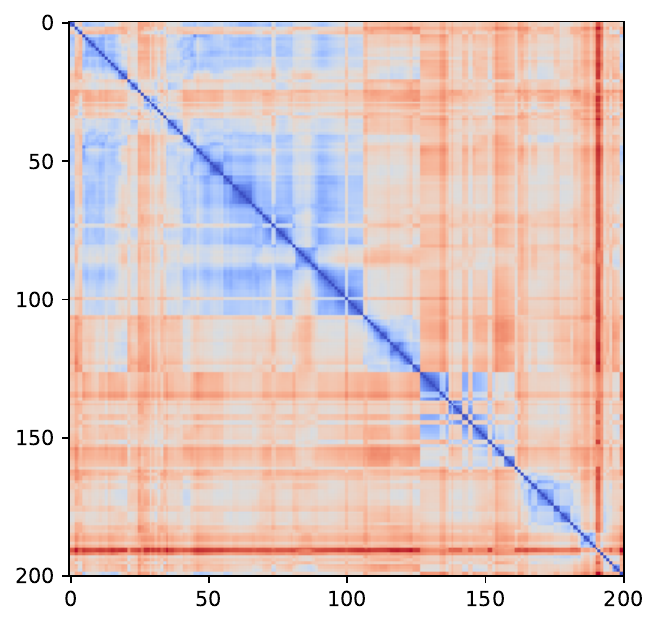}
    \caption{Wasserstein Distance}
    \end{subfigure}
    \begin{subfigure}[t]{0.2\linewidth}
    \includegraphics[width=\linewidth]{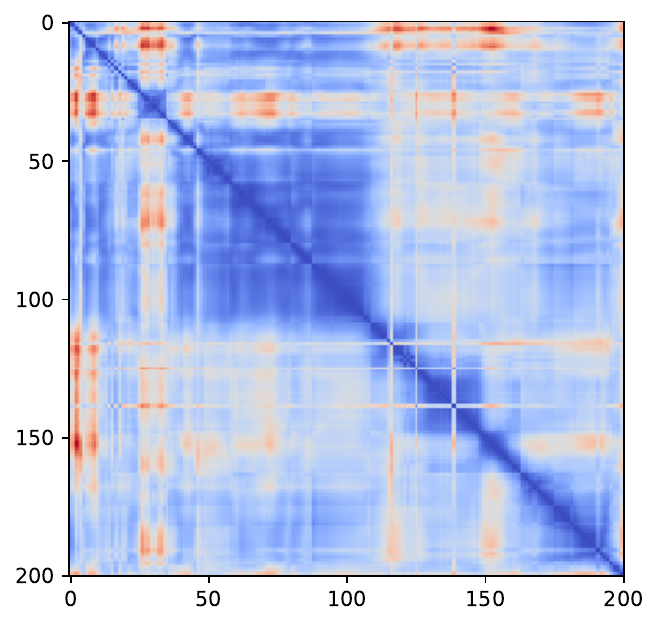}
    \caption{Merge Tree Edit Distance}
    \end{subfigure}
    \begin{subfigure}[t]{0.2\linewidth}
    \includegraphics[width=\linewidth]{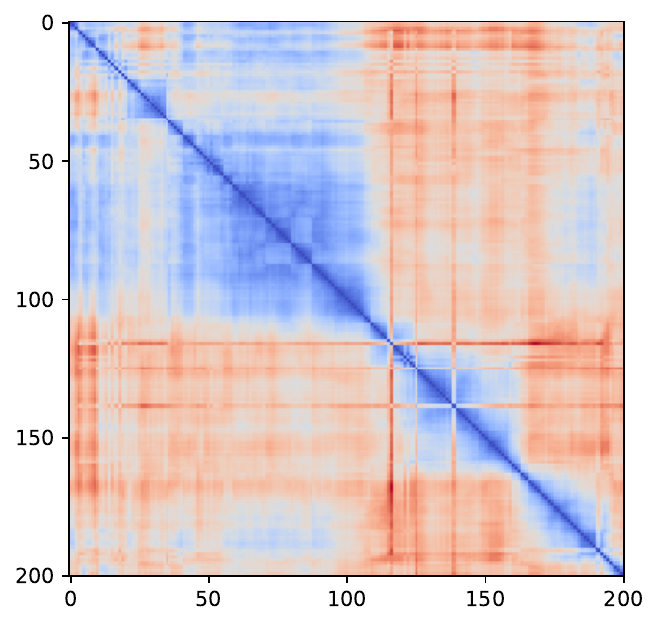}
    \caption{Look-ahead 0}
    \end{subfigure}
    \begin{subfigure}[t]{0.2\linewidth}
    \includegraphics[width=\linewidth]{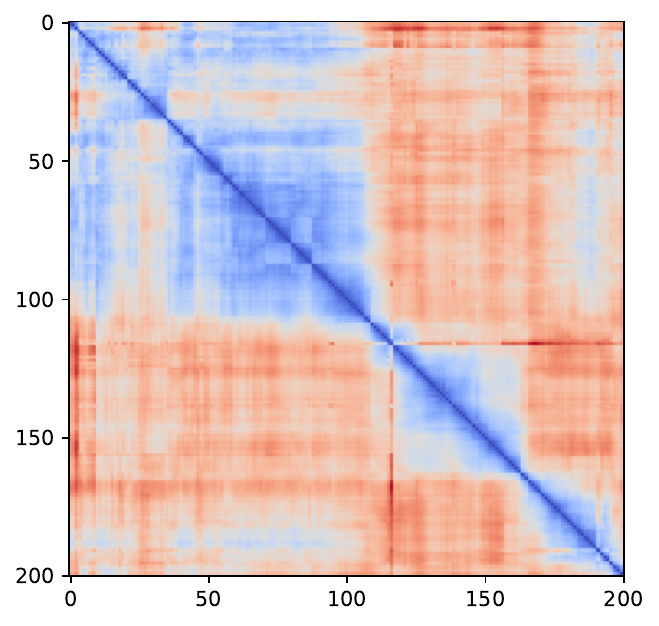}
    \caption{Look-ahead 1}
    \end{subfigure}
    
    \raisebox{10pt}{\includegraphics[angle=90,width=0.013\linewidth]{figures/legend_dist.pdf}}
    \begin{subfigure}[t]{0.2\linewidth}
    \includegraphics[width=\linewidth]{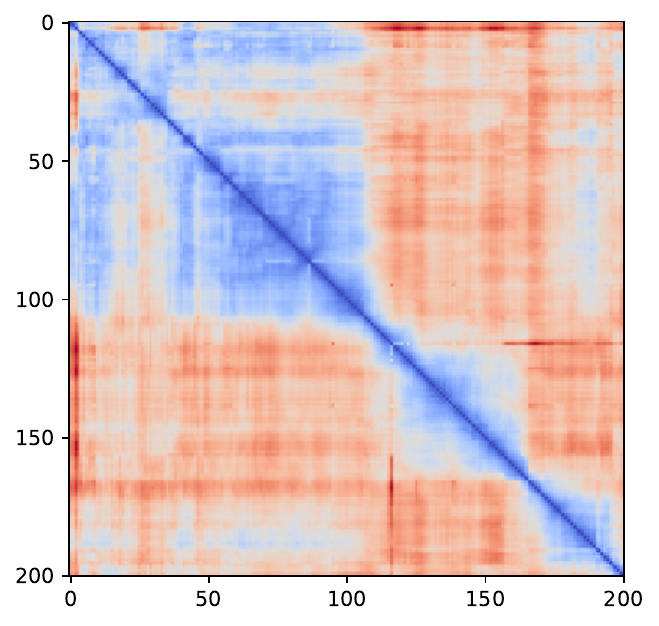}
    \caption{Look-ahead 2}
    \end{subfigure}
    \begin{subfigure}[t]{0.2\linewidth}
    \includegraphics[width=\linewidth]{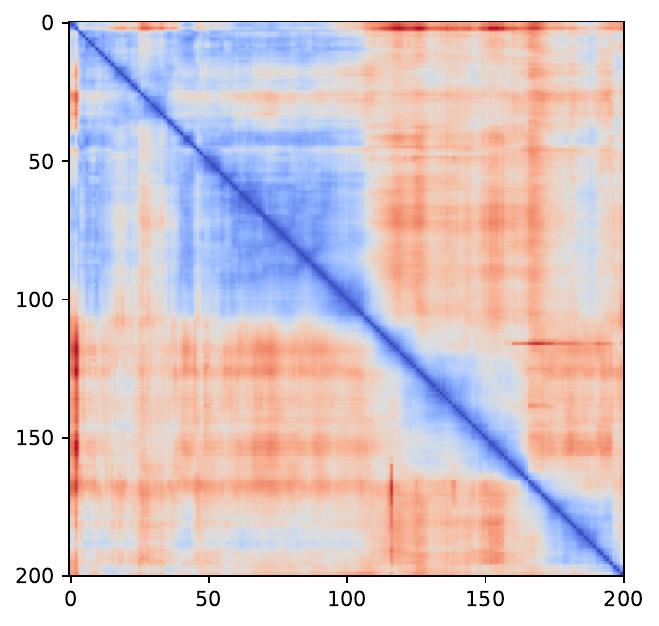}
    \caption{Look-ahead 3}
    \end{subfigure}
    \begin{subfigure}[t]{0.2\linewidth}
    \includegraphics[width=\linewidth]{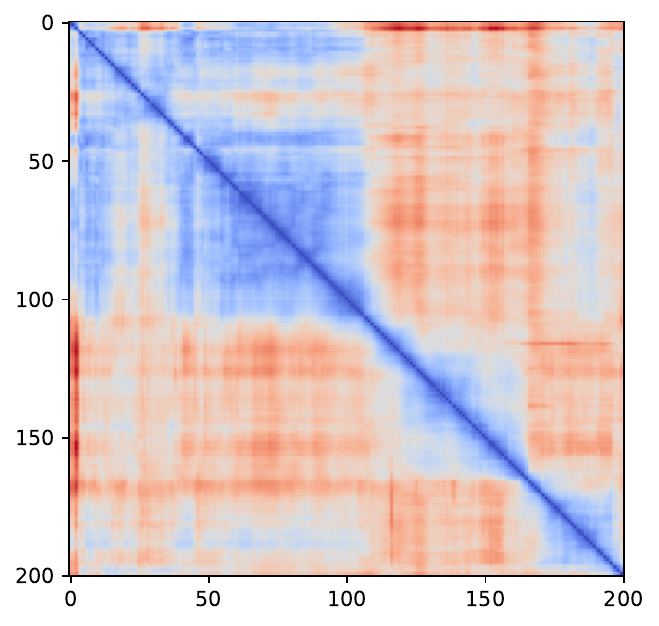}
    \caption{Look-ahead 4}
    \end{subfigure}
    \vspace{-6pt}
    \caption{Distance matrices on the ionization front time series: comparison of all computed distances.}
    \label{fig:ionization_mat_app}
\end{figure*}

\subsection{Generation Algorithm}
\label{sec:app_alg}

In this section, we provide a more detailed description of the generation procedure in Algorithm~\ref{alg:collapse} of the main paper.

The original path mapping algorithm iterates all combinations of nodes $n_1 \in V(T_1),n_2 \in V(T_2)$ with ancestor nodes $p_1 \in V(T_1),p_2 \in V(T_2)$ in a bottom-up fashion.
For each node combination, we extend the recursion by generating the sets $L_{T_1}(\sces_h(n_1,T_1))$ and $L_{T_2}(\sces_h(n_2,T_2))$ as described below.
Note that we generate the leaf sets of $\sces_h(n_1,T_1)$ and $\sces_h(n_2,T_2)$ directly.

We maintain a worklist of candidate contractions stored in the following way: instead of keeping track of the contracted set $E$, we only maintain the costs $c_E$ of contracting $E$; we also maintain the leafs $L_E$ computed until now (here leaf refers to the leaves of $E$, not the full tree); and lastly we maintain a reference to the next edge $e$ to either keep or contract.
We begin with one entry: costs~$0$, an empty set of leafs and the first child edge of $n_1$/$n_2$ as the next edge.
Note that we assume some left-to-right ordering of the children of a node in the tree data structure, even though the merge tree is interpreted as an unordered tree.

When handling such a tuple $(c_E,L_E,e)$ with $e=(x,y)$, we either contract $e$ or retain it.
If a leaf edge $e$ is contracted (i.e.\ added to $E$), its deletion cost $\cost(e,\bot)$ added to $c_E$ and the next edge is set to $(x',x)$ where $x'$ is the left-most child of $x$.
If it is not collapsed (i.e.\ not added to $E$), the costs $c_E$ remain unchanged, $e$ is added to $L_E$ and the next edge is set to $(x'',P_T(x''))$, where $x''$ is the right sibling of $y'$ with $y'$ being the lowest node that is an ancestor of $x$ and a descendant of $n_1$/$n_2$.
If such a $y'$ does not exist, the candidate tuple is finished.
Both cases are potentially pushed back onto the worklist: we add $(c_E+\cost(e,\bot),L_E,(x',x))$ and $(c_E,L_E\cup\{e\},(x'',P_T(x'')))$.
If $e$ is outside the look-ahead range of $n_1$/$n_2$, we do not consider the contraction of $e$.
If no next edge $(x'',y')$ exists, we do not continue but add $L_E$ to the generated set.

\section{TOSCA Ensemble}
\label{sec:app_tosca}

\begin{figure*}
    \centering
    \captionsetup[subfigure]{aboveskip=-1pt,belowskip=-1pt}
    
    \begin{subfigure}[t]{0.24\linewidth}
    \includegraphics[width=\linewidth]{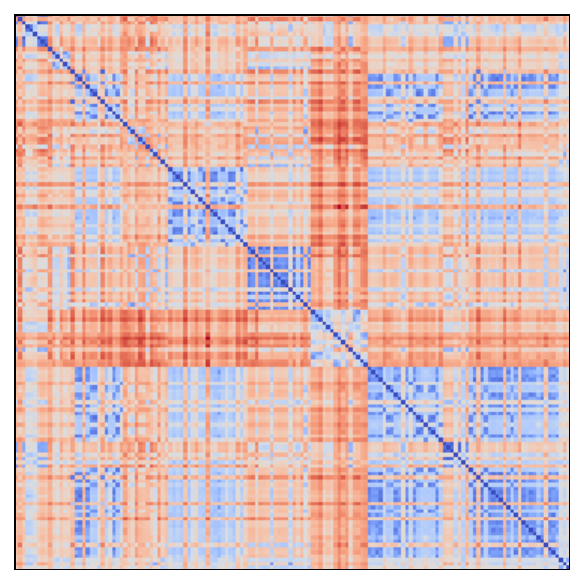}
    \caption{Wasserstein Distance}
    \end{subfigure}
    \begin{subfigure}[t]{0.24\linewidth}
    \includegraphics[width=\linewidth]{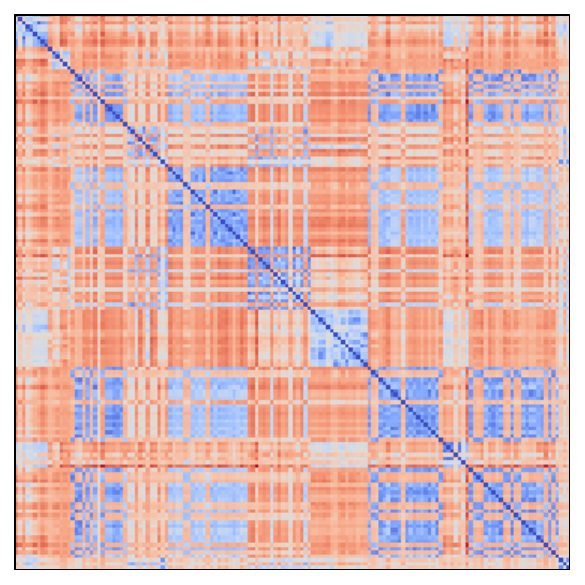}
    \caption{Merge Tree Edit Distance}
    \end{subfigure}
    \begin{subfigure}[t]{0.24\linewidth}
    \includegraphics[width=\linewidth]{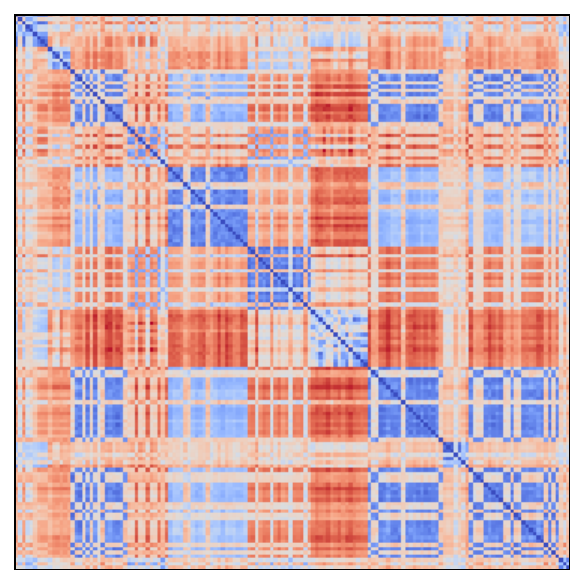}
    \caption{Path Mapping Distance}
    \end{subfigure}
    \begin{subfigure}[t]{0.24\linewidth}
    \includegraphics[width=\linewidth]{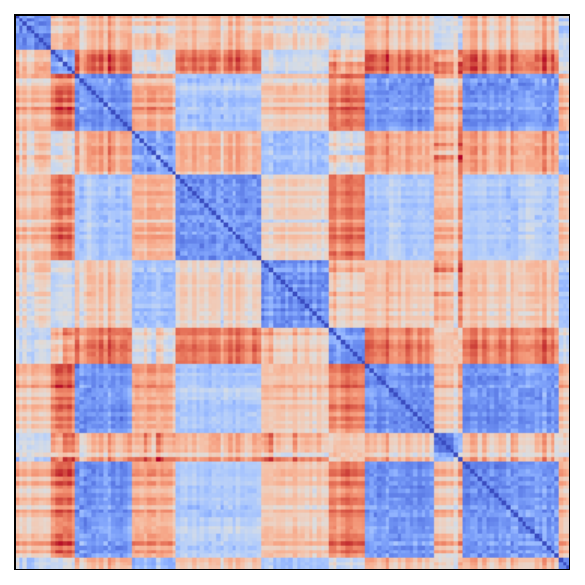}
    \caption{Unconstrained Deformation Distance}
    \end{subfigure}
    
    \begin{subfigure}[t]{0.24\linewidth}
    \includegraphics[width=\linewidth]{figures/tosca/dm_pmd1.pdf}
    \caption{Look-ahead 1}
    \end{subfigure}
    \begin{subfigure}[t]{0.24\linewidth}
    \includegraphics[width=\linewidth]{figures/tosca/dm_pmd2.pdf}
    \caption{Look-ahead 2}
    \end{subfigure}
    \begin{subfigure}[t]{0.24\linewidth}
    \includegraphics[width=\linewidth]{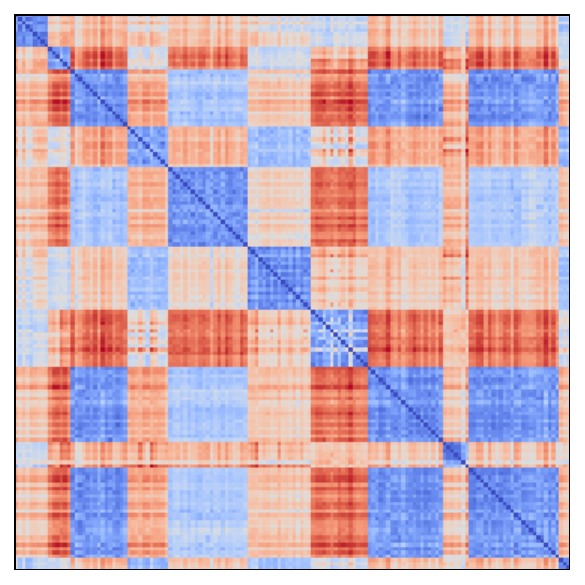}
    \caption{Look-ahead 3}
    \end{subfigure}
    \begin{subfigure}[t]{0.24\linewidth}
    \includegraphics[width=\linewidth]{figures/tosca/dm_pmd4.pdf}
    \caption{Look-ahead 4}
    \end{subfigure}
    
    \begin{subfigure}[t]{0.24\linewidth}
    \includegraphics[width=\linewidth]{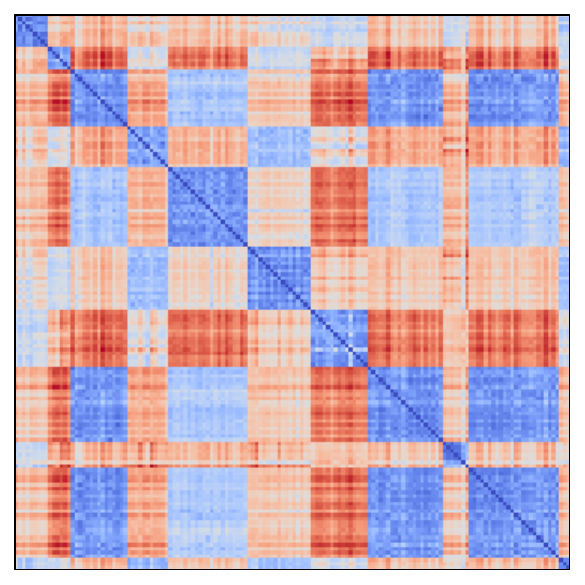}
    \caption{Look-ahead 5}
    \end{subfigure}
    \begin{subfigure}[t]{0.24\linewidth}
    \includegraphics[width=\linewidth]{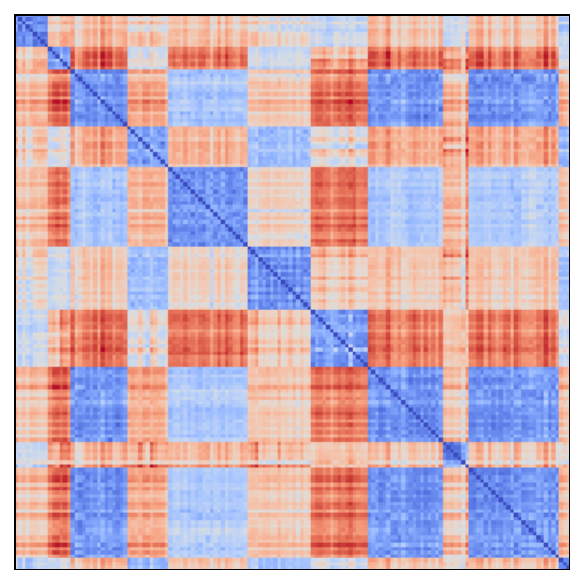}
    \caption{Look-ahead 6}
    \end{subfigure}
    \begin{subfigure}[t]{0.24\linewidth}
    \includegraphics[width=\linewidth]{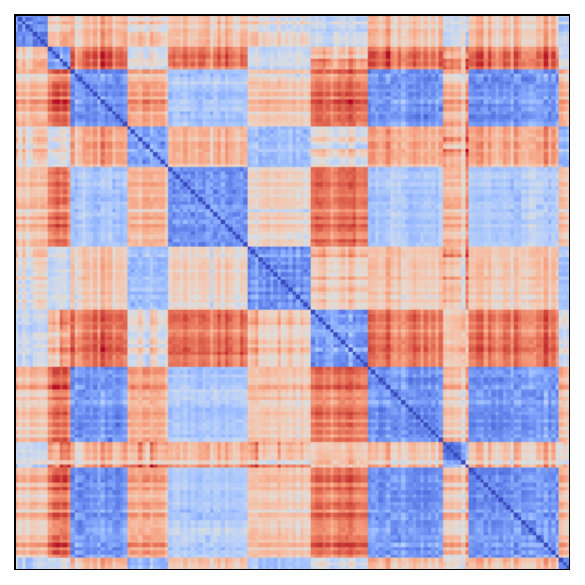}
    \caption{Look-ahead 7}
    \end{subfigure}
    \begin{subfigure}[t]{0.24\linewidth}
    \includegraphics[width=\linewidth]{figures/tosca/dm_pmd8.pdf}
    \caption{Look-ahead 8}
    \end{subfigure}
    \caption{Distance matrices on the TOSCA ensemble using various different distance metrics.}
    \label{fig:tosca_mat_all}
\end{figure*}

\begin{figure*}
    \centering
    \includegraphics[width=0.7\linewidth]{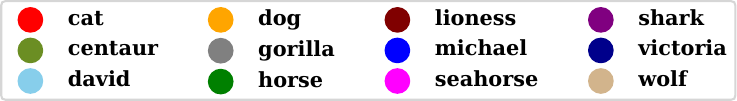}
    
    \captionsetup[subfigure]{aboveskip=-1pt,belowskip=-1pt}
    \begin{subfigure}[t]{0.24\linewidth}
    \includegraphics[width=\linewidth]{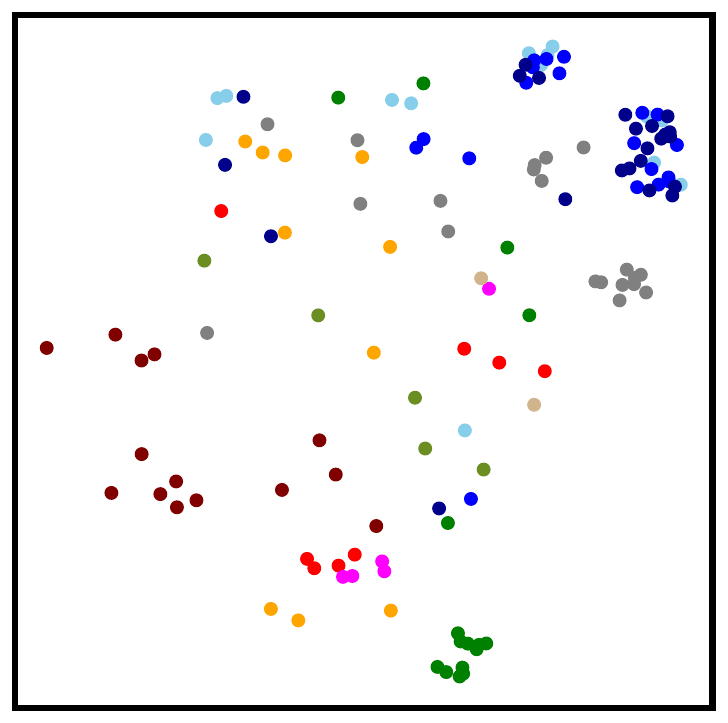}
    \caption{Wasserstein Distance}
    \end{subfigure}
    \begin{subfigure}[t]{0.24\linewidth}
    \includegraphics[width=\linewidth]{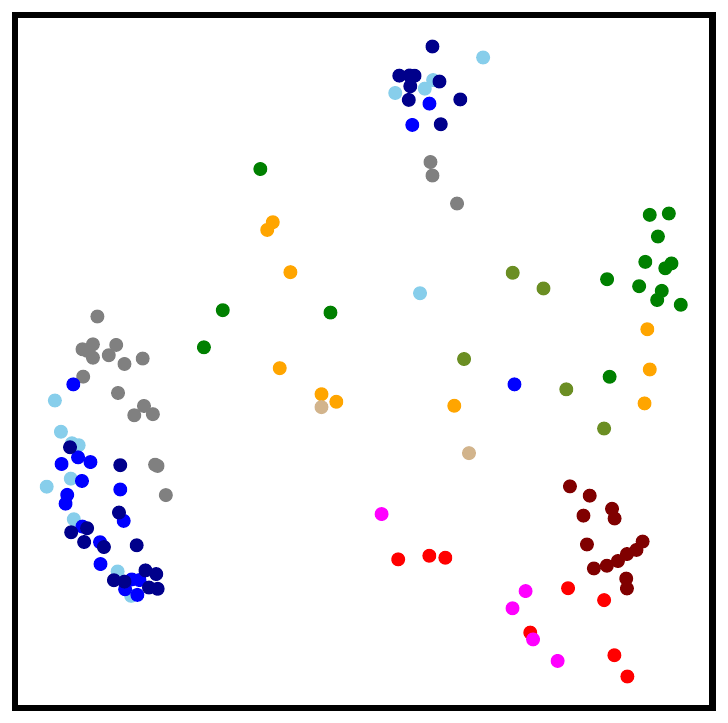}
    \caption{Merge Tree Edit Distance}
    \end{subfigure}
    \begin{subfigure}[t]{0.24\linewidth}
    \includegraphics[width=\linewidth]{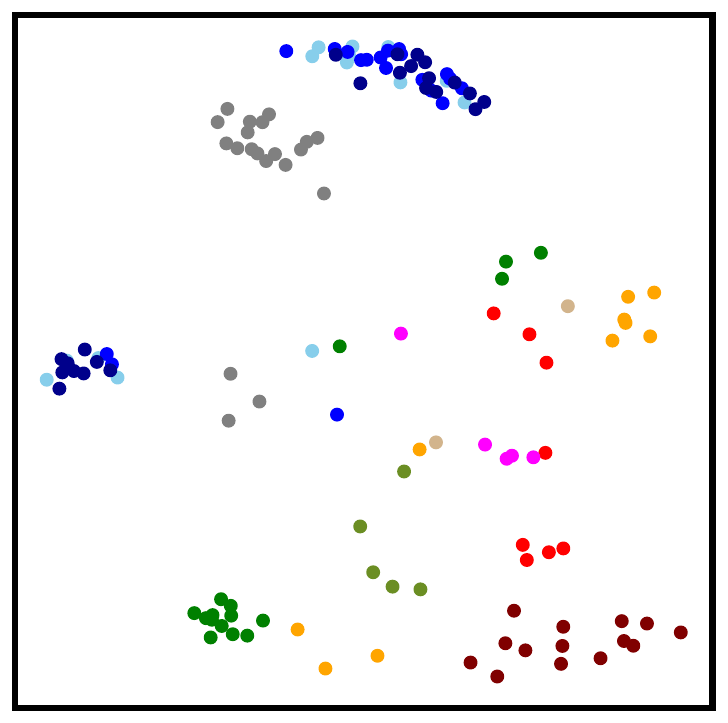}
    \caption{Path Mapping Distance}
    \end{subfigure}
    \begin{subfigure}[t]{0.24\linewidth}
    \includegraphics[width=\linewidth]{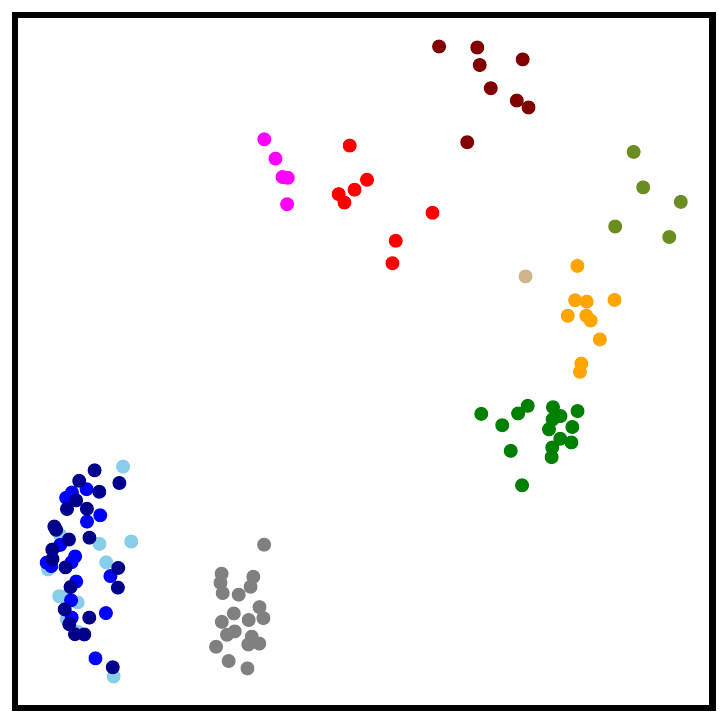}
    \caption{Unconstrained Deformation Distance}
    \end{subfigure}
    
    \begin{subfigure}[t]{0.24\linewidth}
    \includegraphics[width=\linewidth]{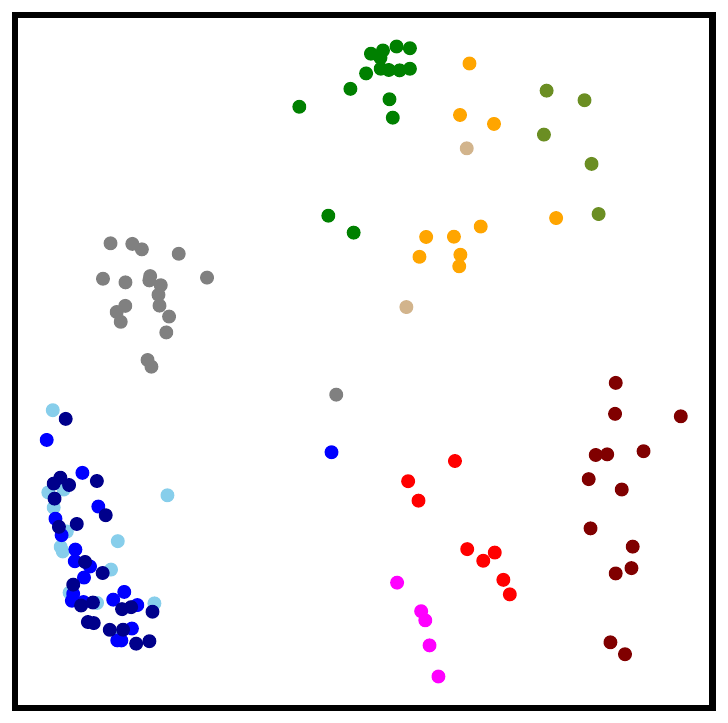}
    \caption{Look-ahead 1}
    \end{subfigure}
    \begin{subfigure}[t]{0.24\linewidth}
    \includegraphics[width=\linewidth]{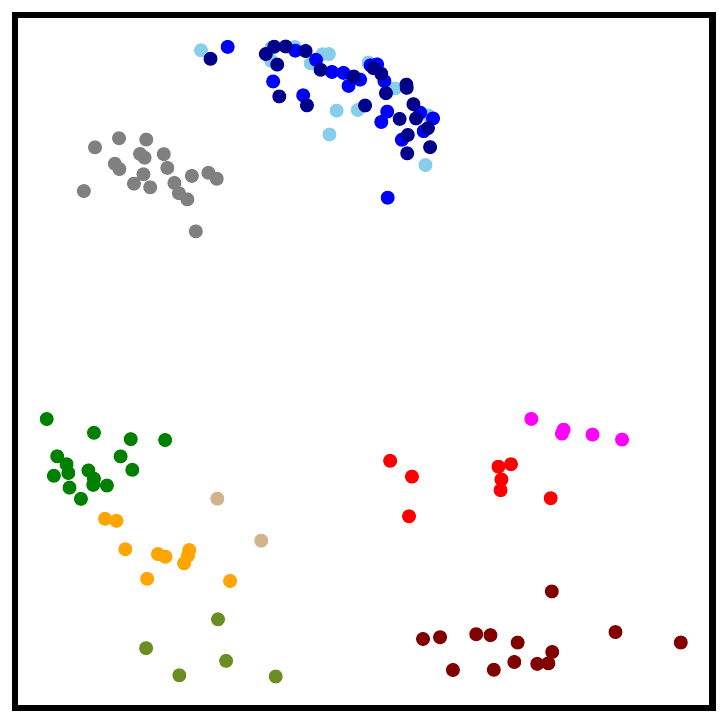}
    \caption{Look-ahead 2}
    \end{subfigure}
    \begin{subfigure}[t]{0.24\linewidth}
    \includegraphics[width=\linewidth]{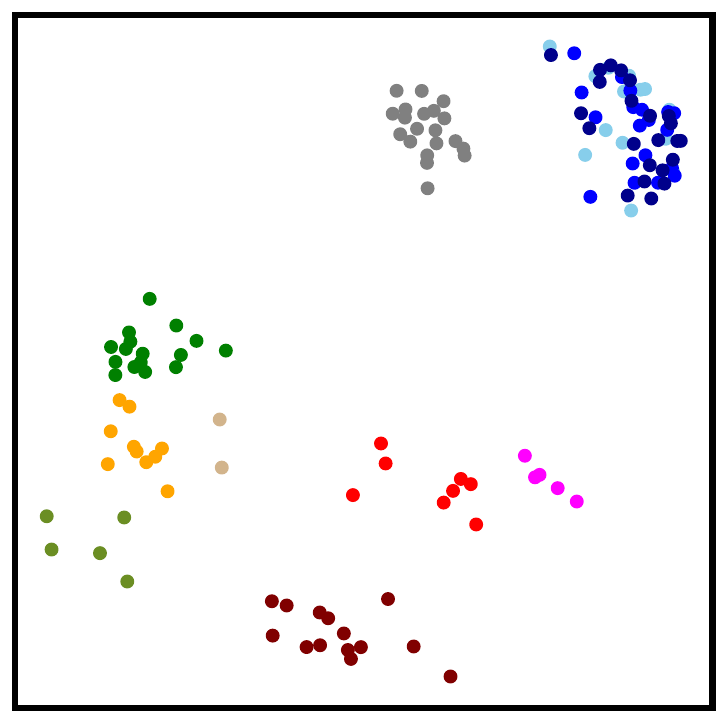}
    \caption{Look-ahead 3}
    \end{subfigure}
    \begin{subfigure}[t]{0.24\linewidth}
    \includegraphics[width=\linewidth]{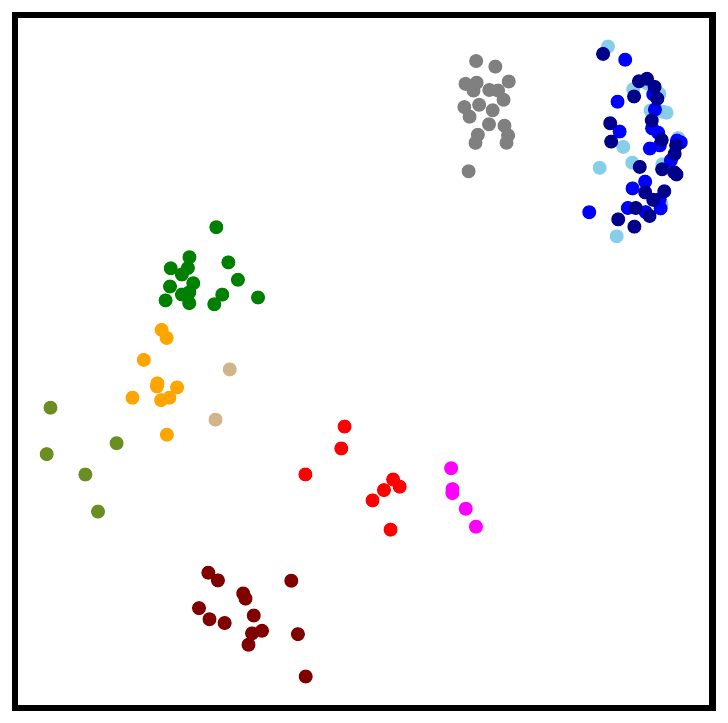}
    \caption{Look-ahead 4}
    \end{subfigure}
    
    \begin{subfigure}[t]{0.24\linewidth}
    \includegraphics[width=\linewidth]{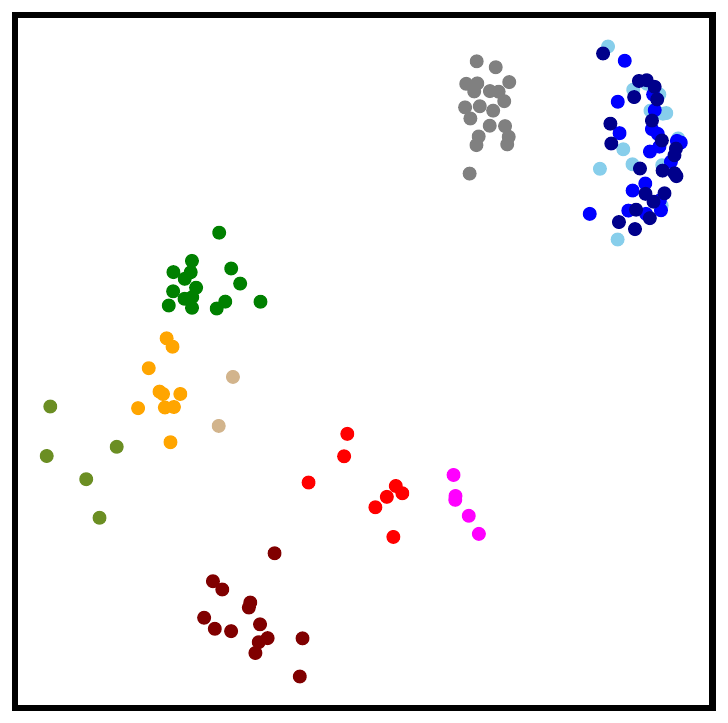}
    \caption{Look-ahead 5}
    \end{subfigure}
    \begin{subfigure}[t]{0.24\linewidth}
    \includegraphics[width=\linewidth]{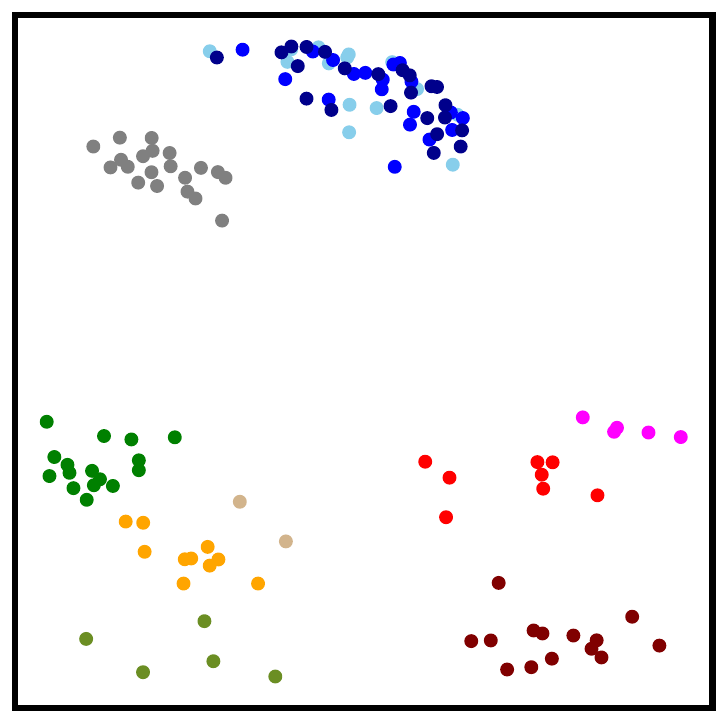}
    \caption{Look-ahead 6}
    \end{subfigure}
    \begin{subfigure}[t]{0.24\linewidth}
    \includegraphics[width=\linewidth]{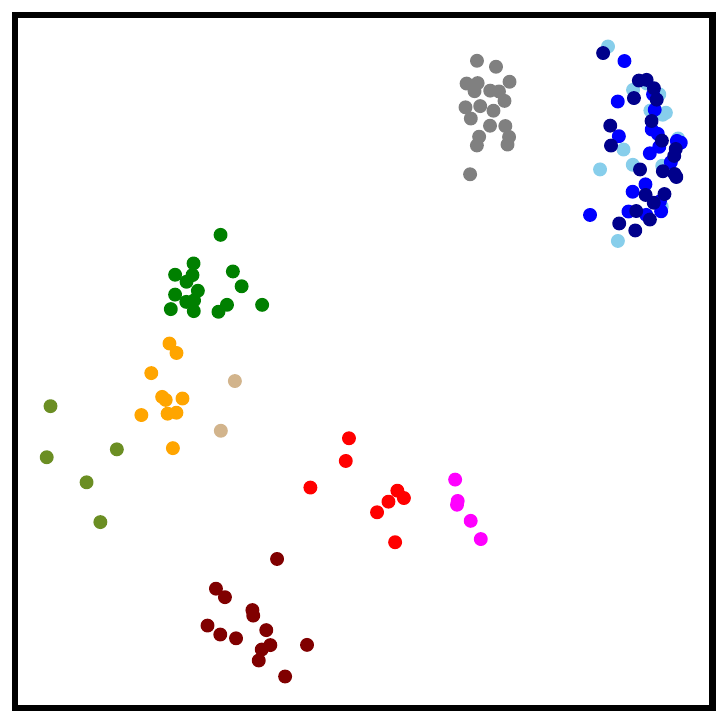}
    \caption{Look-ahead 7}
    \end{subfigure}
    \begin{subfigure}[t]{0.24\linewidth}
    \includegraphics[width=\linewidth]{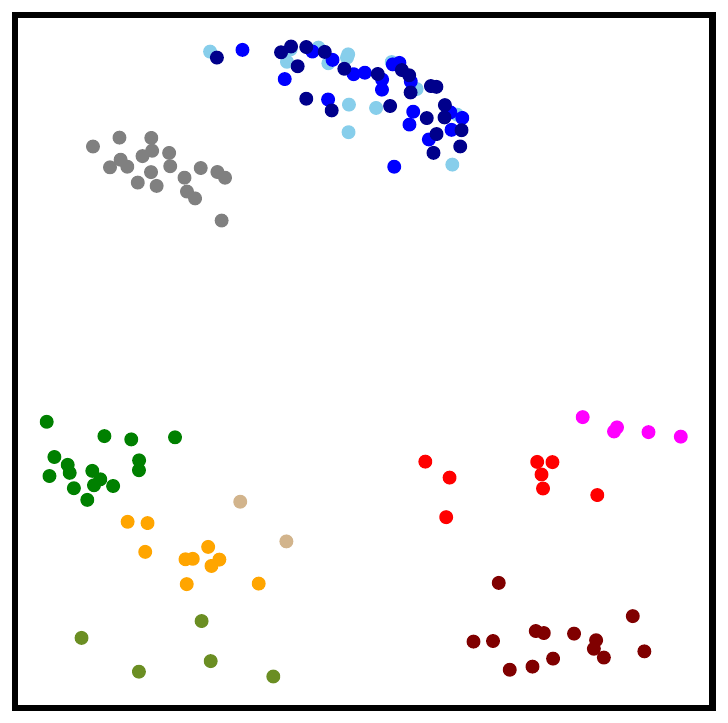}
    \caption{Look-ahead 8}
    \end{subfigure}
    \caption{MDS embeddings of the TOSCA ensemble using various different distance metrics.}
    \label{fig:tosca_mds_all}
\end{figure*}

We now provide the complete results for the TOSCA ensemble.
We computed the distance matrix for the original merge tree edit distance, the merge tree Wasserstein distance, the original path mapping distance and the adapted path mapping distance with look-ahead values of~$1$ up to~$8$.
All matrices are shown in Figure~\ref{fig:tosca_mat_all}.
Figure~\ref{fig:tosca_mds_all} shows the corresponding MDS embeddings for all distance measures.

\begin{figure*}
    \centering
    \captionsetup[subfigure]{aboveskip=-1pt,belowskip=-1pt}
    
    \begin{subfigure}[t]{0.24\linewidth}
    \includegraphics[width=\linewidth]{figures/tosca/dm_pmd0.pdf}
    \caption{LA=0,simpl=6\%}
    \end{subfigure}
    \begin{subfigure}[t]{0.24\linewidth}
    \includegraphics[width=\linewidth]{figures/tosca/dm_pmd2.pdf}
    \caption{LA=2,simpl=6\%}
    \end{subfigure}
    \begin{subfigure}[t]{0.24\linewidth}
    \includegraphics[width=\linewidth]{figures/tosca/dm_pmd3.pdf}
    \caption{LA=3,simpl=6\%}
    \end{subfigure}
    \begin{subfigure}[t]{0.24\linewidth}
    \includegraphics[width=\linewidth]{figures/tosca/dm_pmd6.pdf}
    \caption{LA=6,simpl=6\%}
    \end{subfigure}
    
    \begin{subfigure}[t]{0.24\linewidth}
    \includegraphics[width=\linewidth]{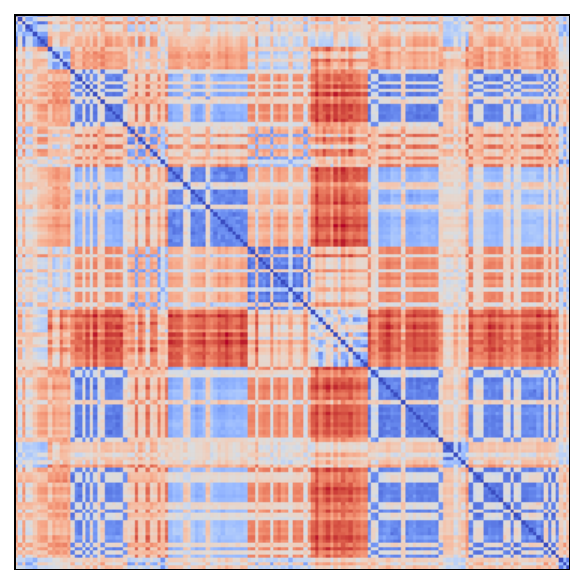}
    \caption{LA=0,simpl=4\%}
    \end{subfigure}
    \begin{subfigure}[t]{0.24\linewidth}
    \includegraphics[width=\linewidth]{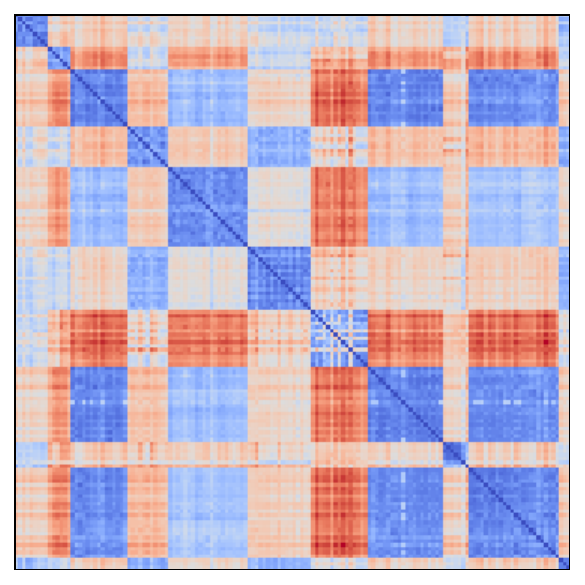}
    \caption{LA=2,simpl=4\%}
    \end{subfigure}
    \begin{subfigure}[t]{0.24\linewidth}
    \includegraphics[width=\linewidth]{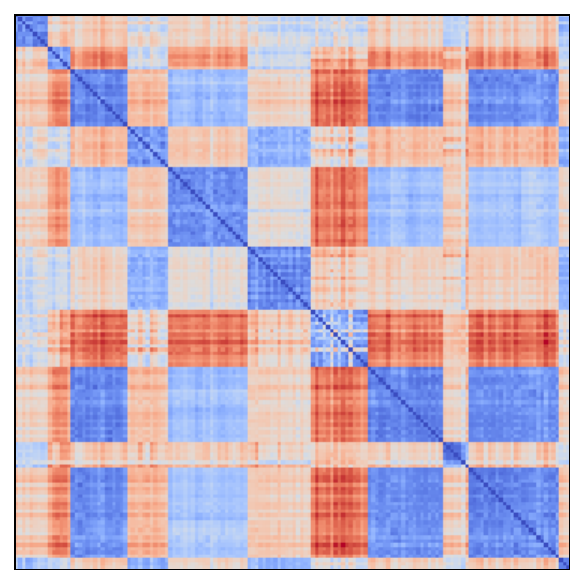}
    \caption{LA=3,simpl=4\%}
    \end{subfigure}
    \begin{subfigure}[t]{0.24\linewidth}
    \includegraphics[width=\linewidth]{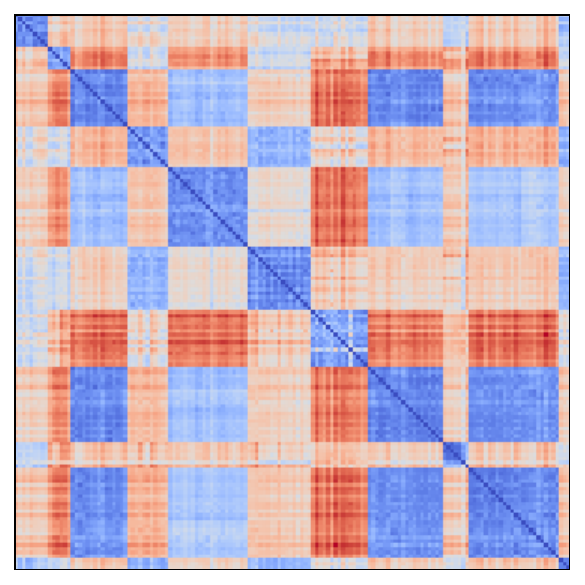}
    \caption{LA=6,simpl=4\%}
    \end{subfigure}
    
    \begin{subfigure}[t]{0.24\linewidth}
    \includegraphics[width=\linewidth]{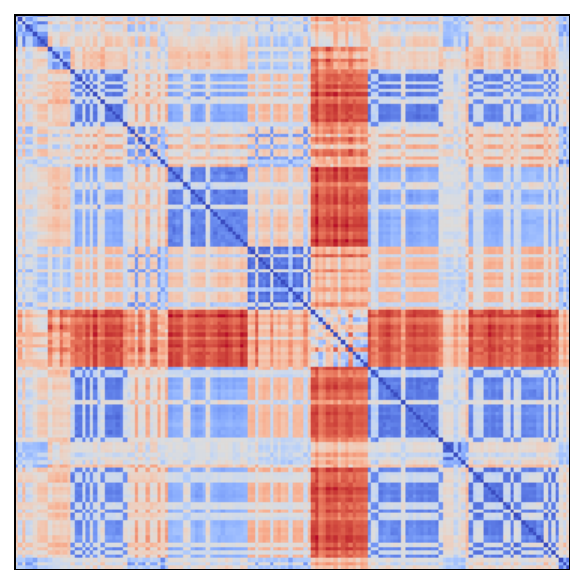}
    \caption{LA=0,simpl=2\%}
    \end{subfigure}
    \begin{subfigure}[t]{0.24\linewidth}
    \includegraphics[width=\linewidth]{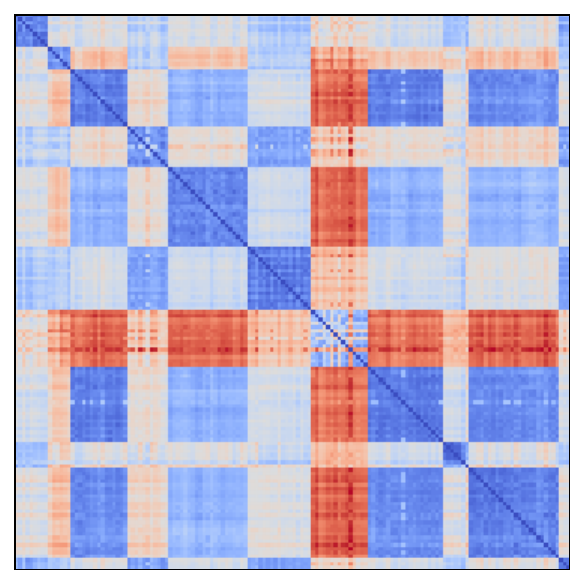}
    \caption{LA=2,simpl=2\%}
    \end{subfigure}
    \begin{subfigure}[t]{0.24\linewidth}
    \includegraphics[width=\linewidth]{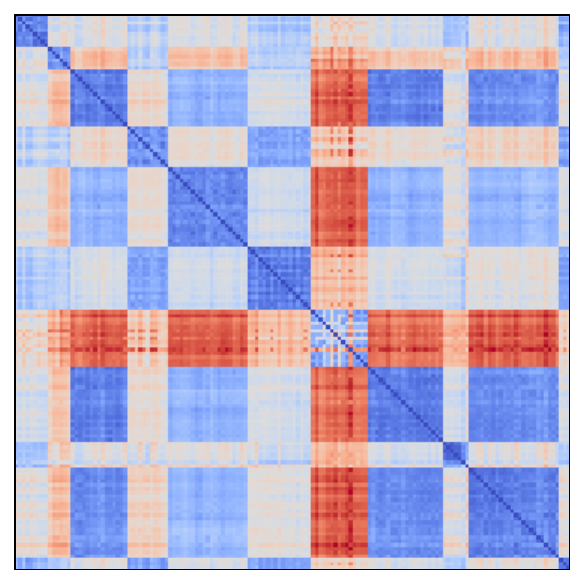}
    \caption{LA=3,simpl=2\%}
    \end{subfigure}
    \begin{subfigure}[t]{0.24\linewidth}
    \includegraphics[width=\linewidth]{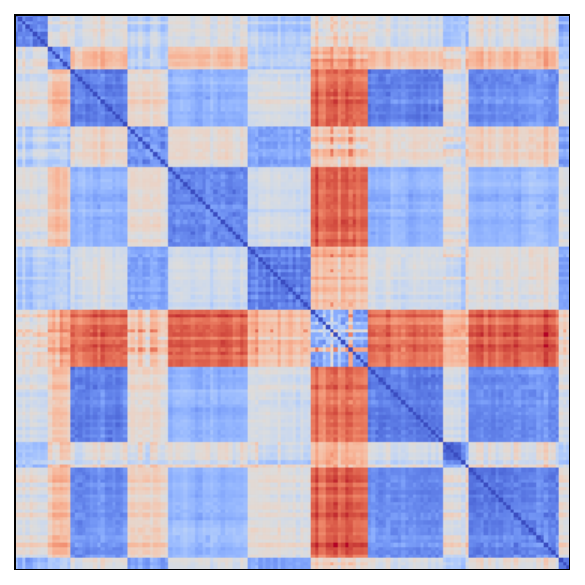}
    \caption{LA=6,simpl=2\%}
    \end{subfigure}
    \caption{Comparison of distance matrices on the TOSCA ensemble using varying look-ahead values and simplification thresholds.}
    \label{fig:tosca_mat_simpl}
\end{figure*}

Furthermore, we show the distance matrices for the TOSCA ensemble with lower simplification thresholds in Figure~\ref{fig:tosca_mat_simpl}, to showcase the unlocked potential of the look-ahead heuristic.
Using the unconstrained deformation-based edit distance, the limit was a simplification threshold of~6\%.
Even then, we had to filter out large merge trees.
In contrast, even with a look-ahead of~$6$, much smaller thresholds are possible.
For thresholds up to~2\%, we computation times the corresponding matrices were in the range of seconds or minutes.

In Figure~\ref{fig:tosca_mat_simpl}, we can observe that the lower the simplification threshold, the more the cluster of the lion shape forms an outlier.
This is due to the fact that the lion shapes have significantly more low-persistence features representing facial details than other shapes.
Due to the increased ``noise'' in the merge tree, this cluster contains several instabilities, even for high look-ahead values.
In contrast, other clusters seem to be more pronounced for lower simplification thresholds.
This could be due to the stretched color map because of the stronger outlier behavior, or that the shapes are actually more distinct because of less features like fingers or toes being dropped in the simplification.

\section{MVK Time Series}
\label{sec:app_mvk}

\begin{figure}
    \centering
    \captionsetup[subfigure]{aboveskip=-1pt,belowskip=-1pt}
    
    \begin{subfigure}[t]{0.49\linewidth}
    \includegraphics[width=\linewidth]{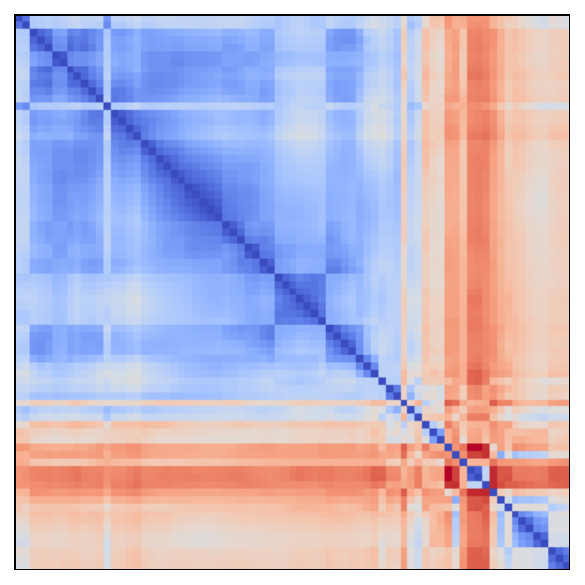}
    \caption{Wasserstein Distance}
    \end{subfigure}
    \begin{subfigure}[t]{0.49\linewidth}
    \includegraphics[width=\linewidth]{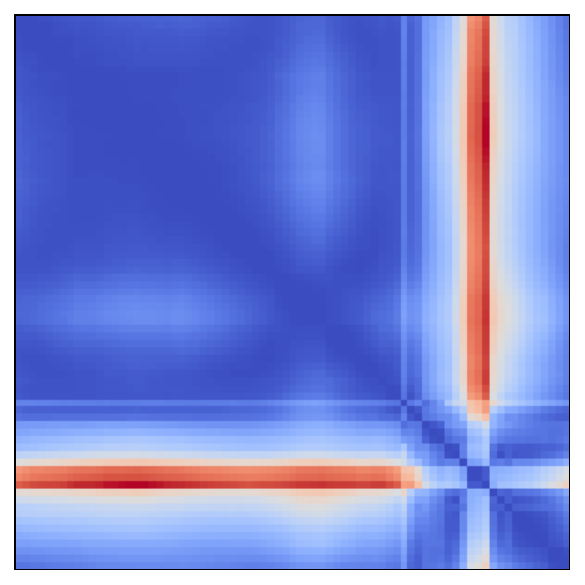}
    \caption{Merge Tree Edit Distance}
    \end{subfigure}
    
    \begin{subfigure}[t]{0.49\linewidth}
    \includegraphics[width=\linewidth]{figures/mvk/dm_s1_orb01_pmd0.pdf}
    \caption{Path Mapping Distance}
    \end{subfigure}
    \begin{subfigure}[t]{0.49\linewidth}
    \includegraphics[width=\linewidth]{figures/mvk/dm_s1_orb01_pmd1.pdf}
    \caption{Look-ahead 1}
    \end{subfigure}
    
    \begin{subfigure}[t]{0.49\linewidth}
    \includegraphics[width=\linewidth]{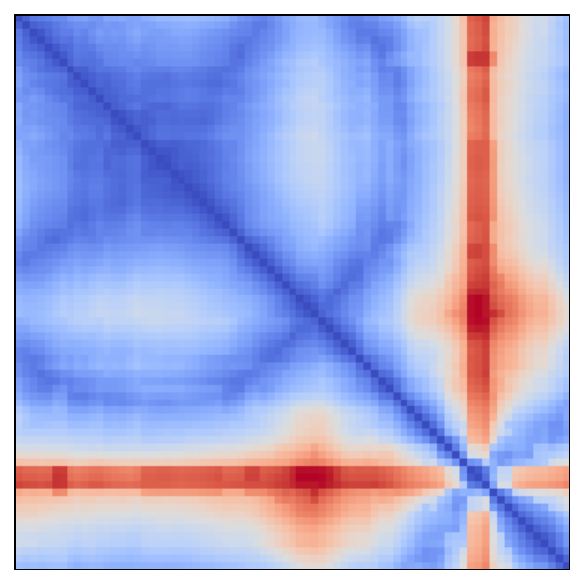}
    \caption{Look-ahead 2}
    \end{subfigure}
    \begin{subfigure}[t]{0.49\linewidth}
    \includegraphics[width=\linewidth]{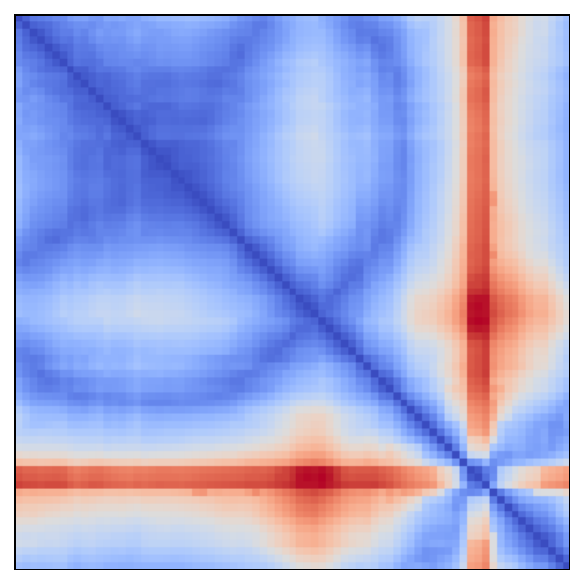}
    \caption{Look-ahead 3}
    \end{subfigure}
    
    \begin{subfigure}[t]{0.49\linewidth}
    \includegraphics[width=\linewidth]{figures/mvk/dm_s1_orb01_pmd4.pdf}
    \caption{Look-ahead 4}
    \end{subfigure}
    \begin{subfigure}[t]{0.49\linewidth}
    \includegraphics[width=\linewidth]{figures/mvk/dm_s1_orb01_uted.pdf}
    \caption{Unconstrained Deformation Distance}
    \end{subfigure}
    \caption{Distance matrices for the $S_1$ hole field of the MVK molecule. Stable behavior is reached at look-ahead~$4$.}
    \label{fig:mvk_s1_orb01_mat_all}
\end{figure}

\begin{figure}
    \centering
    \captionsetup[subfigure]{aboveskip=-1pt,belowskip=-1pt}
    
    \begin{subfigure}[t]{0.49\linewidth}
    \includegraphics[width=\linewidth]{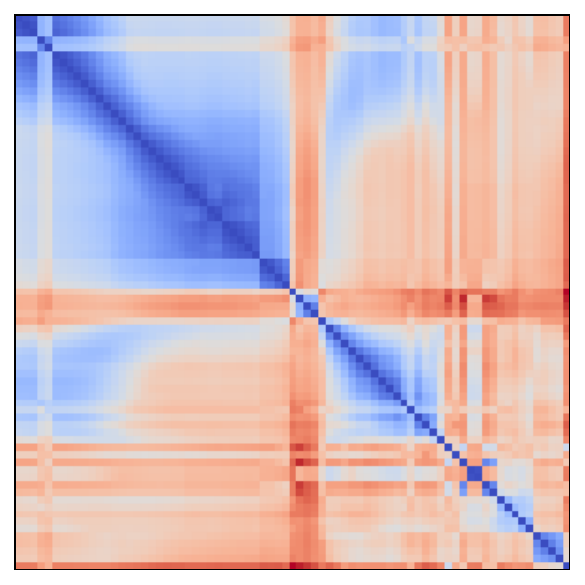}
    \caption{Wasserstein Distance}
    \end{subfigure}
    \begin{subfigure}[t]{0.49\linewidth}
    \includegraphics[width=\linewidth]{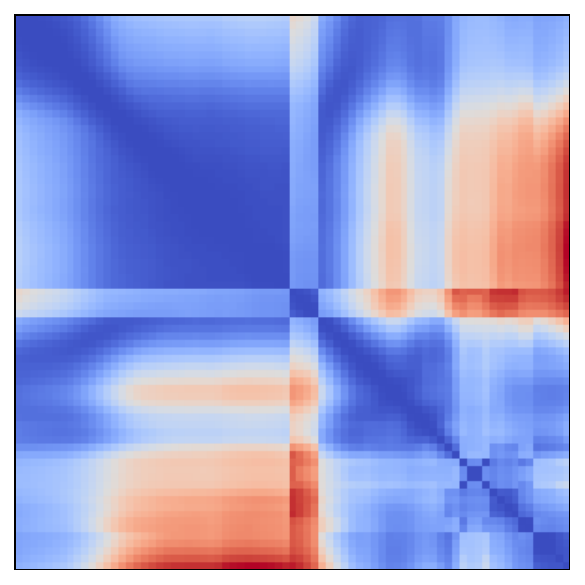}
    \caption{Merge Tree Edit Distance}
    \end{subfigure}
    
    \begin{subfigure}[t]{0.49\linewidth}
    \includegraphics[width=\linewidth]{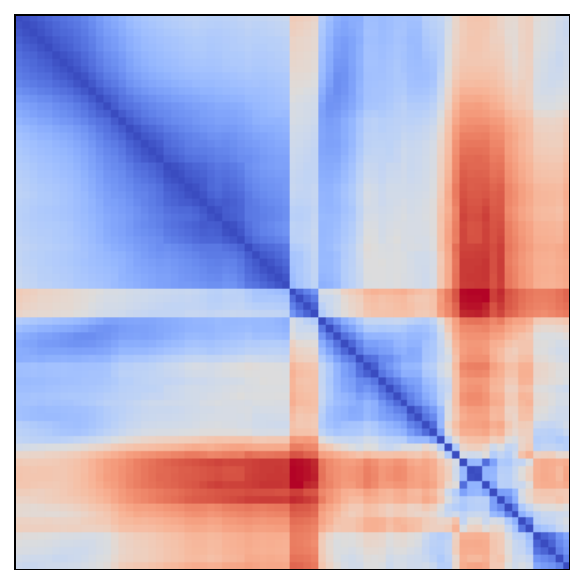}
    \caption{Path Mapping Distance}
    \end{subfigure}
    \begin{subfigure}[t]{0.49\linewidth}
    \includegraphics[width=\linewidth]{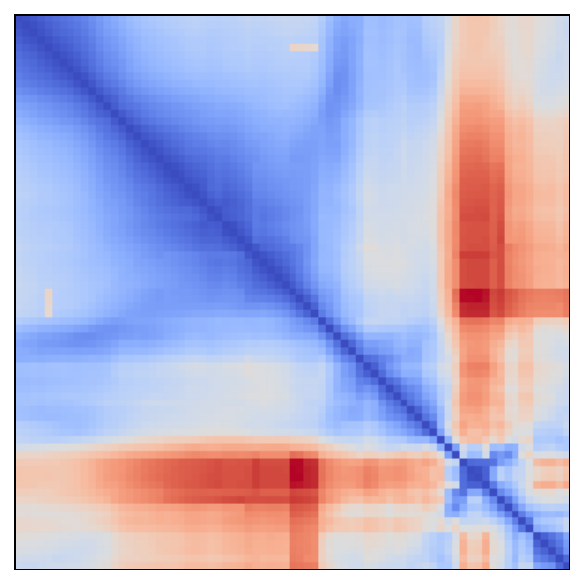}
    \caption{Look-ahead 1}
    \end{subfigure}
    
    \begin{subfigure}[t]{0.49\linewidth}
    \includegraphics[width=\linewidth]{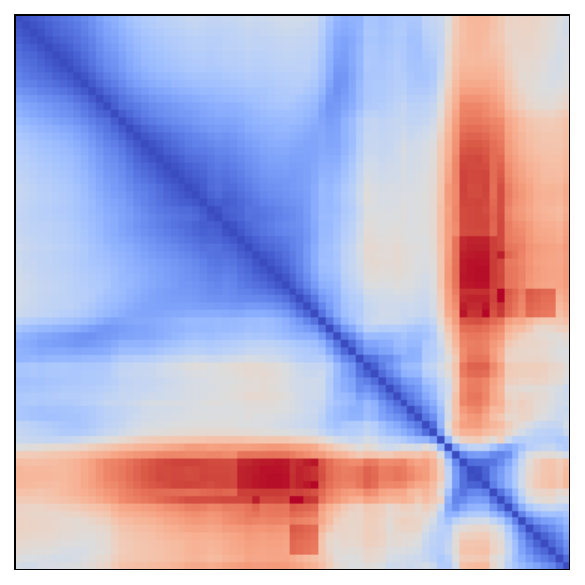}
    \caption{Look-ahead 2}
    \end{subfigure}
    \begin{subfigure}[t]{0.49\linewidth}
    \includegraphics[width=\linewidth]{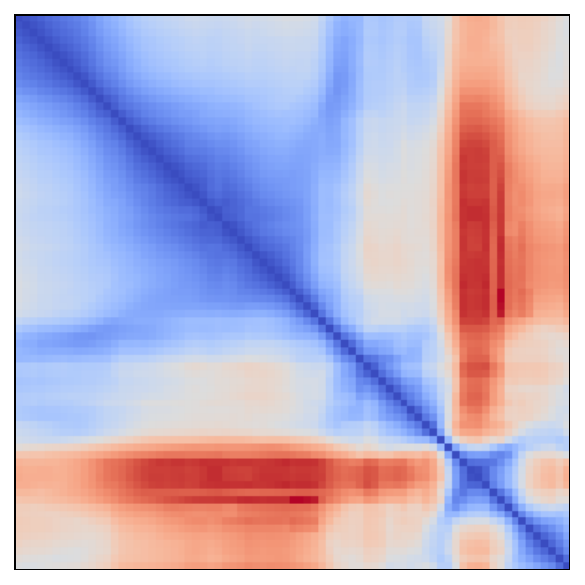}
    \caption{Look-ahead 3}
    \end{subfigure}
    
    \begin{subfigure}[t]{0.49\linewidth}
    \includegraphics[width=\linewidth]{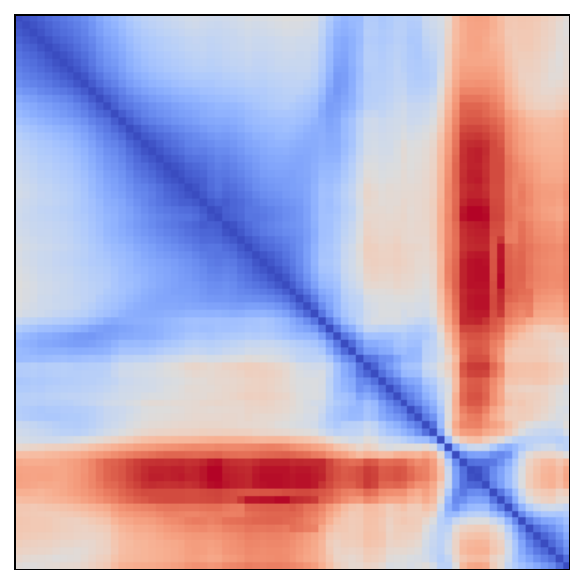}
    \caption{Look-ahead 4}
    \end{subfigure}
    \begin{subfigure}[t]{0.49\linewidth}
    \includegraphics[width=\linewidth]{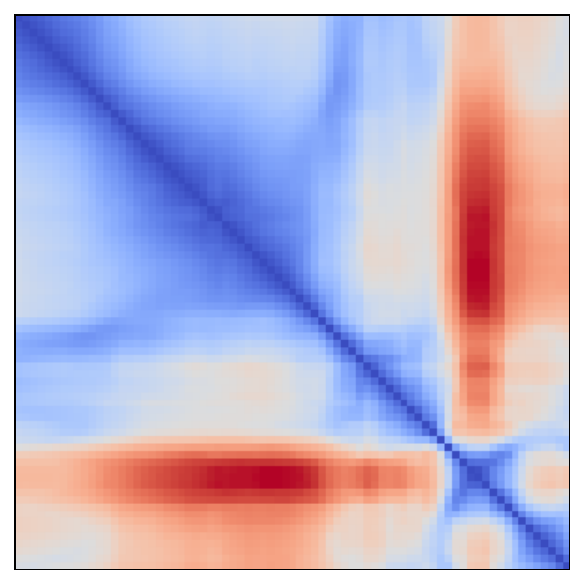}
    \caption{Unconstrained Deformation Distance}
    \end{subfigure}
    \caption{Distance matrices for the $S_2$ hole field of the MVK molecule. Stable behavior is reached at look-ahead~$4$.}
    \label{fig:mvk_s2_orb01_mat_all}
\end{figure}

\begin{figure}
    \centering
    \captionsetup[subfigure]{aboveskip=-1pt,belowskip=-1pt}
    
    \begin{subfigure}[t]{0.49\linewidth}
    \includegraphics[width=\linewidth]{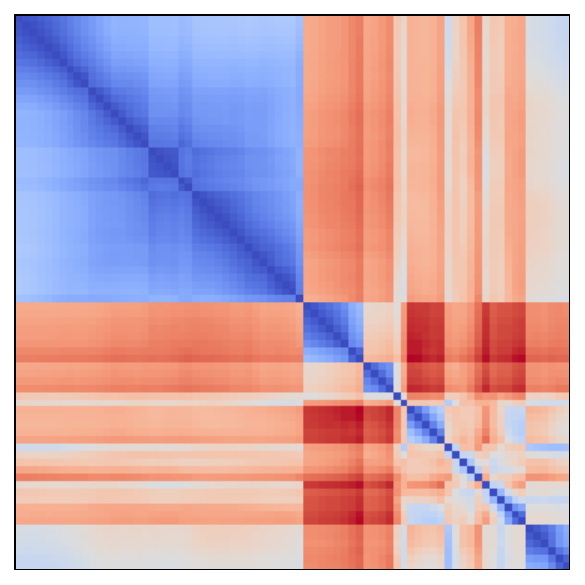}
    \caption{Wasserstein Distance}
    \end{subfigure}
    \begin{subfigure}[t]{0.49\linewidth}
    \includegraphics[width=\linewidth]{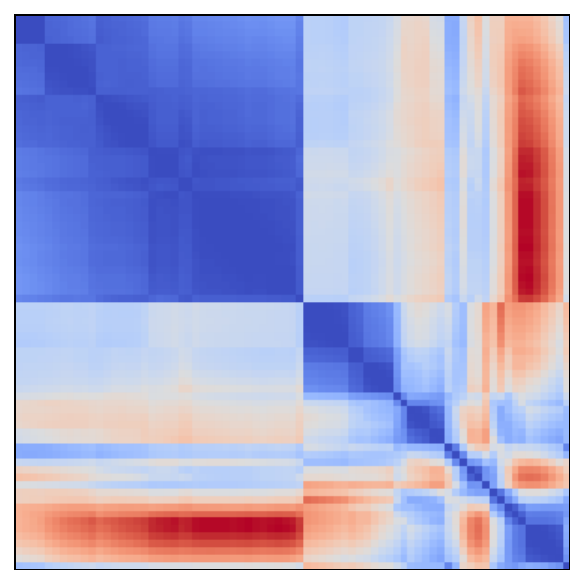}
    \caption{Merge Tree Edit Distance}
    \end{subfigure}
    
    \begin{subfigure}[t]{0.49\linewidth}
    \includegraphics[width=\linewidth]{figures/mvk/dm_s1_orb00_pmd0.pdf}
    \caption{Path Mapping Distance}
    \end{subfigure}
    \begin{subfigure}[t]{0.49\linewidth}
    \includegraphics[width=\linewidth]{figures/mvk/dm_s1_orb00_pmd1.pdf}
    \caption{Look-ahead 1}
    \end{subfigure}
    
    \begin{subfigure}[t]{0.49\linewidth}
    \includegraphics[width=\linewidth]{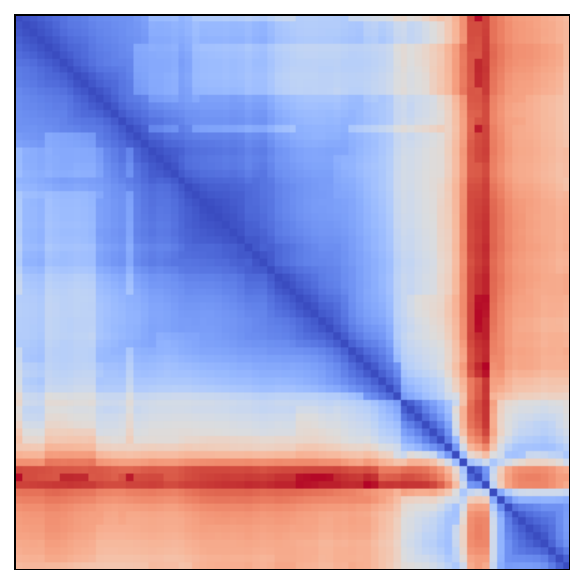}
    \caption{Look-ahead 2}
    \end{subfigure}
    \begin{subfigure}[t]{0.49\linewidth}
    \includegraphics[width=\linewidth]{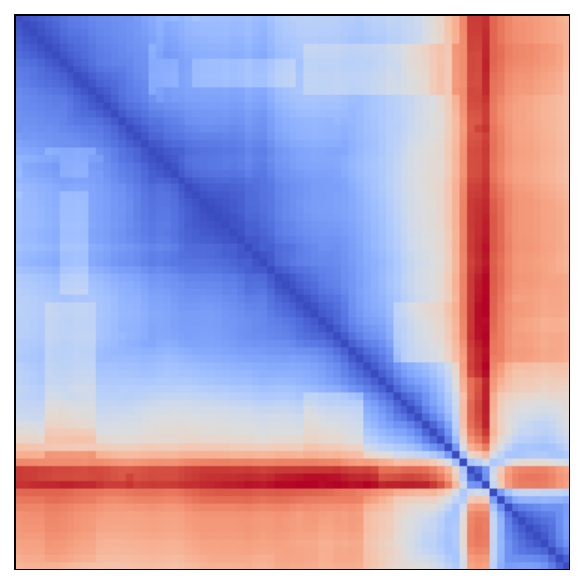}
    \caption{Look-ahead 3}
    \end{subfigure}
    
    \begin{subfigure}[t]{0.49\linewidth}
    \includegraphics[width=\linewidth]{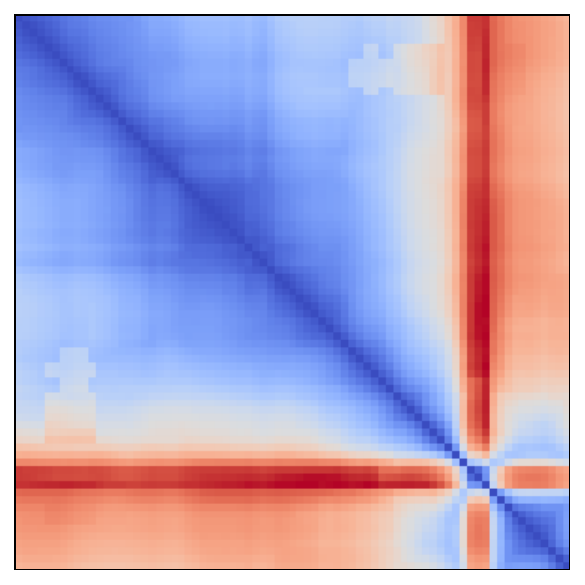}
    \caption{Look-ahead 4}
    \end{subfigure}
    \begin{subfigure}[t]{0.49\linewidth}
    \includegraphics[width=\linewidth]{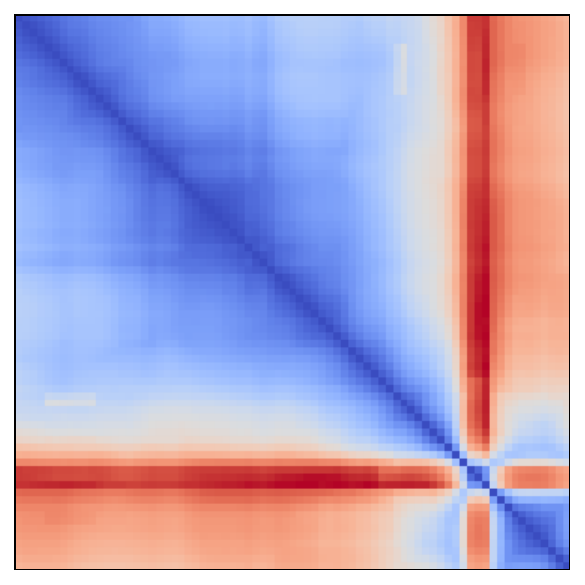}
    \caption{Look-ahead 5}
    \end{subfigure}
    
    \begin{subfigure}[t]{0.49\linewidth}
    \includegraphics[width=\linewidth]{figures/mvk/dm_s1_orb00_pmd6.pdf}
    \caption{Look-ahead 6}
    \end{subfigure}
    \begin{subfigure}[t]{0.49\linewidth}
    \includegraphics[width=\linewidth]{figures/mvk/dm_s1_orb00_uted.pdf}
    \caption{Unconstrained Deformation Distance}
    \end{subfigure}
    \caption{Distance matrices for the $S_1$ particle field of the MVK molecule. Stable behavior is reached at look-ahead~$6$.}
    \label{fig:mvk_s1_orb00_mat_all}
\end{figure}

\begin{figure}
    \centering
    \captionsetup[subfigure]{aboveskip=-1pt,belowskip=-1pt}
    
    \begin{subfigure}[t]{0.49\linewidth}
    \includegraphics[width=\linewidth]{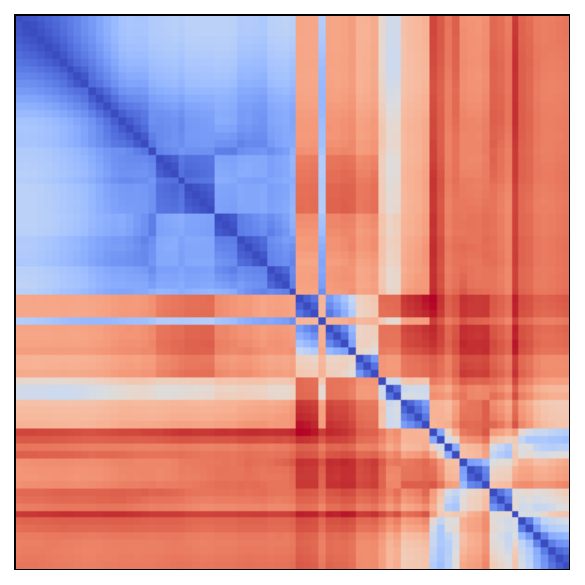}
    \caption{Wasserstein Distance}
    \end{subfigure}
    \begin{subfigure}[t]{0.49\linewidth}
    \includegraphics[width=\linewidth]{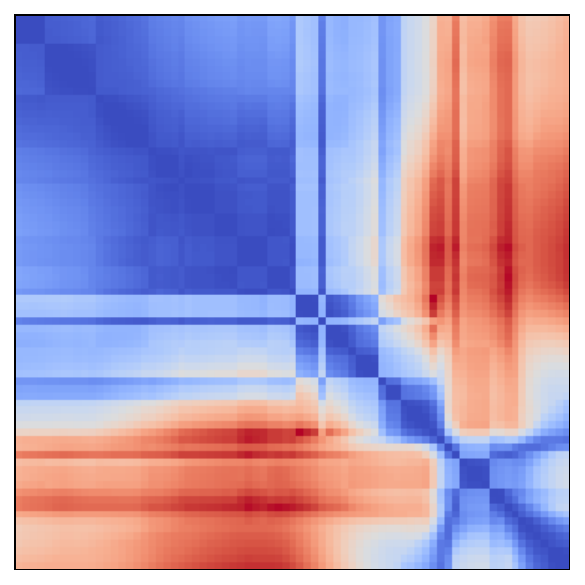}
    \caption{Merge Tree Edit Distance}
    \end{subfigure}
    
    \begin{subfigure}[t]{0.49\linewidth}
    \includegraphics[width=\linewidth]{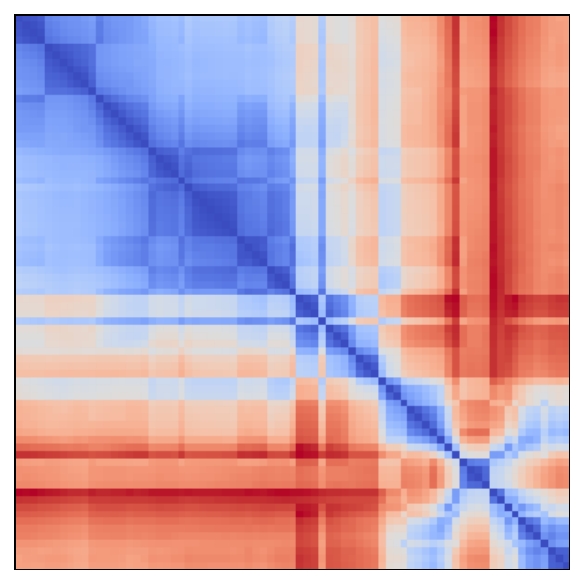}
    \caption{Path Mapping Distance}
    \end{subfigure}
    \begin{subfigure}[t]{0.49\linewidth}
    \includegraphics[width=\linewidth]{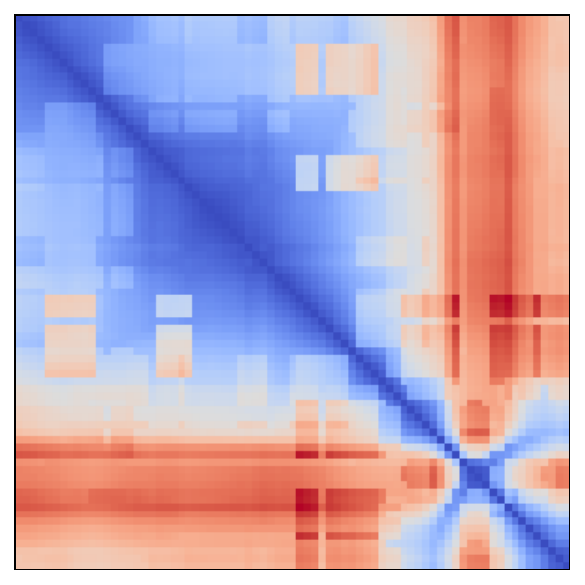}
    \caption{Look-ahead 1}
    \end{subfigure}
    
    \begin{subfigure}[t]{0.49\linewidth}
    \includegraphics[width=\linewidth]{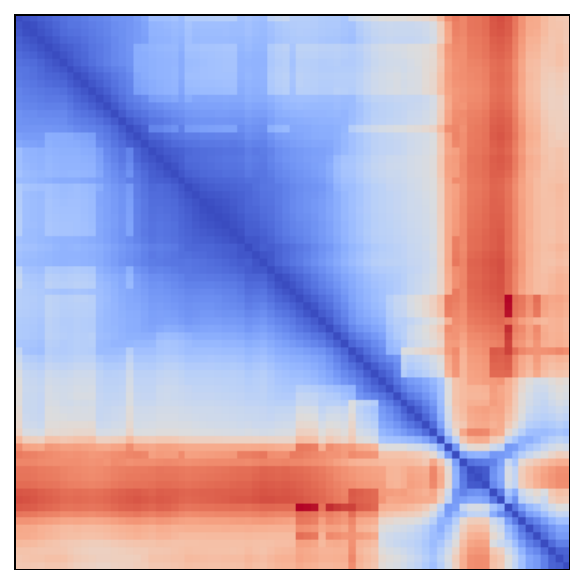}
    \caption{Look-ahead 2}
    \end{subfigure}
    \begin{subfigure}[t]{0.49\linewidth}
    \includegraphics[width=\linewidth]{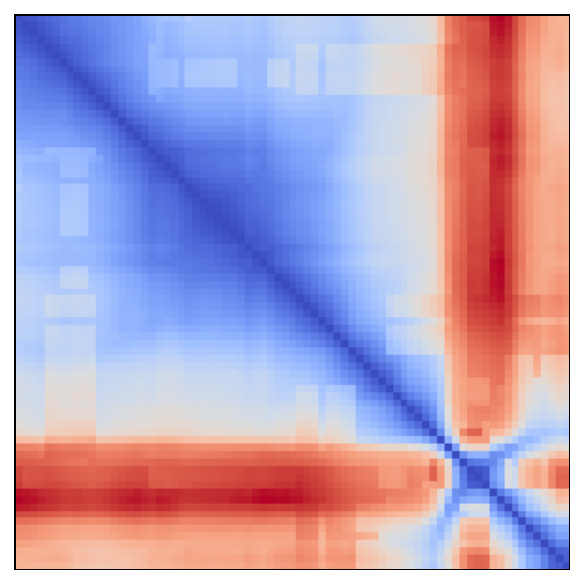}
    \caption{Look-ahead 3}
    \end{subfigure}
    
    \begin{subfigure}[t]{0.49\linewidth}
    \includegraphics[width=\linewidth]{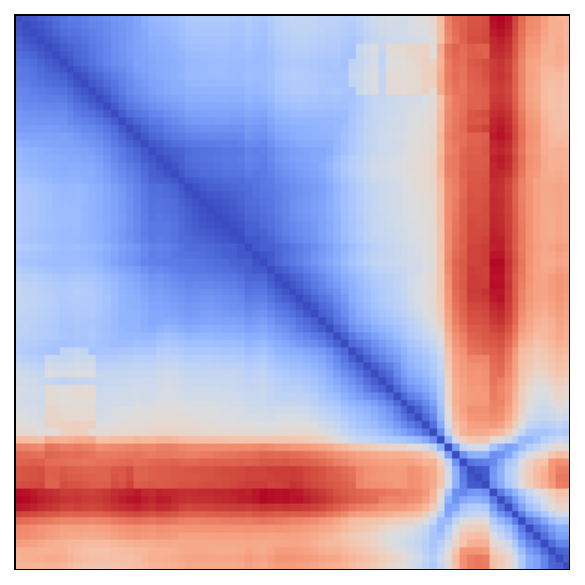}
    \caption{Look-ahead 4}
    \end{subfigure}
    \begin{subfigure}[t]{0.49\linewidth}
    \includegraphics[width=\linewidth]{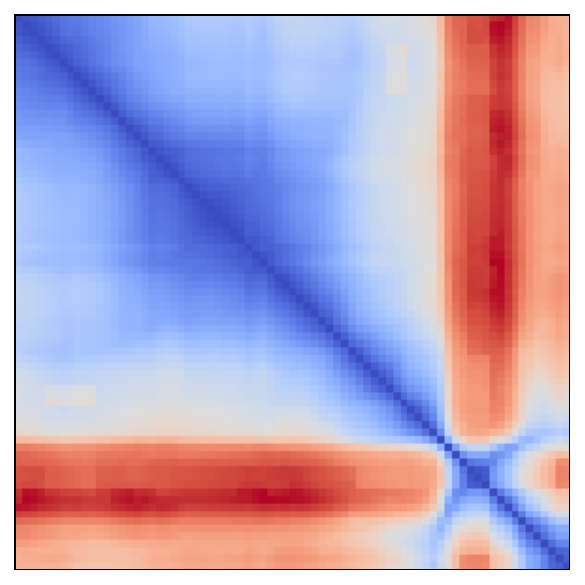}
    \caption{Look-ahead 5}
    \end{subfigure}
    
    \begin{subfigure}[t]{0.49\linewidth}
    \includegraphics[width=\linewidth]{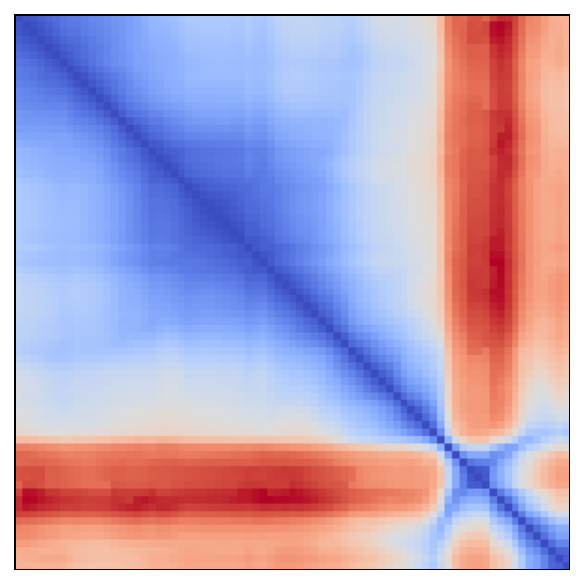}
    \caption{Look-ahead 6}
    \end{subfigure}
    \begin{subfigure}[t]{0.49\linewidth}
    \includegraphics[width=\linewidth]{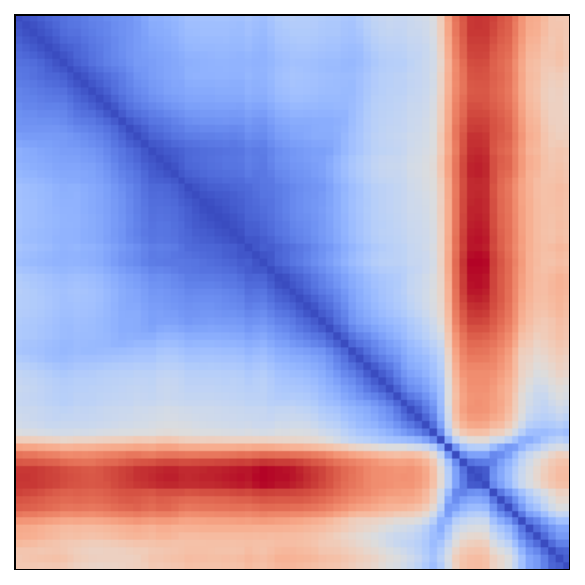}
    \caption{Unconstrained Deformation Distance}
    \end{subfigure}
    \caption{Distance matrices for the $S_2$ particle field of the MVK molecule. Stable behavior is reached at look-ahead~$6$.}
    \label{fig:mvk_s2_orb00_mat_all}
\end{figure}

Next, we provide further results on the MVK time series.
As explained in the main paper, four different electron density fields are given in the dataset.
For each, we computed the distance matrices with the novel heuristic, as well as previous distances.
We increased the look-ahead value until we were unable to detect any instabilities visually.
Furthermore, we also computed the matrices for the merge tree Wasserstein distance and the merge tree edit distance, to show that the instability is common among all constrained edit distances.
Note that for the latter two distances, we did not implement the geometric constraints.
However, removing constraints only improves stability, so the comparison remains reasonable.
The matrices can be found in Figures~\ref{fig:mvk_s1_orb01_mat_all},~\ref{fig:mvk_s2_orb01_mat_all},~\ref{fig:mvk_s1_orb00_mat_all} and ~\ref{fig:mvk_s2_orb00_mat_all}. 

\section{Ionization Front}
\label{sec:app_ionization}

We now discuss further results on the ionization front dataset.
We begin by providing additional distance matrices for the merge tree Wasserstein distance and the merge tree edit distance, to show that instabilities appear for all kinds of constrained edit distances.
Furthermore, we also show the distance matrix for the look-ahead values missing in the main paper.
The comparison of all distances can be found in Figure~\ref{fig:ionization_mat_app}.
With look-ahead~$1$ or~$2$, some instabilities remain, as described in the main paper.
For look-ahead~$4$, the matrix looks almost identical to the one for look-ahead~$3$.

\begin{figure}
    \centering
    \captionsetup[subfigure]{aboveskip=-1pt,belowskip=-1pt}
    
    \begin{subfigure}[t]{0.48\linewidth}
    \includegraphics[width=\linewidth]{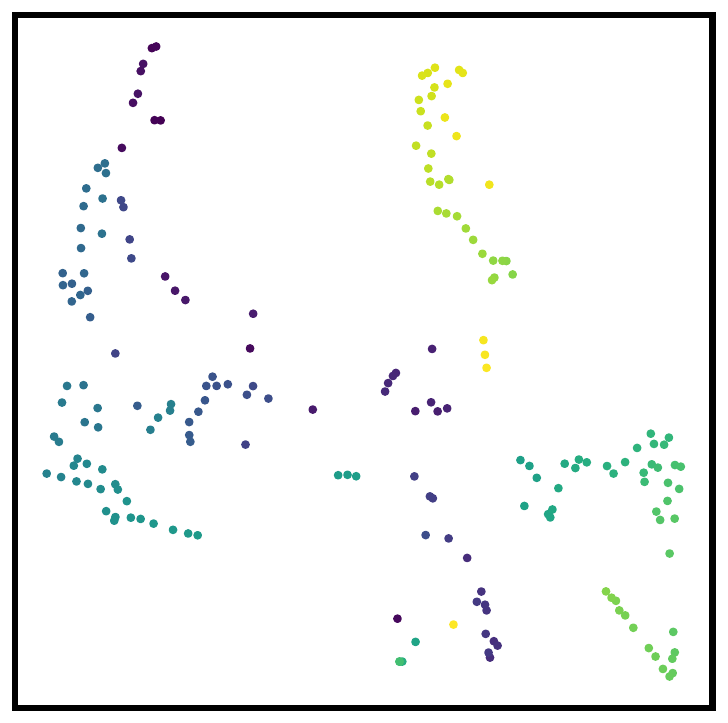}
    \caption{Merge Tree Edit Distance}
    \end{subfigure}
    \begin{subfigure}[t]{0.48\linewidth}
    \includegraphics[width=\linewidth]{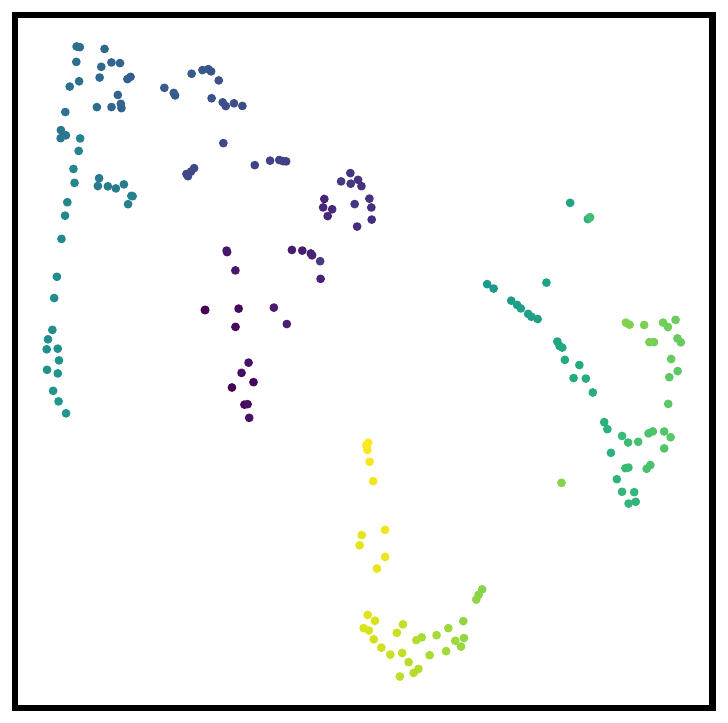}
    \caption{Path Mapping Distance}
    \end{subfigure}
    
    \begin{subfigure}[t]{0.48\linewidth}
    \includegraphics[width=\linewidth]{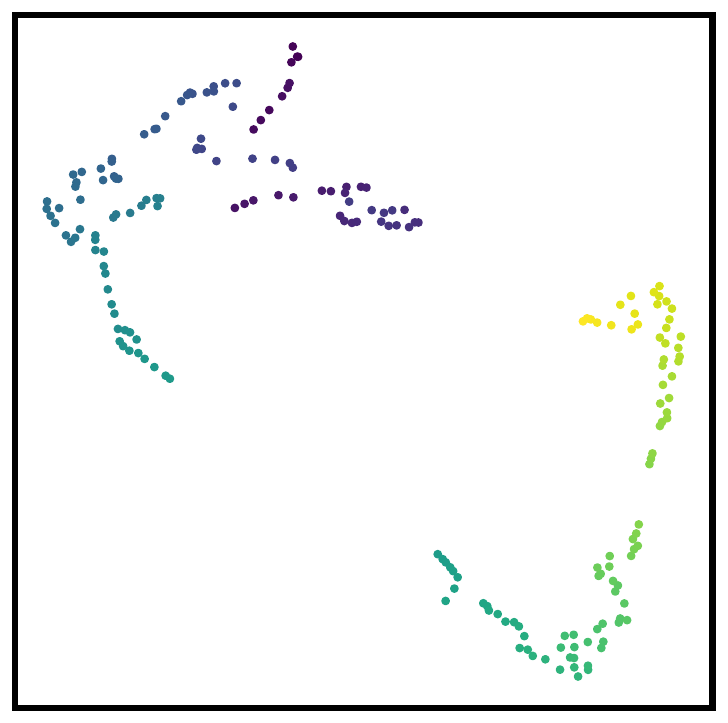}
    \caption{Look-ahead 1}
    \end{subfigure}
    \begin{subfigure}[t]{0.48\linewidth}
    \includegraphics[width=\linewidth]{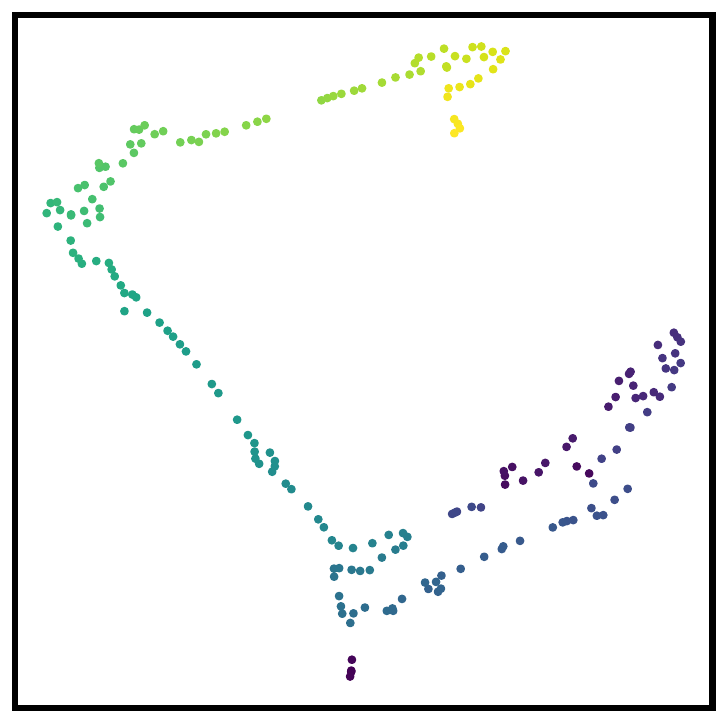}
    \caption{Look-ahead 3}
    \end{subfigure}
    \caption{Embeddings of the ionization time series with t-SNE for perplexity 30 and early exaggeration 12.}
    \label{fig:ionization_tsne_30_12}
\end{figure}

\begin{figure}
    \centering
    \captionsetup[subfigure]{aboveskip=-1pt,belowskip=-1pt}
    
    \begin{subfigure}[t]{0.48\linewidth}
    \includegraphics[width=\linewidth]{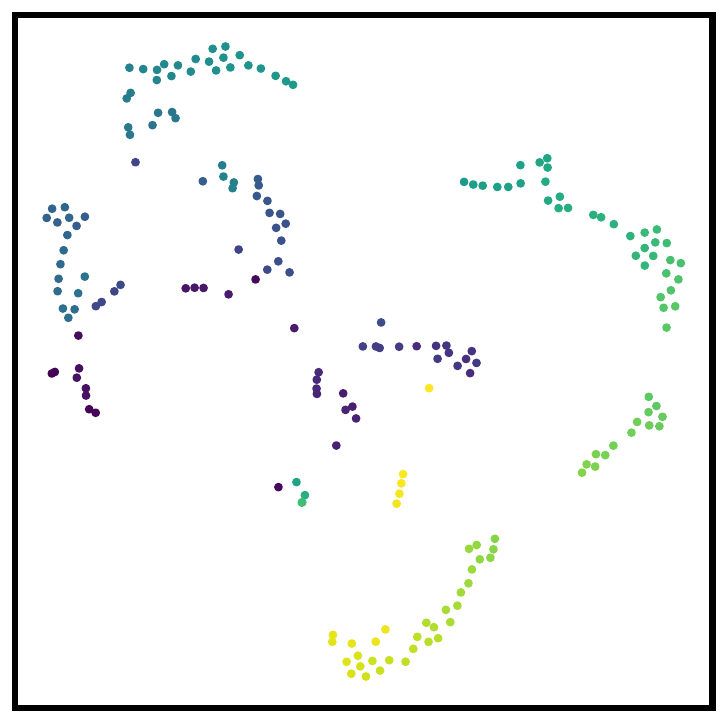}
    \caption{Merge Tree Edit Distance}
    \end{subfigure}
    \begin{subfigure}[t]{0.48\linewidth}
    \includegraphics[width=\linewidth]{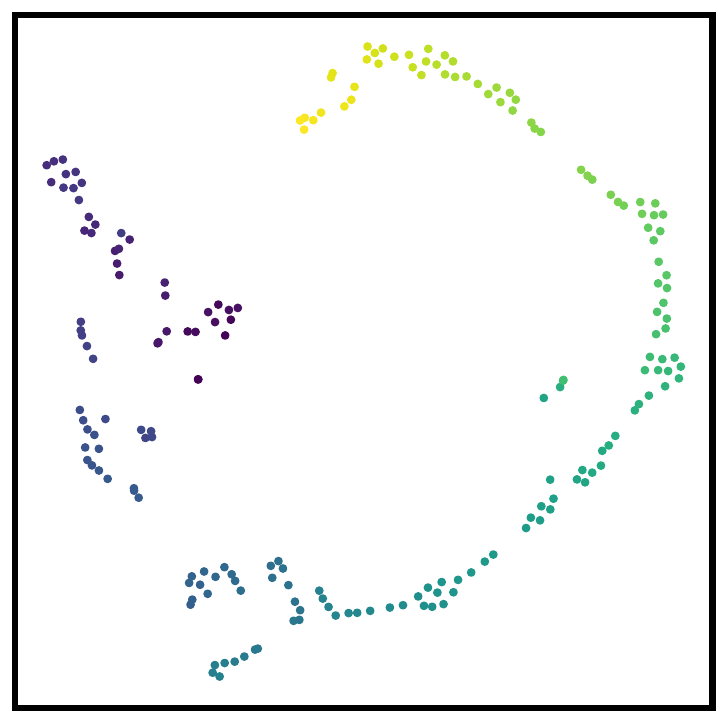}
    \caption{Path Mapping Distance}
    \end{subfigure}
    
    \begin{subfigure}[t]{0.48\linewidth}
    \includegraphics[width=\linewidth]{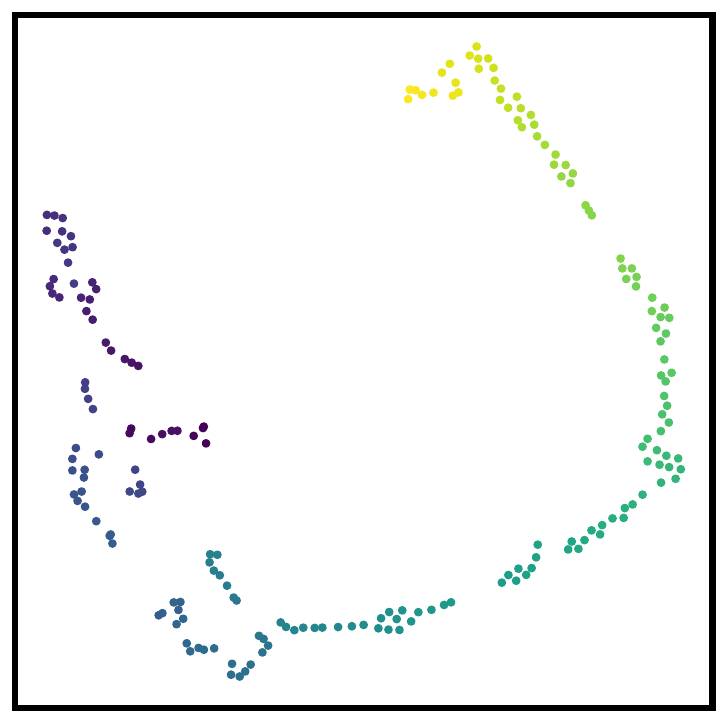}
    \caption{Look-ahead 1}
    \end{subfigure}
    \begin{subfigure}[t]{0.48\linewidth}
    \includegraphics[width=\linewidth]{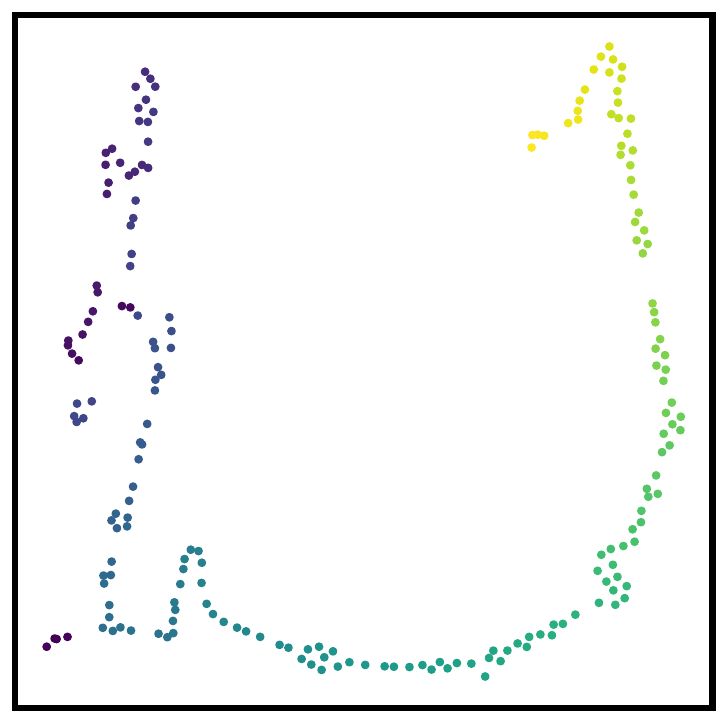}
    \caption{Look-ahead 3}
    \end{subfigure}
    \caption{Embeddings of the ionization time series with t-SNE for perplexity 5 and early exaggeration 12.}
    \label{fig:ionization_tsne_15_12}
\end{figure}

\begin{figure}
    \centering
    \captionsetup[subfigure]{aboveskip=-1pt,belowskip=-1pt}
    
    \begin{subfigure}[t]{0.48\linewidth}
    \includegraphics[width=\linewidth]{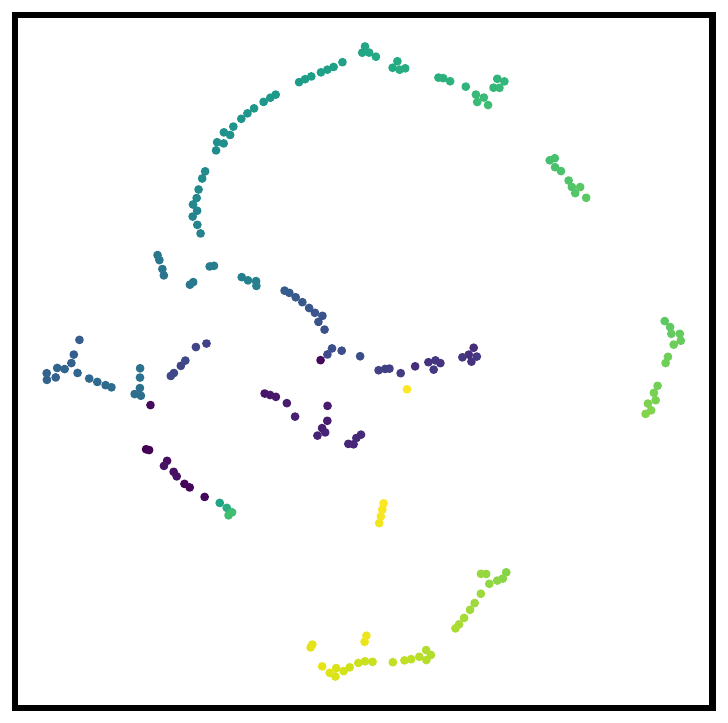}
    \caption{Merge Tree Edit Distance}
    \end{subfigure}
    \begin{subfigure}[t]{0.48\linewidth}
    \includegraphics[width=\linewidth]{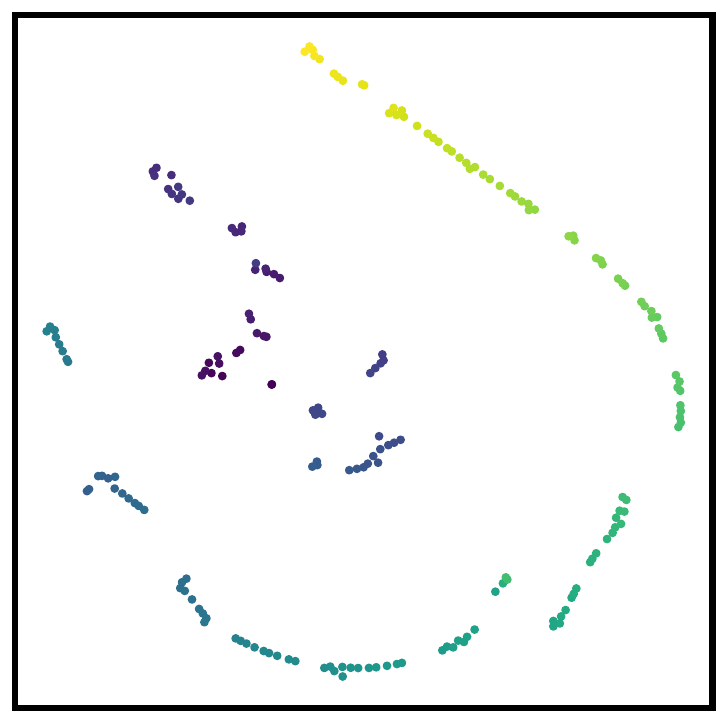}
    \caption{Path Mapping Distance}
    \end{subfigure}
    
    \begin{subfigure}[t]{0.48\linewidth}
    \includegraphics[width=\linewidth]{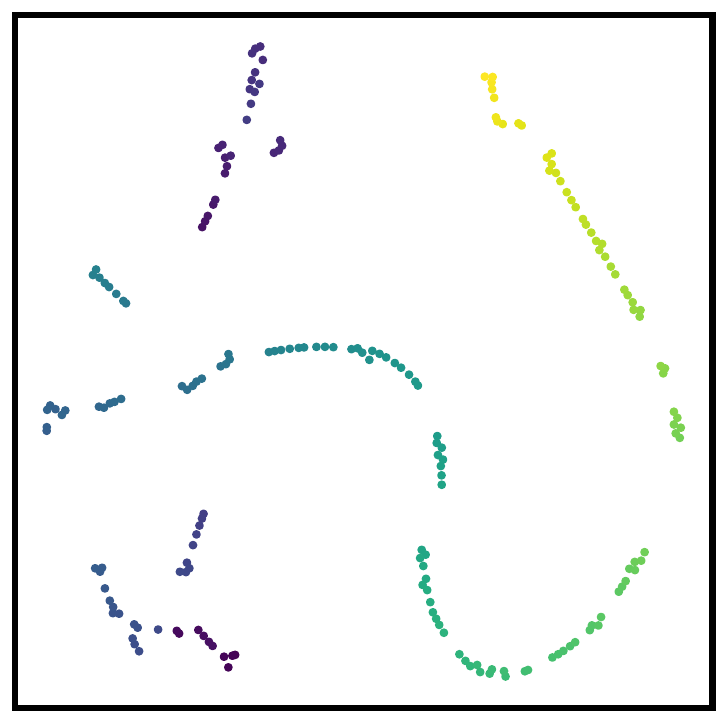}
    \caption{Look-ahead 1}
    \end{subfigure}
    \begin{subfigure}[t]{0.48\linewidth}
    \includegraphics[width=\linewidth]{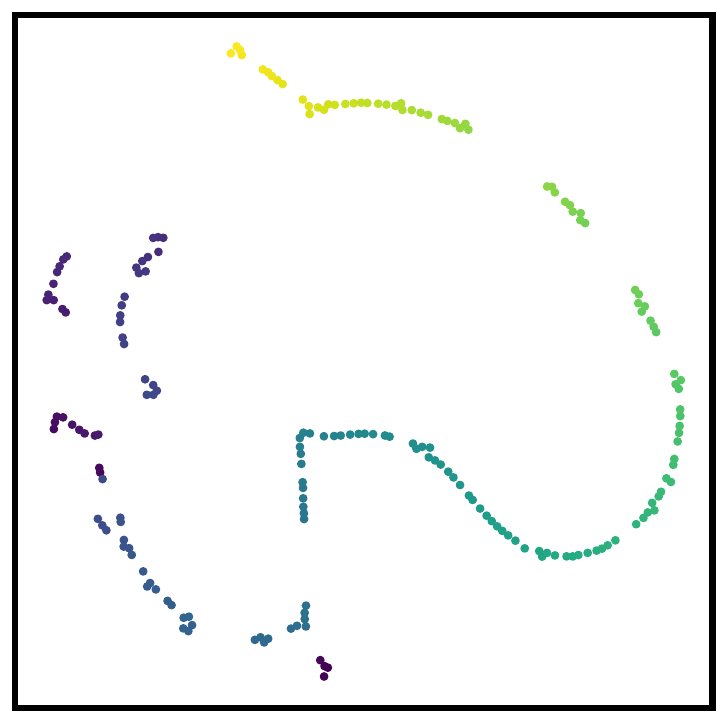}
    \caption{Look-ahead 3}
    \end{subfigure}
    \caption{Embeddings of the ionization time series with t-SNE for perplexity 5 and early exaggeration 12.}
    \label{fig:ionization_tsne_5_12}
\end{figure}

\begin{figure}
    \centering
    \captionsetup[subfigure]{aboveskip=-1pt,belowskip=-1pt}
    
    \begin{subfigure}[t]{0.48\linewidth}
    \includegraphics[width=\linewidth]{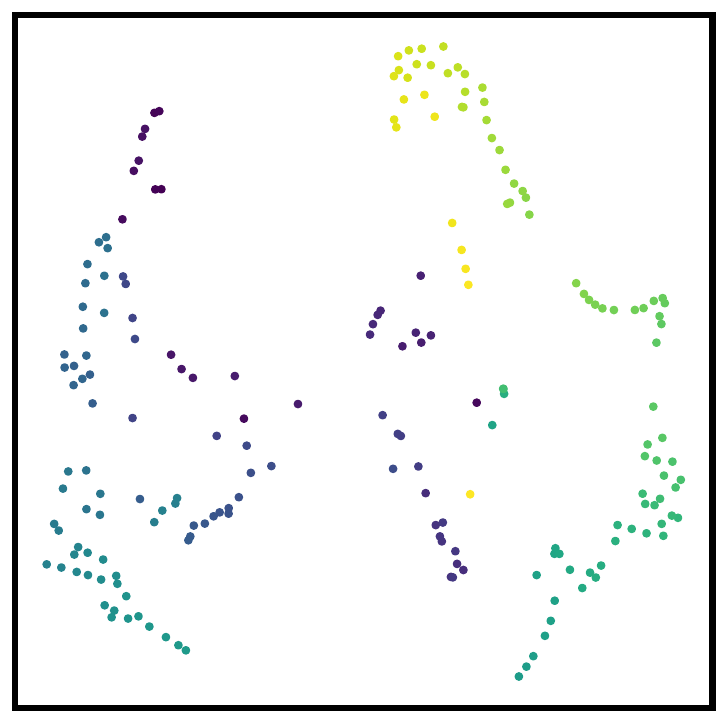}
    \caption{Merge Tree Edit Distance}
    \end{subfigure}
    \begin{subfigure}[t]{0.48\linewidth}
    \includegraphics[width=\linewidth]{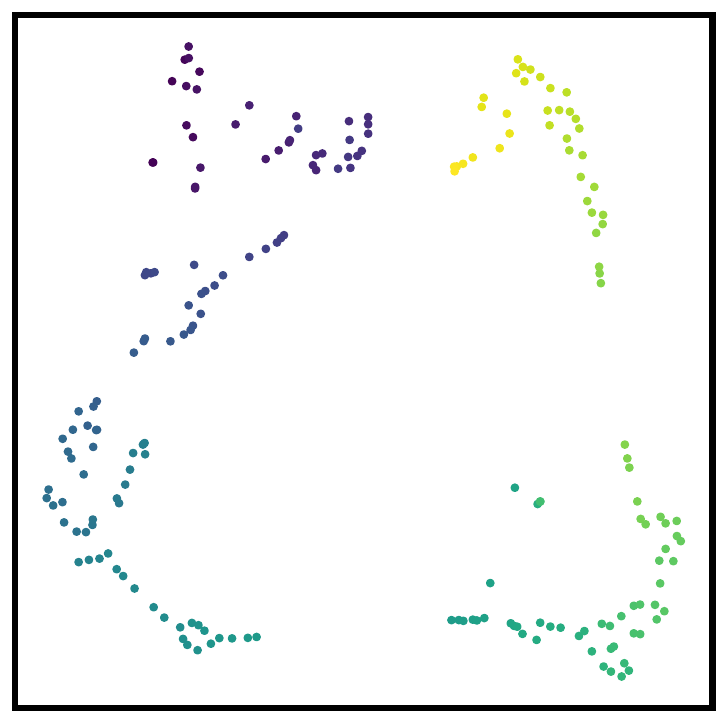}
    \caption{Path Mapping Distance}
    \end{subfigure}
    
    \begin{subfigure}[t]{0.48\linewidth}
    \includegraphics[width=\linewidth]{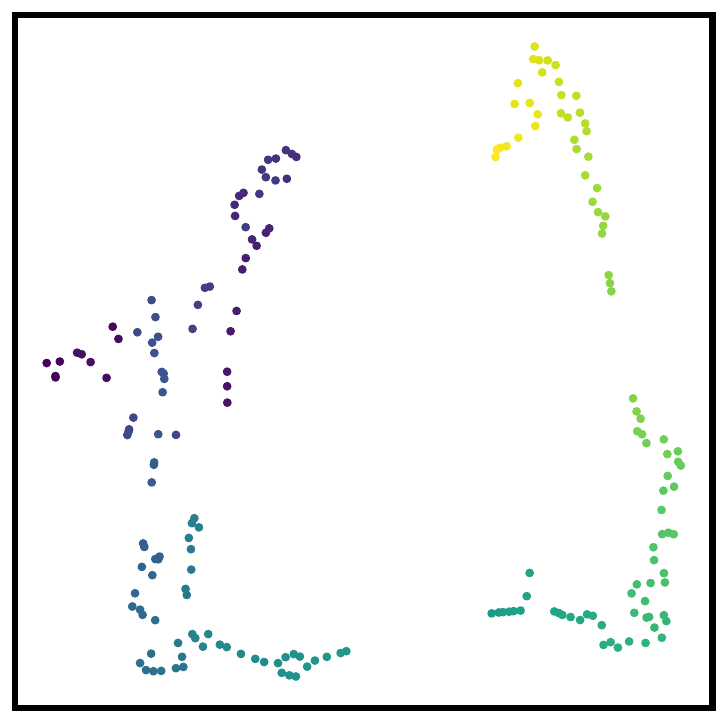}
    \caption{Look-ahead 1}
    \end{subfigure}
    \begin{subfigure}[t]{0.48\linewidth}
    \includegraphics[width=\linewidth]{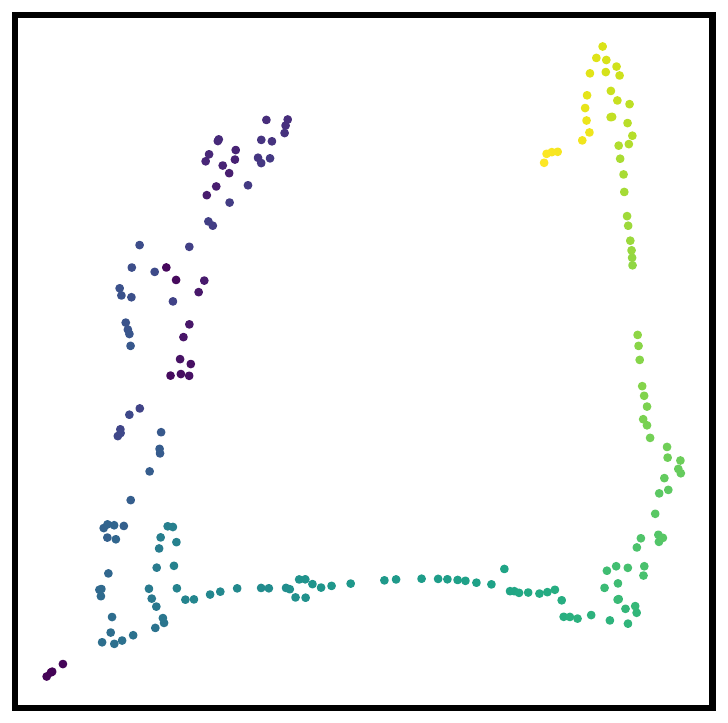}
    \caption{Look-ahead 3}
    \end{subfigure}
    \caption{Embeddings of the ionization time series with t-SNE for perplexity 30 and early exaggeration 5.}
    \label{fig:ionization_tsne_30_5}
\end{figure}

\begin{figure}
    \centering
    \captionsetup[subfigure]{aboveskip=-1pt,belowskip=-1pt}
    
    \begin{subfigure}[t]{0.48\linewidth}
    \includegraphics[width=\linewidth]{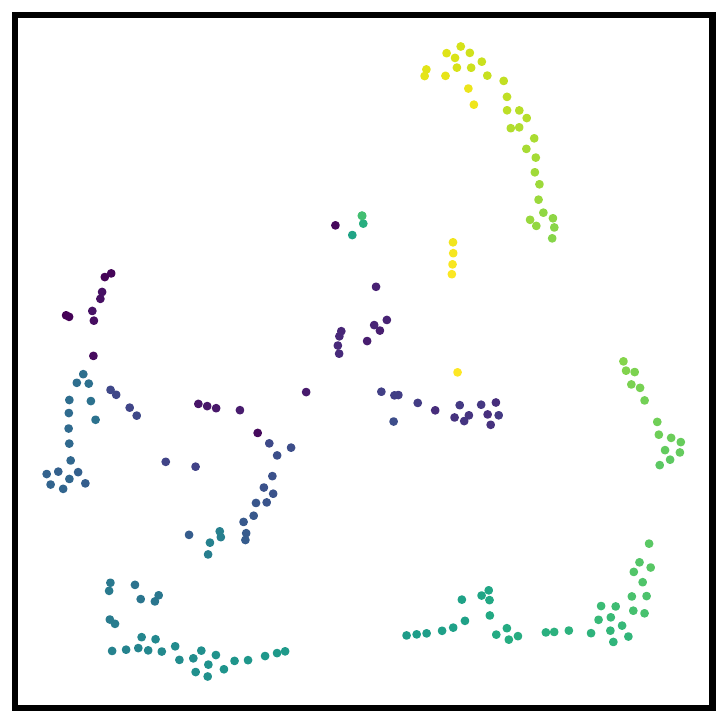}
    \caption{Merge Tree Edit Distance}
    \end{subfigure}
    \begin{subfigure}[t]{0.48\linewidth}
    \includegraphics[width=\linewidth]{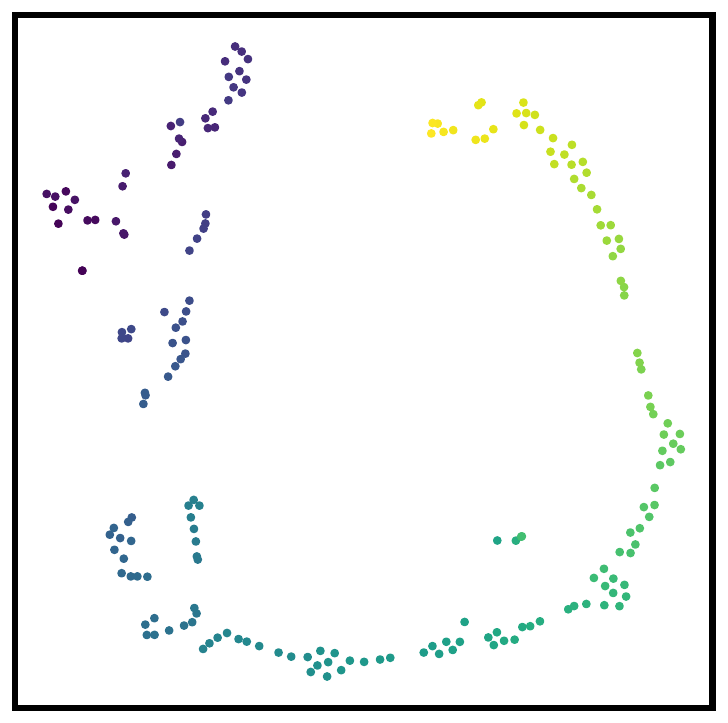}
    \caption{Path Mapping Distance}
    \end{subfigure}
    
    \begin{subfigure}[t]{0.48\linewidth}
    \includegraphics[width=\linewidth]{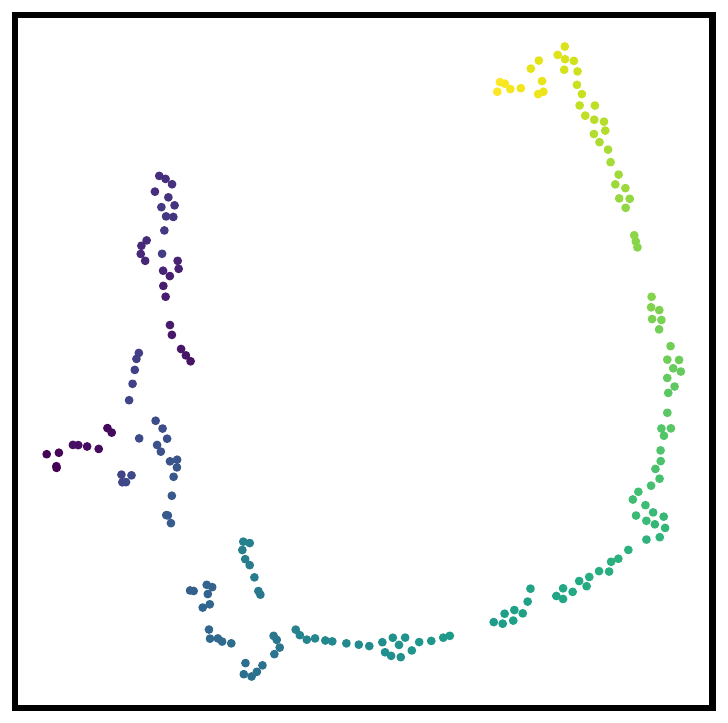}
    \caption{Look-ahead 1}
    \end{subfigure}
    \begin{subfigure}[t]{0.48\linewidth}
    \includegraphics[width=\linewidth]{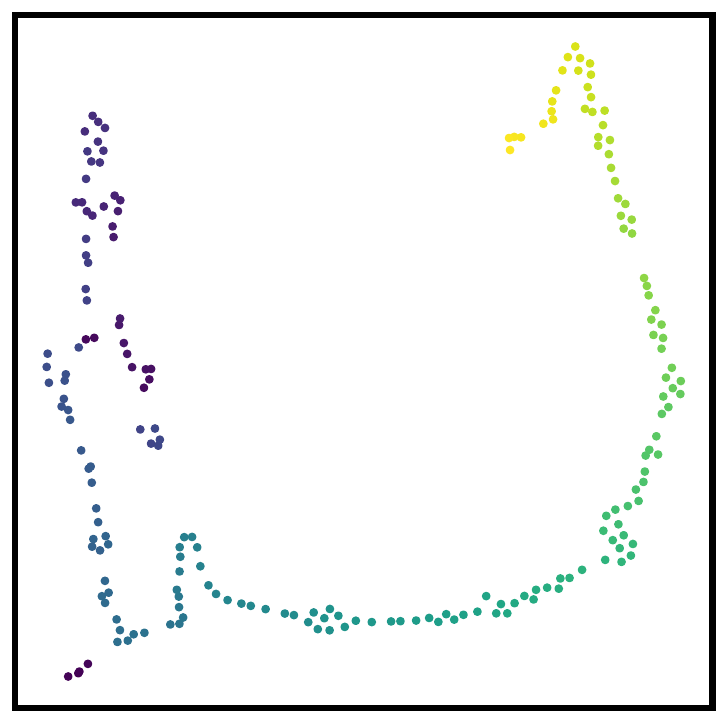}
    \caption{Look-ahead 3}
    \end{subfigure}
    \caption{Embeddings of the ionization time series with t-SNE for perplexity 15 and early exaggeration 5.}
    \label{fig:ionization_tsne_15_5}
\end{figure}

Next, we consider the t-SNE embeddings of the ionization front time series.
Two parameters mainly influence the outcome of the embedding algorithm, perplexity and early exaggeration.
For both parameters, higher values increase the tendency towards more pronounced clusters in our experiments whereas lower values favor a more path-like result.
The default parameters are 30 for perplexity and 12 for early exaggeration, for which the embeddings can be seen in Figure~\ref{fig:ionization_tsne_30_12}.
First, we look at lowering them individually: Figures~\ref{fig:ionization_tsne_15_12} and~\ref{fig:ionization_tsne_5_12} show embeddings for lower perplexity, Figure~\ref{fig:ionization_tsne_30_5} for lower early exaggeration.
We observe the described tendency towards path-like embeddings.
The same holds for decreasing both parameters, see Figure~\ref{fig:ionization_tsne_15_5}.

\begin{figure}
    \centering
    \captionsetup[subfigure]{aboveskip=-1pt,belowskip=-1pt}
    
    \begin{subfigure}[t]{0.48\linewidth}
    \includegraphics[width=\linewidth]{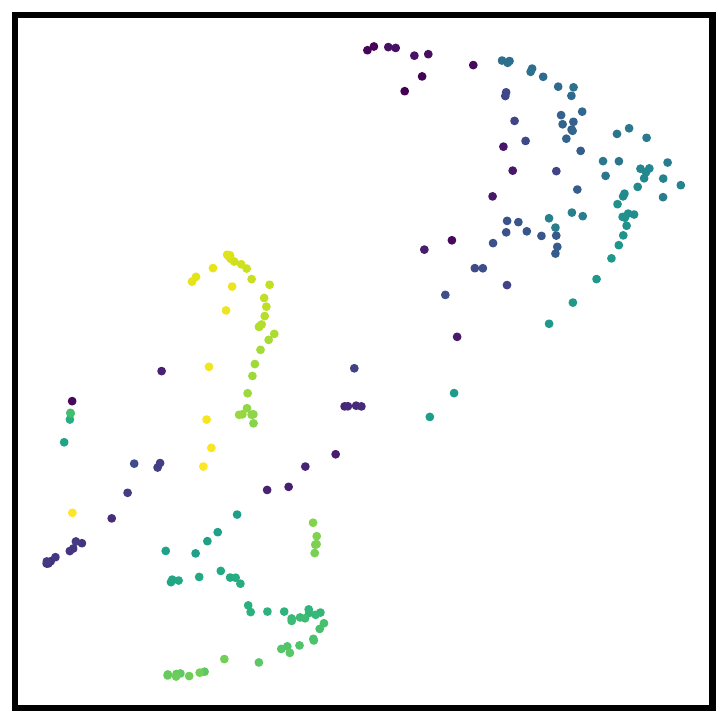}
    \caption{Merge Tree Edit Distance}
    \end{subfigure}
    \begin{subfigure}[t]{0.48\linewidth}
    \includegraphics[width=\linewidth]{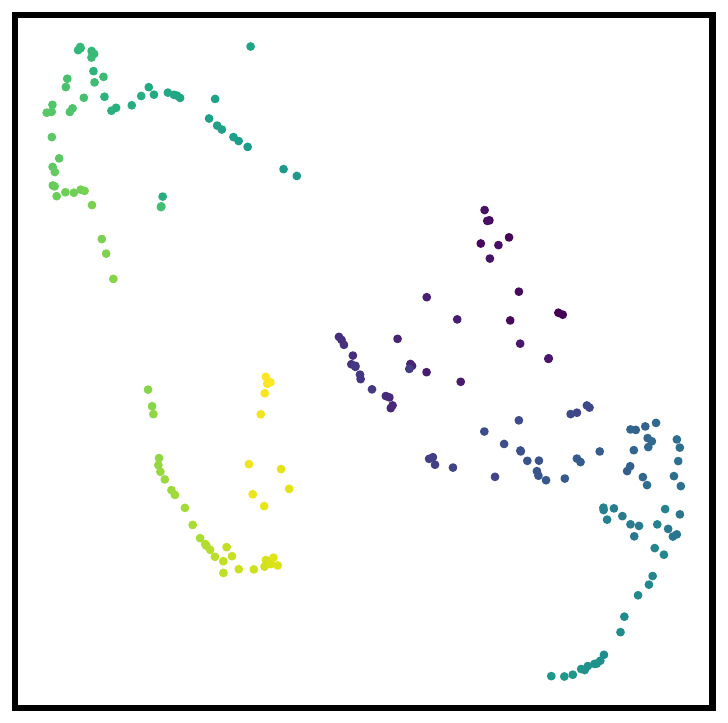}
    \caption{Path Mapping Distance}
    \end{subfigure}
    
    \begin{subfigure}[t]{0.48\linewidth}
    \includegraphics[width=\linewidth]{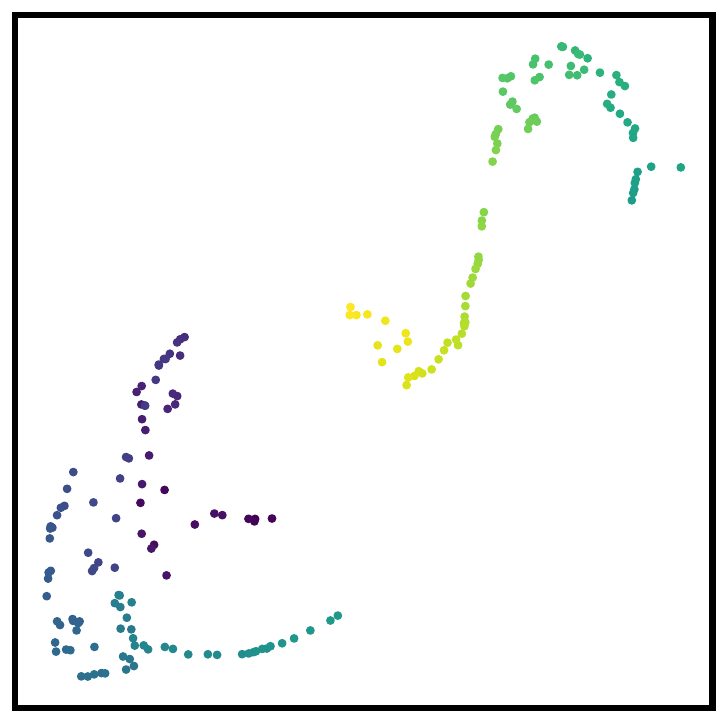}
    \caption{Look-ahead 1}
    \end{subfigure}
    \begin{subfigure}[t]{0.48\linewidth}
    \includegraphics[width=\linewidth]{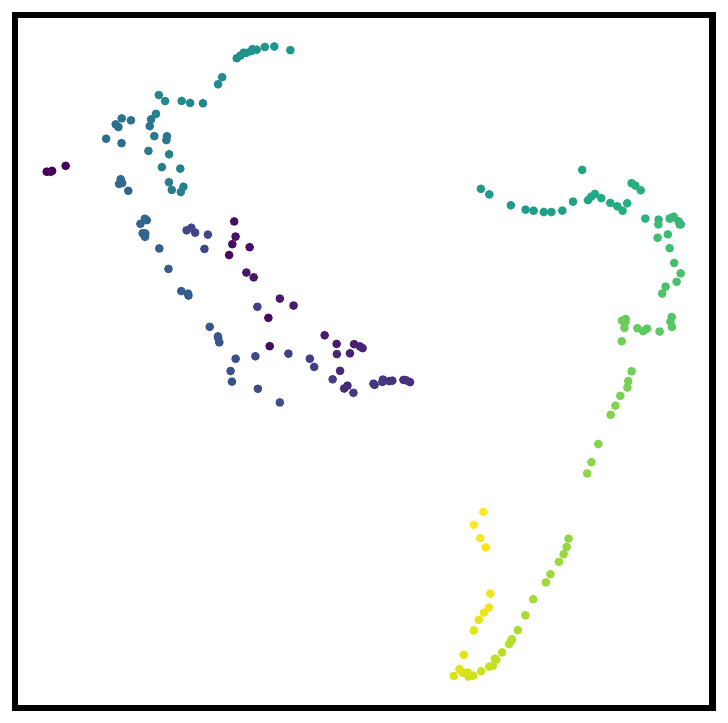}
    \caption{Look-ahead 3}
    \end{subfigure}
    \caption{Embeddings of the ionization time series with t-SNE for perplexity 60 and early exaggeration 12.}
    \label{fig:ionization_tsne_60_12}
\end{figure}

\begin{figure}
    \centering
    \captionsetup[subfigure]{aboveskip=-1pt,belowskip=-1pt}
    
    \begin{subfigure}[t]{0.48\linewidth}
    \includegraphics[width=\linewidth]{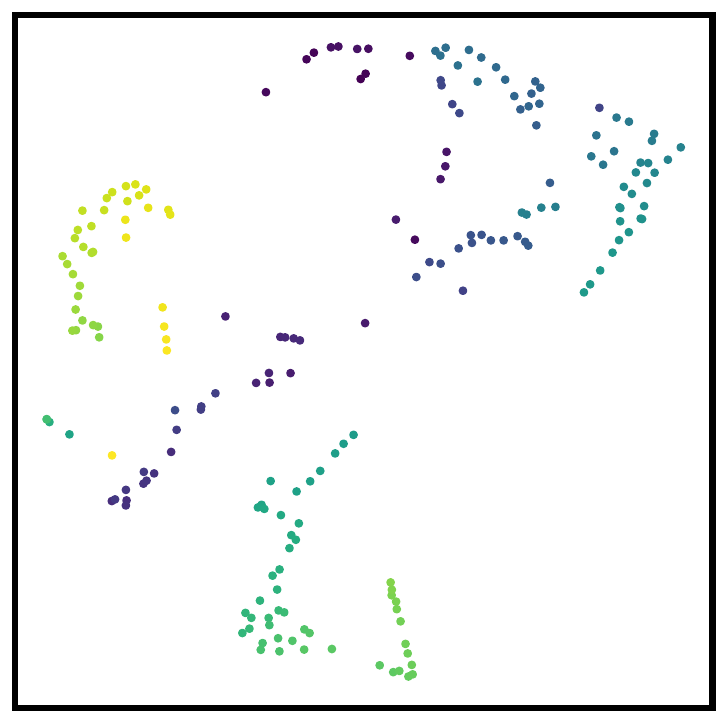}
    \caption{Merge Tree Edit Distance}
    \end{subfigure}
    \begin{subfigure}[t]{0.48\linewidth}
    \includegraphics[width=\linewidth]{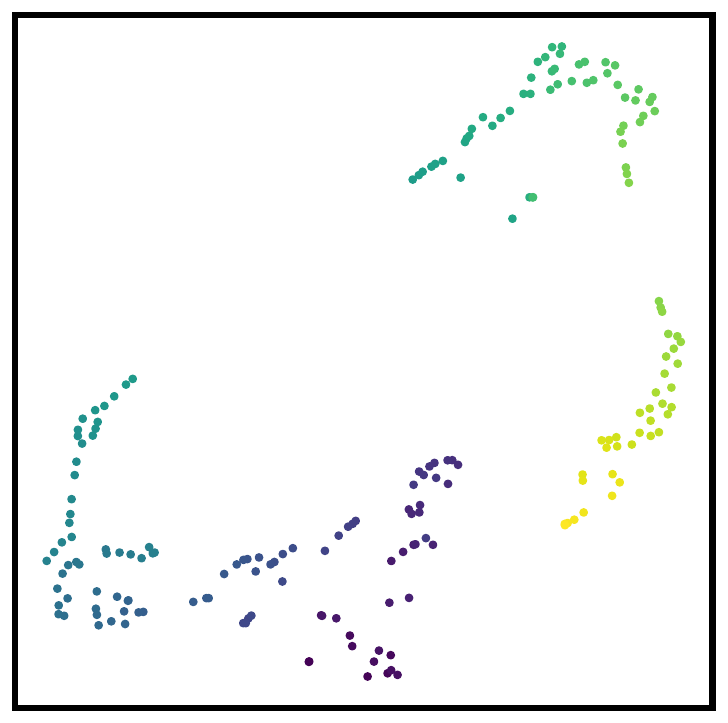}
    \caption{Path Mapping Distance}
    \end{subfigure}
    
    \begin{subfigure}[t]{0.48\linewidth}
    \includegraphics[width=\linewidth]{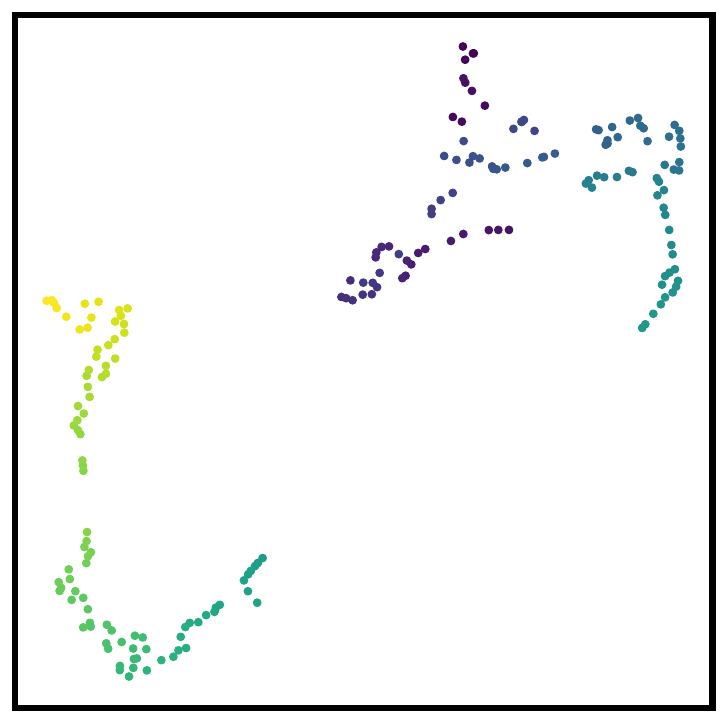}
    \caption{Look-ahead 1}
    \end{subfigure}
    \begin{subfigure}[t]{0.48\linewidth}
    \includegraphics[width=\linewidth]{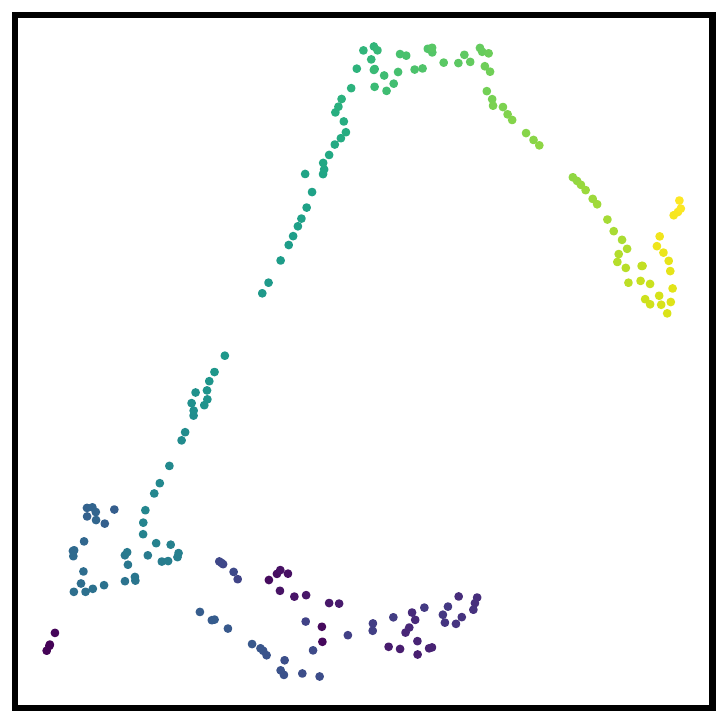}
    \caption{Look-ahead 3}
    \end{subfigure}
    \caption{Embeddings of the ionization time series with t-SNE for perplexity 30 and early exaggeration 30.}
    \label{fig:ionization_tsne_30_30}
\end{figure}

\begin{figure}
    \centering
    \captionsetup[subfigure]{aboveskip=-1pt,belowskip=-1pt}
    
    \begin{subfigure}[t]{0.48\linewidth}
    \includegraphics[width=\linewidth]{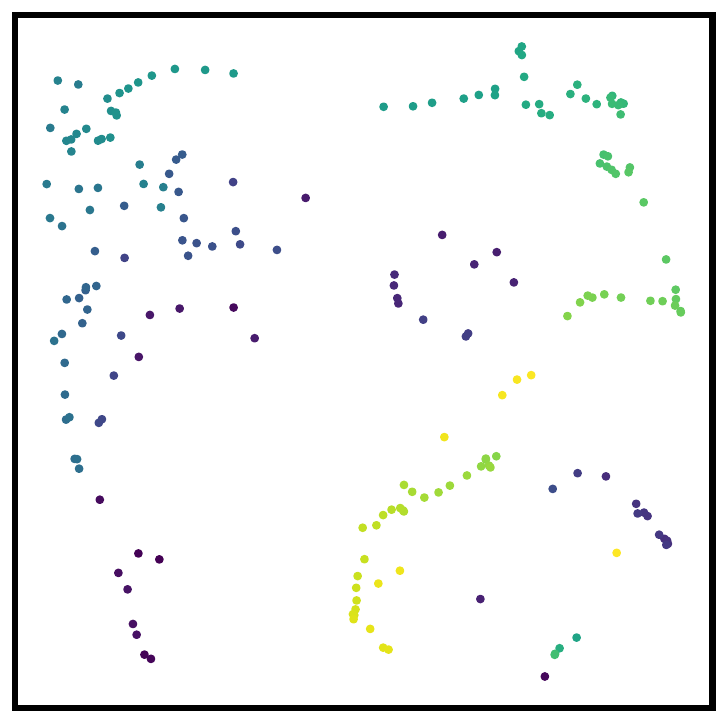}
    \caption{Merge Tree Edit Distance}
    \end{subfigure}
    \begin{subfigure}[t]{0.48\linewidth}
    \includegraphics[width=\linewidth]{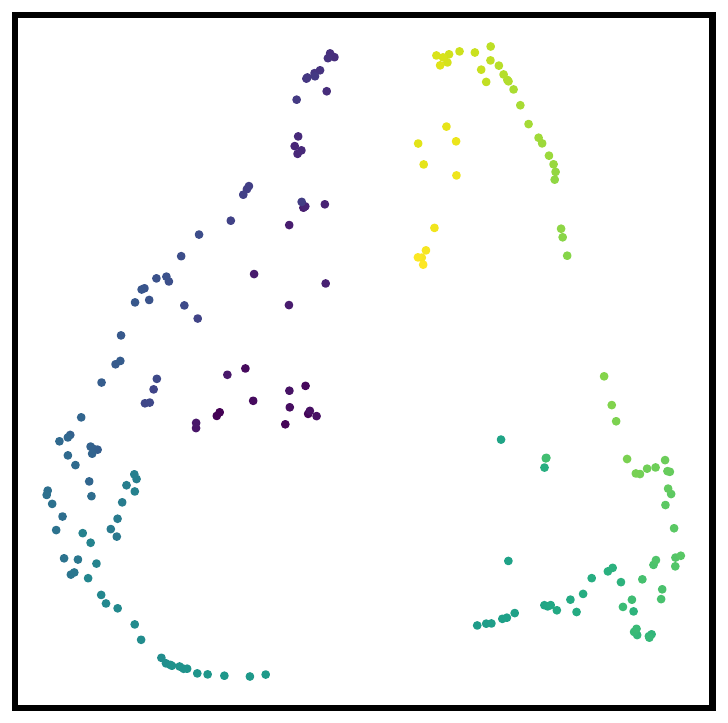}
    \caption{Path Mapping Distance}
    \end{subfigure}
    
    \begin{subfigure}[t]{0.48\linewidth}
    \includegraphics[width=\linewidth]{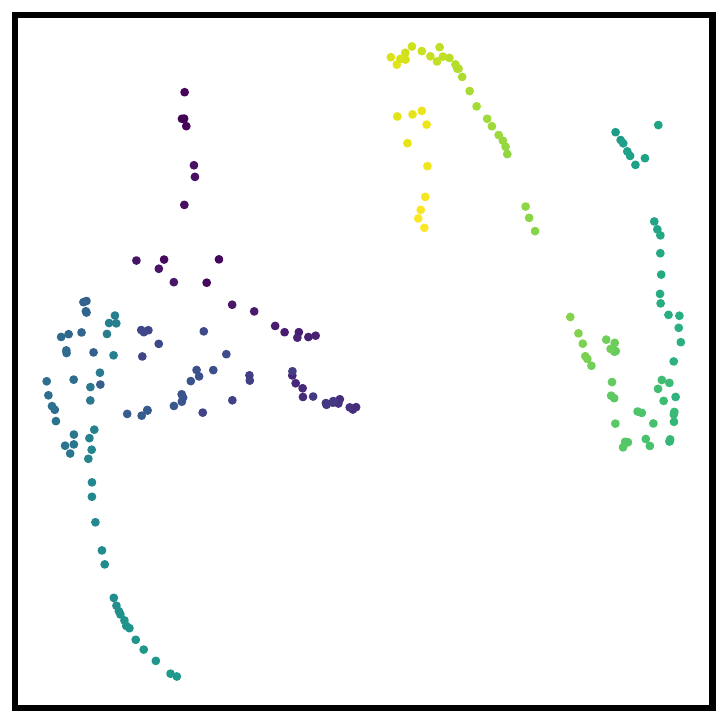}
    \caption{Look-ahead 1}
    \end{subfigure}
    \begin{subfigure}[t]{0.48\linewidth}
    \includegraphics[width=\linewidth]{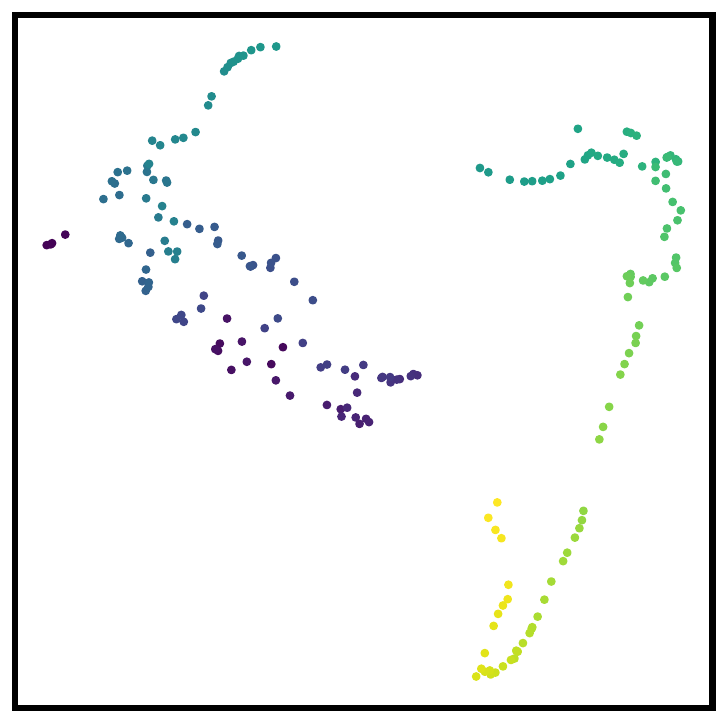}
    \caption{Look-ahead 3}
    \end{subfigure}
    \caption{Embeddings of the ionization time series with t-SNE for perplexity 60 and early exaggeration 5.}
    \label{fig:ionization_tsne_60_5}
\end{figure}

\begin{figure}
    \centering
    \captionsetup[subfigure]{aboveskip=-1pt,belowskip=-1pt}
    
    \begin{subfigure}[t]{0.48\linewidth}
    \includegraphics[width=\linewidth]{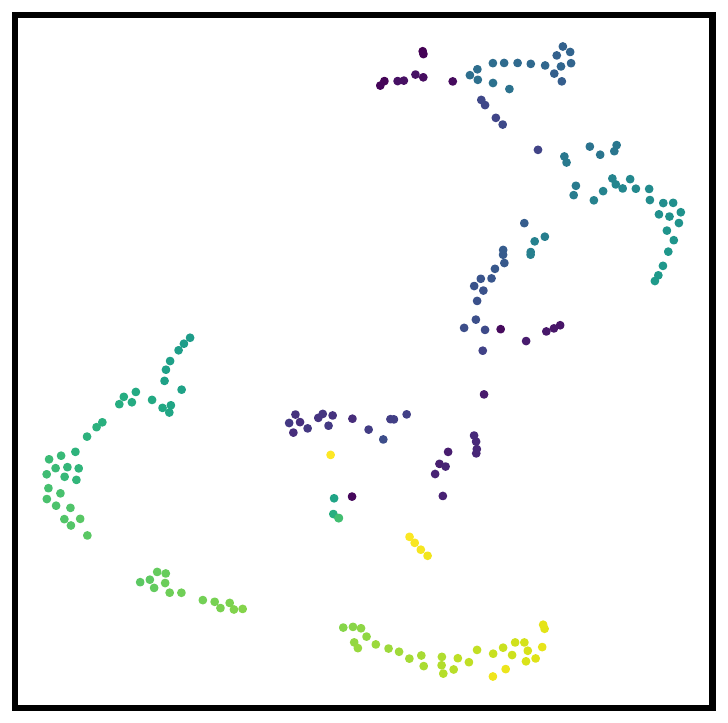}
    \caption{Merge Tree Edit Distance}
    \end{subfigure}
    \begin{subfigure}[t]{0.48\linewidth}
    \includegraphics[width=\linewidth]{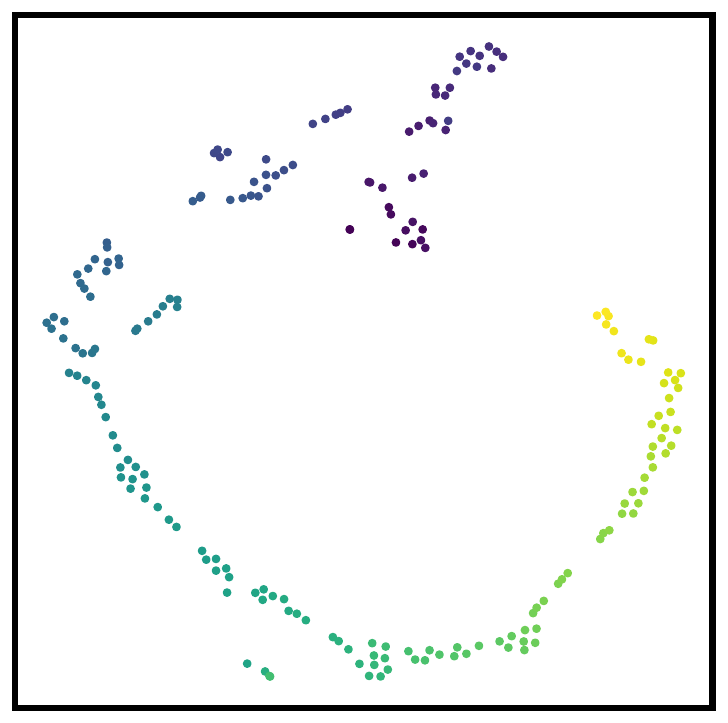}
    \caption{Path Mapping Distance}
    \end{subfigure}
    
    \begin{subfigure}[t]{0.48\linewidth}
    \includegraphics[width=\linewidth]{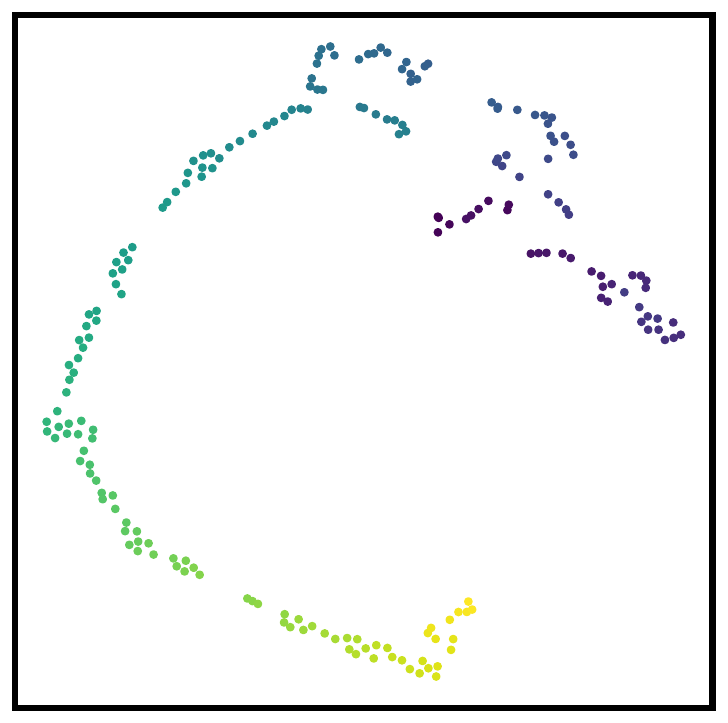}
    \caption{Look-ahead 1}
    \end{subfigure}
    \begin{subfigure}[t]{0.48\linewidth}
    \includegraphics[width=\linewidth]{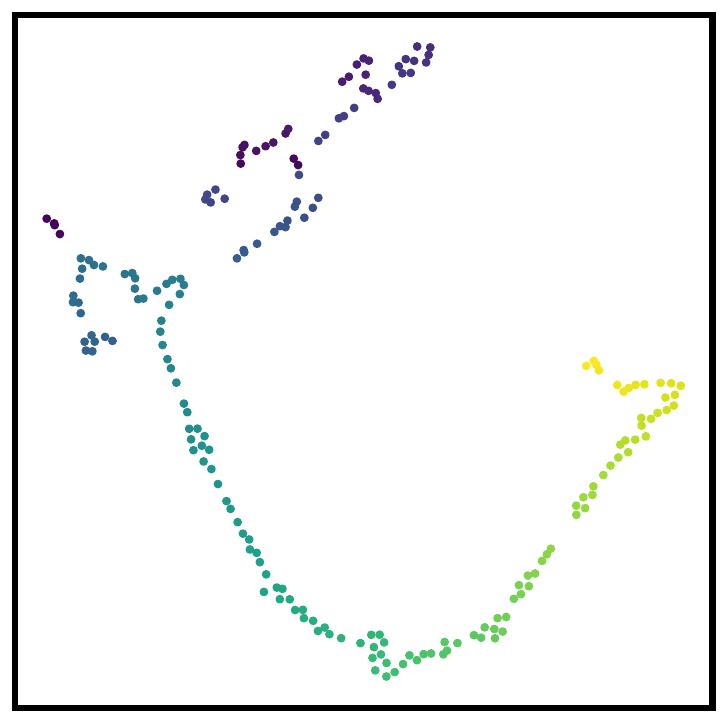}
    \caption{Look-ahead 3}
    \end{subfigure}
    \caption{Embeddings of the ionization time series with t-SNE for perplexity 15 and early exaggeration 30.}
    \label{fig:ionization_tsne_15_30}
\end{figure}

The opposite tendency holds for increasing the parameters: the results are less path like.
However, the stable distance matrices computed with the look-ahead heuristic are more robust against these effects and more frequently remain in path-like embeddings than with unstable distances.
Figures~\ref{fig:ionization_tsne_60_12} and~\ref{fig:ionization_tsne_30_30} show corresponding results.
Similar observations hold for mixing high and low values for perplexity and early exaggeration, see Figures~\ref{fig:ionization_tsne_15_30} and~\ref{fig:ionization_tsne_60_5}, though exceptions exists.
Overall, we conclude that the look-ahead heuristic gives more consistent results and overall represents an intuitive understanding of the original data better.

Note that we did not compute MDS embeddings on this dataset, as results were generally poor.

\section{Vortex Street}
\label{sec:app_vortex}

In this section, we provide additional matrices for the vortex street time series as well as a more detailed discussion of the provided 2D embeddings.

\begin{figure}
    \centering
    \captionsetup[subfigure]{aboveskip=-1pt,belowskip=-1pt}
    
    \begin{subfigure}[t]{0.9\linewidth}
    \includegraphics[width=\linewidth]{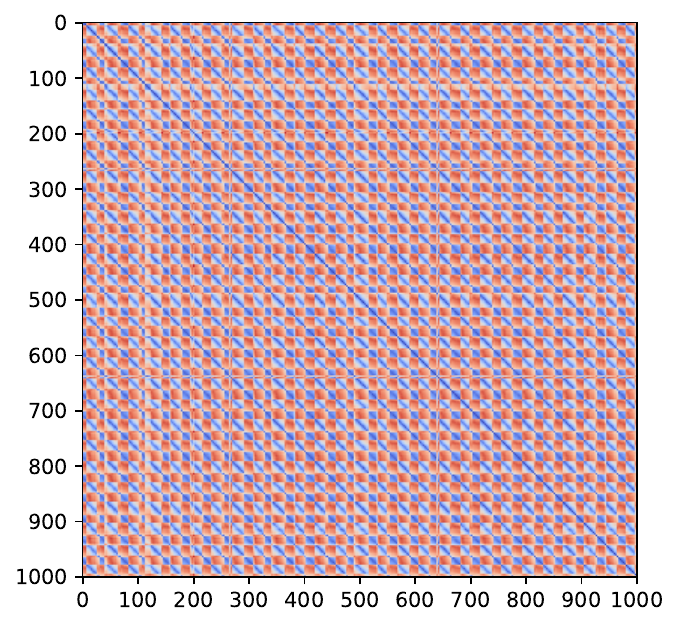}
    \caption{Look-ahead 0}
    \end{subfigure}
    
    \begin{subfigure}[t]{0.9\linewidth}
    \includegraphics[width=\linewidth]{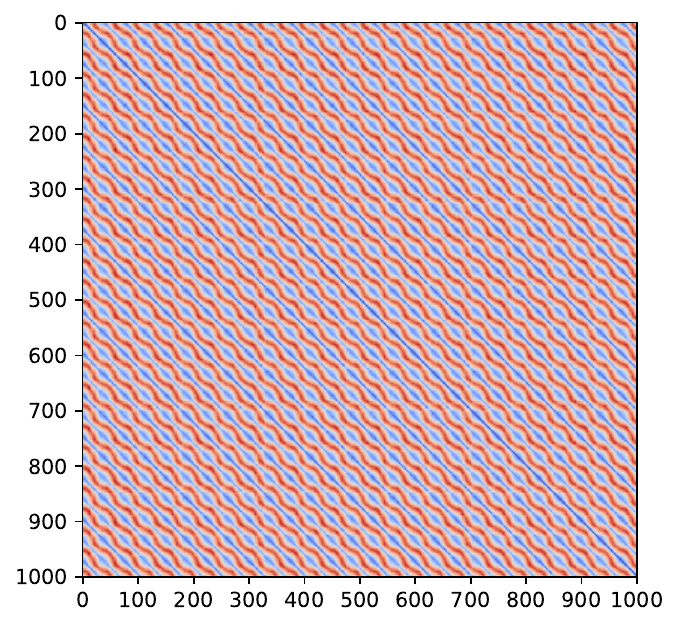}
    \caption{Look-ahead 4}
    \end{subfigure}
    \caption{Distance matrices for the complete vortex street time series with the highest and lowest look-ahead values.}
    \label{fig:vortex_mat_full}
\end{figure}

\begin{figure}
    \centering
    \captionsetup[subfigure]{aboveskip=-1pt,belowskip=-1pt}
    
    \begin{subfigure}[t]{0.49\linewidth}
    \includegraphics[width=\linewidth]{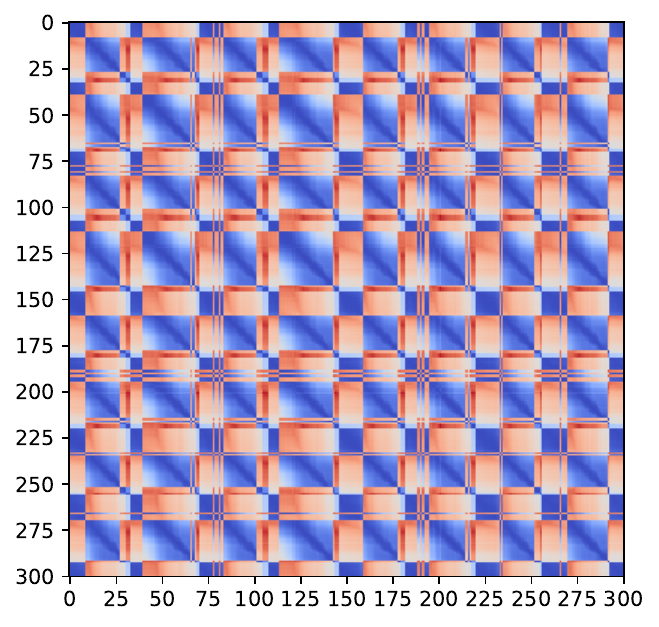}
    \caption{Merge Tree Edit Distance}
    \end{subfigure}
    \begin{subfigure}[t]{0.49\linewidth}
    \includegraphics[width=\linewidth]{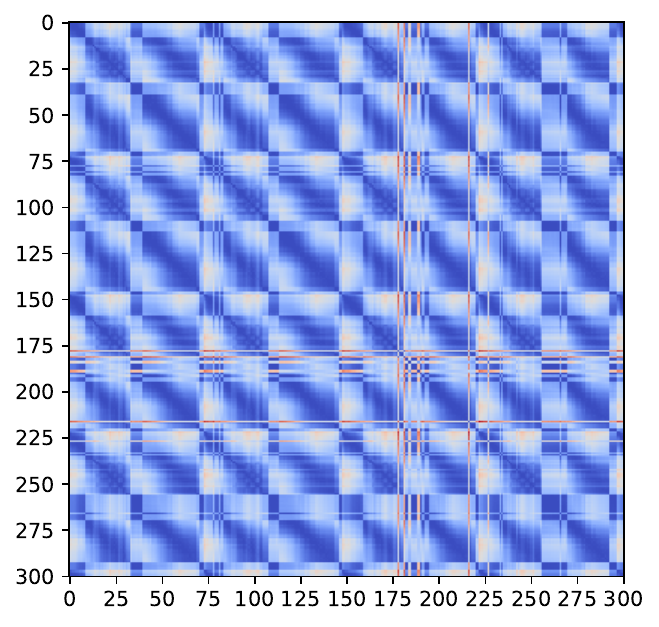}
    \caption{Merge Tree Wasserstein Distance}
    \end{subfigure}
    \caption{Distance matrices for vortex street time series using branch decomposition-based edit distances.}
    \label{fig:vortex_mat_wsd_mted}
\end{figure}

We begin with the full distance matrix for all 1000 time steps, using look-ahead~$0$ (unstable) and~$4$ (stable).
It is shown in Figure~\ref{fig:vortex_mat_full}.
The (reduced) $300\times300$ distance matrices for the original merge tree edit distance and the merge tree Wasserstein distance are shown in Figure~\ref{fig:vortex_mat_wsd_mted}.
They look very similar to the path mapping distance, however, the Wasserstein distance has more distinct outliers.

\begin{figure*}
    \centering
    \captionsetup[subfigure]{aboveskip=-1pt,belowskip=-1pt}
    
    \begin{subfigure}[t]{0.24\linewidth}
    \includegraphics[width=\linewidth]{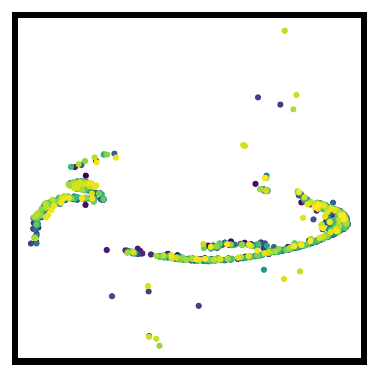}
    \caption{Wasserstein Distance}
    \end{subfigure}
    \begin{subfigure}[t]{0.24\linewidth}
    \includegraphics[width=\linewidth]{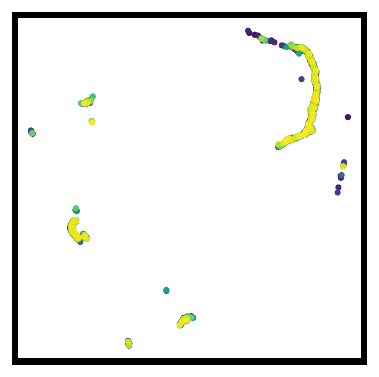}
    \caption{Merge Tree Edit Distance}
    \end{subfigure}
    \begin{subfigure}[t]{0.24\linewidth}
    \includegraphics[width=\linewidth]{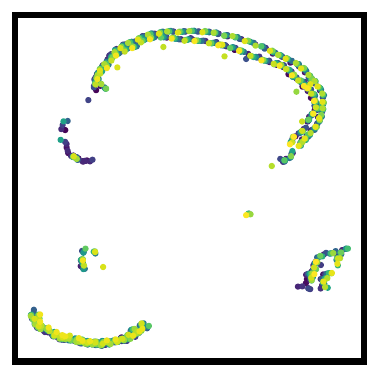}
    \caption{Path Mapping Distance}
    \end{subfigure}
    
    \begin{subfigure}[t]{0.24\linewidth}
    \includegraphics[width=\linewidth]{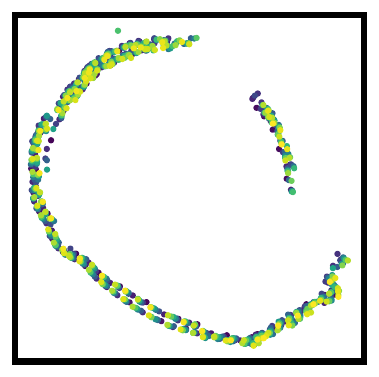}
    \caption{Look-ahead 1}
    \end{subfigure}
    \begin{subfigure}[t]{0.24\linewidth}
    \includegraphics[width=\linewidth]{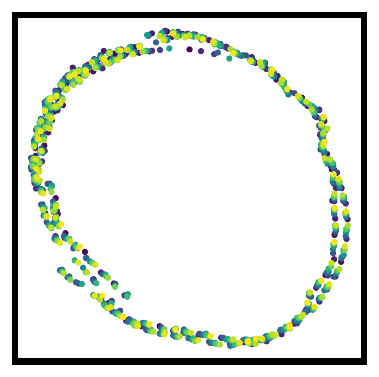}
    \caption{Look-ahead 2}
    \end{subfigure}
    \begin{subfigure}[t]{0.24\linewidth}
    \includegraphics[width=\linewidth]{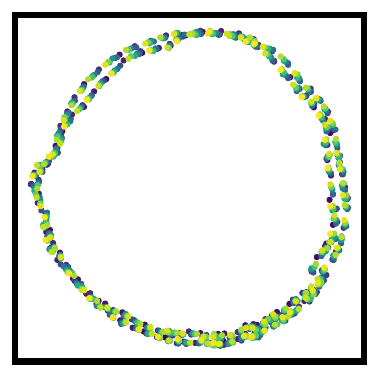}
    \caption{Look-ahead 3}
    \end{subfigure}
    \begin{subfigure}[t]{0.24\linewidth}
    \includegraphics[width=\linewidth]{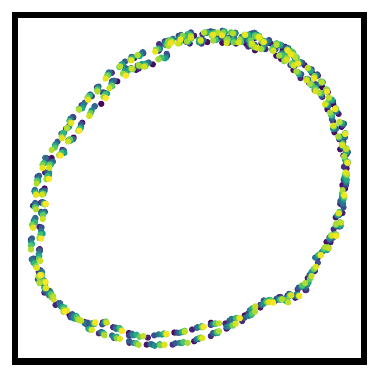}
    \caption{Look-ahead 4}
    \end{subfigure}
    \caption{MDS embeddings for the vortex street dataset with different distance metrics.}
    \label{fig:vortex_mds}
\end{figure*}

\begin{figure*}
    \centering
    \captionsetup[subfigure]{aboveskip=-1pt,belowskip=-1pt}
    
    \begin{subfigure}[t]{0.24\linewidth}
    \includegraphics[width=\linewidth]{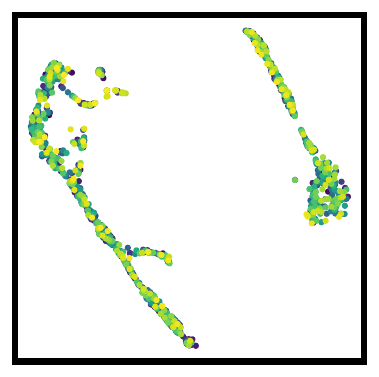}
    \caption{Wasserstein Distance}
    \end{subfigure}
    \begin{subfigure}[t]{0.24\linewidth}
    \includegraphics[width=\linewidth]{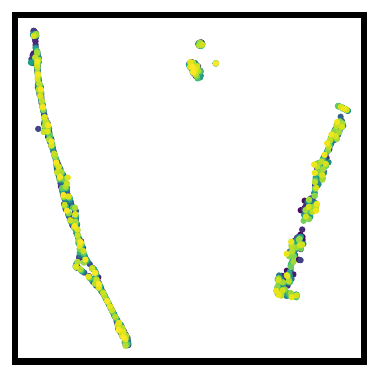}
    \caption{Merge Tree Edit Distance}
    \end{subfigure}
    \begin{subfigure}[t]{0.24\linewidth}
    \includegraphics[width=\linewidth]{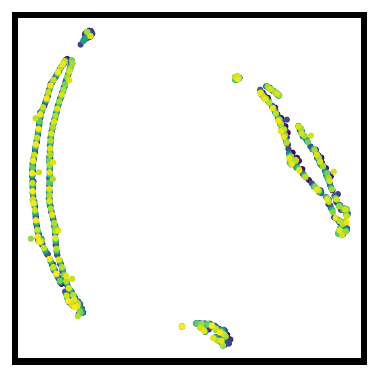}
    \caption{Path Mapping Distance}
    \end{subfigure}
    
    \begin{subfigure}[t]{0.24\linewidth}
    \includegraphics[width=\linewidth]{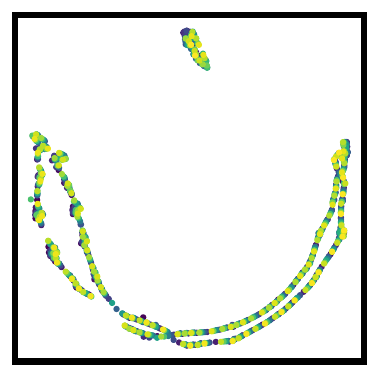}
    \caption{Look-ahead 1}
    \end{subfigure}
    \begin{subfigure}[t]{0.24\linewidth}
    \includegraphics[width=\linewidth]{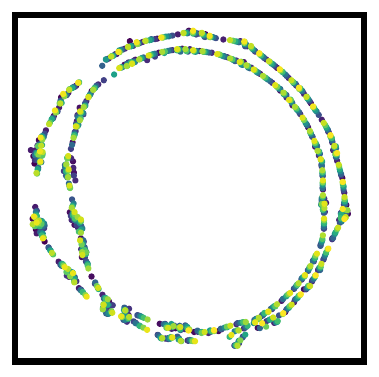}
    \caption{Look-ahead 2}
    \end{subfigure}
    \begin{subfigure}[t]{0.24\linewidth}
    \includegraphics[width=\linewidth]{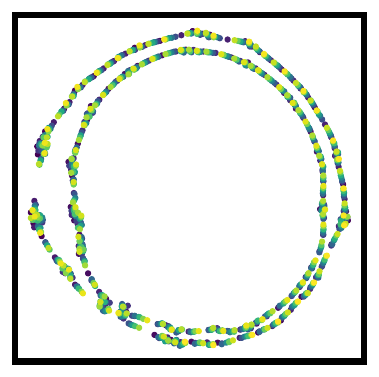}
    \caption{Look-ahead 3}
    \end{subfigure}
    \begin{subfigure}[t]{0.24\linewidth}
    \includegraphics[width=\linewidth]{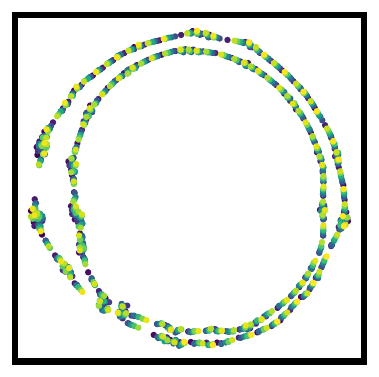}
    \caption{Look-ahead 4}
    \end{subfigure}
    \caption{t-SNE embeddings using high perplexity (60) for the vortex street dataset with different distance metrics.}
    \label{fig:vortex_tsne1}
\end{figure*}

\begin{figure*}
    \centering
    \captionsetup[subfigure]{aboveskip=-1pt,belowskip=-1pt}
    
    \begin{subfigure}[t]{0.24\linewidth}
    \includegraphics[width=\linewidth]{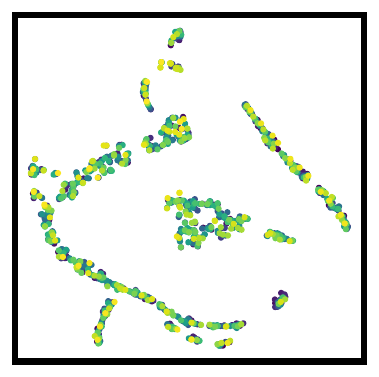}
    \caption{Wasserstein Distance}
    \end{subfigure}
    \begin{subfigure}[t]{0.24\linewidth}
    \includegraphics[width=\linewidth]{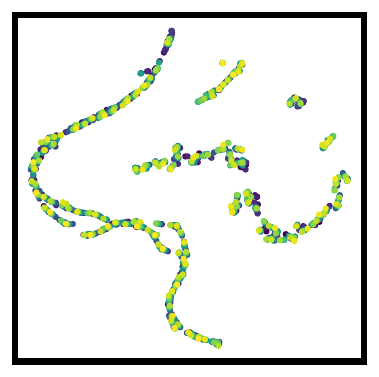}
    \caption{Merge Tree Edit Distance}
    \end{subfigure}
    \begin{subfigure}[t]{0.24\linewidth}
    \includegraphics[width=\linewidth]{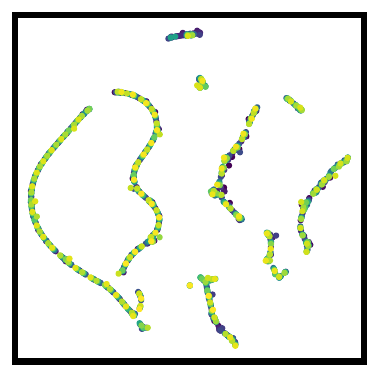}
    \caption{Path Mapping Distance}
    \end{subfigure}
    
    \begin{subfigure}[t]{0.24\linewidth}
    \includegraphics[width=\linewidth]{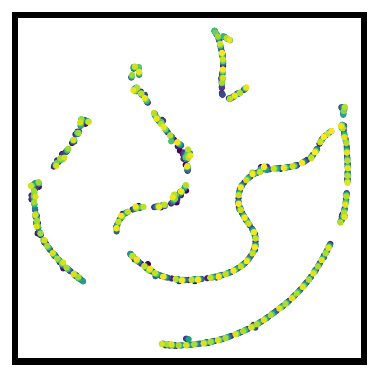}
    \caption{Look-ahead 1}
    \end{subfigure}
    \begin{subfigure}[t]{0.24\linewidth}
    \includegraphics[width=\linewidth]{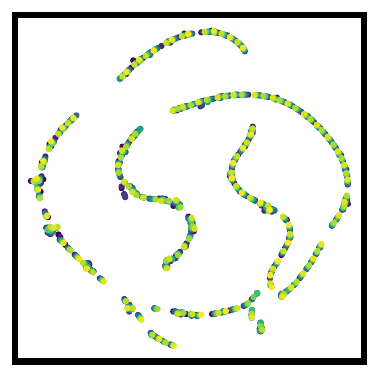}
    \caption{Look-ahead 2}
    \end{subfigure}
    \begin{subfigure}[t]{0.24\linewidth}
    \includegraphics[width=\linewidth]{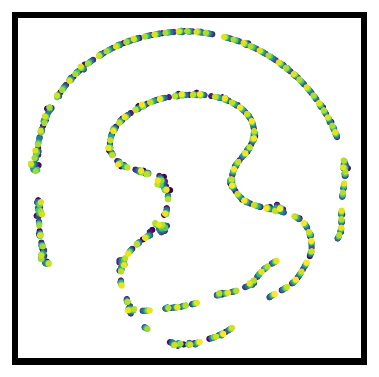}
    \caption{Look-ahead 3}
    \end{subfigure}
    \begin{subfigure}[t]{0.24\linewidth}
    \includegraphics[width=\linewidth]{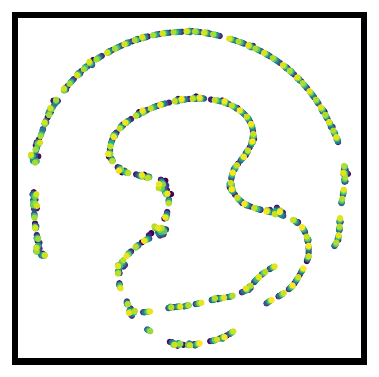}
    \caption{Look-ahead 4}
    \end{subfigure}
    \caption{t-SNE embeddings using low perplexity (15) for the vortex street dataset with different distance metrics.}
    \label{fig:vortex_tsne2}
\end{figure*}

We computed MDS and t-SNE embeddings based on all $1000\times1000$ distance matrices.
They are shown in Figures~\ref{fig:vortex_mds},~\ref{fig:vortex_tsne1} and ~\ref{fig:vortex_tsne2}.
As discussed in the main paper for the reduced time series, the improved stability yields a circle in the embedding, whereas low stability leads to clusters.
This can be observed for both MDS and t-SNE.
However, the t-SNE embeddings are susceptible to changes in the perplexity.
We show our results for high and low perplexity.
While all distance yield a clustered result with low perplexity, we can observe huge differences for high perplexity, very similar to MDS.
In contrast to perplexity, the early exaggeration parameter does not have any significant impact on this dataset.
Therefore, we left it on default.

\section{Source Code}
\label{sec:app_code}

In addition to this document, our supplementary material also contains the source code of the TTK implementation.
It is provided as a zip file containing the full TTK source code and can be compiled following the usual installation instructions given on the TTK website.
The module \texttt{ttkMergeTreeDistanceMatrix} allows to choose the path mapping distance as metric and the look-ahead can be set freely as a simple integer parameter.
We plan to properly integrate the the adapted module into the TTK upon publication. 

\clearpage

\end{document}